\documentclass[structabstract]{aa}  
\usepackage{graphicx,aalongtable}
\usepackage{txfonts,natbib}
\bibpunct{(}{)}{;}{a}{}{,}
\begin{document}
\title{CNONa and $^{12}$C/$^{13}$C in giant stars 
of 10 open clusters
\thanks{Observations collected at ESO, La Silla, Chile (programmes 
56.A-0285 and 65.L-0026A).}}

\author{R. Smiljanic\inst{1}
\and
R. Gauderon\inst{2}
\and
P. North\inst{2}
\and
B. Barbuy\inst{1}
\and
C. Charbonnel\inst{3,4}
\and
N. Mowlavi\inst{5}
}

\offprints{R. Smiljanic}

\institute{Universidade de S\~ao Paulo, IAG, Rua do 
Mat\~ao 1226, Cidade Universit\'aria, 05508-090, 
S\~ao Paulo, SP, Brazil \\  \email{[rodolfo;barbuy]@astro.iag.usp.br}
\and
Laboratoire d'astrophysique, Ecole Polytechnique Federale 
de Lausanne (EPFL) - Observatoire de Sauverny - CH 1290 Versoix, 
Switzerland \\ \email{Pierre.North@epfl.ch}
\and
Geneva Observatory, University of Geneva, Chemin des 
Maillettes 51, CH-1290 Versoix, Switzerland \\
\email{Corinne.Charbonnel@unige.ch}
\and
LATT, CNRS UMR 5572, Universit\'e de Toulouse, 
14 avenue Edouard Belin, F-31400 Toulouse Cedex 04, France
\and
Observatoire de Gen\`eve - Integral Science Data Center, 
Chemin d'Ecogia 16 - CH 1290 Versoix, Switzerland\\ 
\email{Nami.Mowlavi@unige.ch}
}

\date{Received ; accepted }

\abstract
{Evolved low-mass stars (0.8 $\leq$ M/M$_{\odot}$ $\leq$ 2.5) of a wide range of 
metallicity bear signatures of a non-standard mixing event in their surface 
abundances of Li, C, and N, and in their $^{12}$C/$^{13}$C ratio. A Na overabundance has 
also been reported in some giants of open clusters but remains controversial. 
The cause of the extra-mixing has been attributed to thermohaline convection 
that should take place after the RGB bump for low-mass stars and on the 
early-AGB for more massive objects.}
{To track the occurrence of this process over a wide mass 
range, we derive in a homogeneous way the abundances of C, N, O, and Na, 
as well as the $^{12}$C/$^{13}$C ratio in a sample of 31 giants of 10 open 
clusters with turn-off masses from 1.7 to 3.1 M$_{\odot}$. The sample includes 
red giants, clump giants, and early-AGB stars. We study the observational behavior
 of the abundances as well as the possible correlations between different elements 
and between the chemical abundances and stellar mass. }
{A model atmosphere analysis is conducted using high 
signal-to-noise ratio, high-resolution FEROS and EMMI spectra. We derive 
atmospheric parameters using \ion{Fe}{i} and \ion{Fe}{ii} lines. 
We calculate abundances for Na, C, N, and O, as well as the 
$^{12}$C/$^{13}$C ratio using spectral synthesis. 
For the elements Mg, Ca, Si, Sc, Ti, V, Cr, Co, and Ni, 
abundances are derived using equivalent widths.}
{A group of first ascent red giants with M/M$_{\odot}$ $\leq$ 2.5 exhibits 
lower [N/C] ratios than those measured in clump giants of the same 
mass range, suggesting an additional increase in the [N/C] ratio 
after the first dredge-up. The sodium abundances corrected from NLTE 
are found to be about solar. [Na/Fe] shows a slight increase 
of 0.10 dex as a function of stellar mass in the 1.8 to 3.2 M$_{\odot}$ 
range covered by our sample, in agreement with standard 
first dredge-up predictions. Our results do not support previous 
claims of sodium  overabundances as high as +0.60 dex. An 
anti-correlation between $^{12}$C/$^{13}$C and 
turn-off mass is identified and interpreted as being caused by a post-bump thermohaline 
mixing. Moreover, we find low $^{12}$C/$^{13}$C ratios in a few 
intermediate-mass early-AGB stars, confirming that an 
extra-mixing process also operates in stars that do not experienced 
the RGB bump. In this case, the extra-mixing possibly acts on the 
early-AGB, in agreement with theoretical expectations for thermohaline mixing.}
   {}
\maketitle
%

\section{Introduction}

In the standard model of stellar evolution, convection is the only mechanism 
that can drive mixing in stellar interiors. In this context, the only 
expected mixing episode between the main sequence (MS) and the tip of the 
red giant branch (RGB) is the so-called first dredge-up \citep{I65}, a deep 
convective envelope that transports nuclear-processed material to the surface, 
as the star approaches and begins to climb the RGB. 

During the MS, hydrogen burns by means of the pp-chain and the CNO-cycle. 
Proton-capture cycles that require higher temperatures, i.e., the NeNa and MgAl 
cycles, remain ineffective \citep{WC04}, except in stars more massive 
than $\sim 25$~M$_{\odot}$ \citep{Dec07,PCI07}. Thus, in low-mass stars, the 
material mixed by the first dredge-up increases the photospheric abundances 
of $^{3}$He, $^{13}$C, and $^{14}$N and decreases that of $^{12}$C as a function 
of the initial stellar mass and metallicity \citep{Ch94}. The surface abundance 
of these elements can thus be used to test our understanding of the stellar 
evolutionary mixing processes.

The observed surface abundances of subgiants and low-luminosity RGB stars 
have been shown to agree with values predicted by first dredge-up 
models \citep{CBW98,Gr00}. However, accumulating observational 
evidence indicates a further increase in N and decrease in Li, 
C, and $^{12}$C/$^{13}$C for low-mass RGB stars just after the luminosity bump. 
This extra-mixing phenomenon has been detected in giants of both open 
\citep{G89,GB91,L94,T00,T05} and  globular clusters 
\citep[and references therein]{Sh03,Pi03,RBdL07}, as well as in field stars 
\citep{SPVB86,CBW98,Gr00}, including extremely metal-poor giants \citep{Coh06,Sp06}, 
and in extragalactic systems such as the LMC \citep{Smith02} and Sculptor 
\citep{Gei05}. These observations suggest that we are witnessing the 
effects of a universal process that is independent 
of environment, able to operate at all metallicities (although possibly with 
different efficiencies), and connected to the luminosity bump \citep{CN98}. 

Effects connected to rotation have long been suspected as being the origin of the 
extra-mixing, as first suggested by \citet{SM79}. In particular, the most 
probable mechanism has been proposed to be the interaction between meridional 
circulation and turbulence induced by rotation, as derived by \citet{Z92} \citep[see also][]{Ch95}. 
However, the maximum-rotation induced mixing scenario developed by \citet{CPT05}, 
as well as the evolutionary models for low-mass, low-metallicity stars of 
\citet{PC06}, which take into account the transport of both angular 
momentum and chemicals induced by meridional circulation and shear 
turbulence self-consistently, show that these processes alone cannot 
explain the observations.

The 3D modeling of a low-mass RGB star by \citet{Dea06} provided 
fresh insight into the physical mechanism involved. Based on these 
simulations, \citet{EDL06} suggested that the process resbonsible was a 
molecular-weight inversion created by the $^{3}$He($^{3}$He,2p)$^{4}$He 
reaction in the external part of the hydrogen-burning shell. 
\citet{CZ07} identified the related transport mechanism as the double 
diffusive instability often referred to as thermohaline convection 
\citep{St60,Ul72,Kip80}, and showed it to be able to reproduce the 
observed Li, C, and N abundances, as well as the carbon isotopic 
ratio in RGB stars after the bump, while simultaneously destroying most 
of the $^3$He produced on the MS.

In addition to the extra-mixing event in low-mass stars at the RGB bump, 
\citet{ChB00} suggested a possible extra-mixing episode in intermediate-mass 
stars (2.5 $\leq$ M/M$_{\odot}$ $\leq$ 5.0) that undergo the equivalent of 
the bump only during the early-AGB phase, after He-core exhaustion. As 
discussed in \citet{ChB00}, these two instances of extra-mixing would be 
connected to the nature of the lithium-rich giants. \citet{CaLan08} 
reported that thermohaline mixing can indeed be present during core 
helium-burning and beyond in stars that still have a $^3$He reservoir.

On the observational side, it has long been discussed whether RGB 
extra-mixing could modify the abundances of heavier elements, namely 
Na, whose abundance in red giants has received considerable attention 
\citep[and references therein]{Ham00,Jac07,Ses08}. Although 
some works have detected a sodium overabundance, some controversy still exists. For example, 
while \citet{T00} found giants of M67 to be sodium-enriched, \citet{R06} did not 
find these overabundances and, moreover, showed unevolved stars in this same cluster 
to have the same sodium abundance as the evolved stars.

Additionally, few works in the literature determined carbon isotopic ratios 
in giant stars of open clusters\citep{G89,GB91,L94,T00,T05}. Only 
the last three of these studies also derived abundances of C, N, O, and Na. Obviously 
the simultaneous determination of all these elements in a homogeneous analysis is 
important in constraining the mixing mechanisms as well as their possible 
dependence on stellar mass. \citet{G89} and \citet{GB91} found that giant stars in 
open clusters with turn-off masses lower than 2.2M$_{\odot}$ showed a 
decreasing carbon isotopic ratio with decreasing turn-off mass.
On the other hand, \citet{L94} also found some stars with low carbon isotopic ratio 
among clusters with turn-off masses higher than 2.2M$_{\odot}$. The latter stars might 
be connected to the extra-mixing during the early-AGB, as suggested by \citet{ChB00}. 
\citet{T00} found a small difference between the isotopic ratios of clump star and 
red giants in M67, although a similar difference was not found in stars of 
NGC 7789 by \citet{T05}.

It is clear that much work is required both on the theoretical and 
observational sides to improve our knowledge of the mixing processes in low- 
and intermediate-mass giant stars. In the present paper, we increase 
significantly the number of giants in Galactic open clusters analyzed so far. 
We derive abundances of several elements, in particular C, N, O, and Na, as 
well as the $^{12}$C/$^{13}$C ratio for a sample of clump and red giants of 10 open 
clusters. In Sect.\@ \ref{sec:obs}, the observations are described. In Sect.\@ \ref{sec:par}, 
we discuss the determination of the atmospheric parameters, and in Sect.\@ \ref{sec:abu}, 
we present the abundances. In Sect.\@ \ref{sec:evo}, we 
determine the evolutionary state of each sample star, and in Sect.\@ \ref{sec:dis}, we discuss 
the results and their implications. Our conclusions are drawn in Sect.\@ \ref{sec:con}.


\section{Observations}\label{sec:obs}

\begin{table*}
\caption{Log book of the observations.}\label{tab:log}
\centering
\begin{tabular}{ccrrrrrrr}
\hline \\[-4pt]
\multicolumn{1}{c}{Star}&\multicolumn{1}{c}{DM or}&\multicolumn{1}{c}{V}
&\multicolumn{1}{c}{[B-V]}&\multicolumn{1}{c}{B-V}
&\multicolumn{1}{c}{S/N}&\multicolumn{1}{c}{around}& JD & t$_{\rm exp}$ \\
&Eggen No & & & &  & [\AA] & -2450000 &  [s]    \\
\hline
IC 2714\_05 &                 & 11.046 & 0.582 & 1.263 & 184 & 6701 & 1730.511 & 5000 \\
IC 4756\_12 & +05$^\circ$ 3805 & 9.473  & 0.500 & 1.030 & 174 & 6701 & 1731.751 & 1800 \\
IC 4756\_14 & +05$^\circ$ 3808 & 8.813  & 0.737 & 0.860 & 197 & 6701 & 1732.750 & 1800 \\
IC 4756\_28 & +05$^\circ$ 3818 & 8.970  &       & 1.360 & 166 & 6701 & 1730.703 & 1500 \\
IC 4756\_38 & +05$^\circ$ 3829 & 9.756  & 0.416 & 1.100 & 237 & 6701 & 1730.731 & 2700 \\
IC 4756\_69 & +05$^\circ$ 3850 & 9.201  & 0.366 & 1.060 & 271 & 6701 & 1730.755 & 1800 \\
NGC 2360\_7   &   8             & 11.087 & 0.294 & 1.000 & 89  & 6600 &  30.741 & 1800 \\ 
NGC 2360\_50  &  67             & 11.082 & 0.311 & 1.030 & 108 & 6600 &  30.767 & 2700 \\
NGC 2360\_62  &  81             & 11.272 & 0.251 & 0.940 &  73 & 6600 &  31.730 & 3600 \\
NGC 2360\_86  & 110             & 10.787 & 0.309 & 1.020 & 181 & 6600 &  31.822 & 3600 \\
NGC 2447\_28  & -23$^\circ$ 6102 & 9.849 & 0.226 & 0.930  & 170 & 6600 &  30.847 & 1800 \\
NGC 2447\_34  &                 & 10.123 & 0.197 & 0.900  & 236 & 6600 &  32.755  & 3000 \\
NGC 2447\_41  &                 & 10.031 & 0.204 & 0.935 & 226 & 6600 &  32.840  & 2700 \\
NGC 3532\_19  &-58$^\circ$ 3090  & 7.711 & 0.241  & 0.962 & 214 & 6701 & 1730.462 & 600 \\
NGC 3532\_100 &-58$^\circ$ 3092  & 7.483 & 0.404  & 1.098 & 171 & 6701 & 1730.474 & 600 \\
NGC 3532\_122 &-58$^\circ$ 3077  & 8.161 & 0.234  & 0.934 & 371 & 6701 & 1731.464 & 900 \\
NGC 3532\_596 &-58$^\circ$ 2968  & 7.930 &        & 0.991 & 367 & 6701 & 1732.463 & 800 \\
NGC 3532\_670 &-57$^\circ$ 4320  & 7.042 &        & 1.340 & 149 & 6701 & 1732.473 & 500 \\
NGC 3680\_13  &-42$^\circ$ 6963  &10.824 & 0.491  & 1.150 & 200 & 6701 & 1731.503 & 5000 \\
NGC 5822\_01  &                 & 9.061 &  0.653 & 1.286 & 126 & 6701 & 1731.544 & 1500 \\
NGC 5822\_201 &                 &10.242 & 0.341  & 1.052 & 175 & 6701 & 1730.567 & 3600 \\
NGC 5822\_240 & 133519          & 9.468 & 0.728  & 1.336 & 128 & 6701 & 1731.569 & 2100 \\
NGC 5822\_316 &                 &10.455 & 0.375  & 1.031 & 175 & 6701 & 1732.617 & 2700 \\
NGC 5822\_443 &                 & 9.720 &        & 1.220 & 177 & 6701 & 1731.600 & 2700 \\
NGC 6134\_30 & 8320-1928-1     &11.840 &        & 1.270 &  200 & 6701 & 1730.646 & 7200 \\
NGC 6134\_99  & 8320-0960-1     &11.633 &        & 1.357 & 168 & 6701 & 1732.673 & 6300 \\
NGC 6134\_202 & 8320-0965-1     &11.619 &        & 1.464 & 146 & 6701 & 1731.690 & 7200 \\
NGC 6281\_03  & 322660          & 7.959 & 0.436  & 1.115 & 236 & 6701 & 1731.627 & 900 \\
NGC 6281\_04  & 322658          & 8.126 & 0.457  & 1.133 & 232 & 6701 & 1731.640 & 900 \\
NGC 6633\_078 & 170053          & 7.304 & 0.845  & 1.430 & 157 & 6701 & 1732.717 & 600 \\
NGC 6633\_100 & 170174          & 8.307 & 0.434  & 1.110 & 194 & 6701 & 1732.731 & 1000 \\
\hline
\end{tabular}
\end{table*}

 Observations of 24 giants were conducted with
 the FEROS spectrograph \citep{Kau99} at the ESO 1.52m telescope at La
 Silla (Chile). FEROS is a fiber-fed echelle spectrograph that
 provides a full wavelength coverage of {$\lambda\lambda$} 3500--9200
 {\rm \AA} \,over 39 orders at a resolving power of R = 48\,000. All spectra
 were reduced using the FEROS pipeline software. Typical signal-to-noise ratios (S/N) 
ranged between 125 and 370 at 6700 {\rm \AA}.

 We also reanalyzed spectra of the seven red giants of NGC 2360
 and NGC 2447 first analyzed by \citet{Ham00}. These spectra were
 obtained in 1995 using the EMMI spectrograph attached to the ESO NTT
 3.5m telescope at La Silla. The spectra have a wavelength coverage of
 $\lambda\lambda$ 4050--6650 \AA\@ with R = 28\,000. The S/N varies between
 73 to 236 at 6600 {\rm \AA}. The log book of the observations is given in 
Table \ref{tab:log}. The table includes the $V$ magnitude of the Geneva photometry 
when available, and from $UBV$ photometry otherwise, the $[B-V]$ index of the Geneva 
photometry, the $B-V$ of the Johnson $UBV$ photometry, and the signal-to-noise ratio per 
pixel of the extracted spectra for the specified wavelength. The adopted 
numbering follows the $WEBDA$ database\footnote{The WEBDA database is a large database 
for stars in Galactic open clusters developed by Jean-Claude Mermilliod and now
 maintained by Ernst Paunzen of the Institute of Astronomy of the 
 University of Vienna. The database can be accessed in the internet at
 the address: http://www.univie.ac.at/webda/}. We also give 
the HD, DM or Tycho-2 identifications when available, or numbers in the 
system of \citep{E68} for NGC 2360. The Julian Dates refer to the 
middle of the exposure. The exposure times are also listed.

\subsection{Sample clusters}

\begin{figure*}
\centering
\includegraphics[width=13cm,angle=-90]{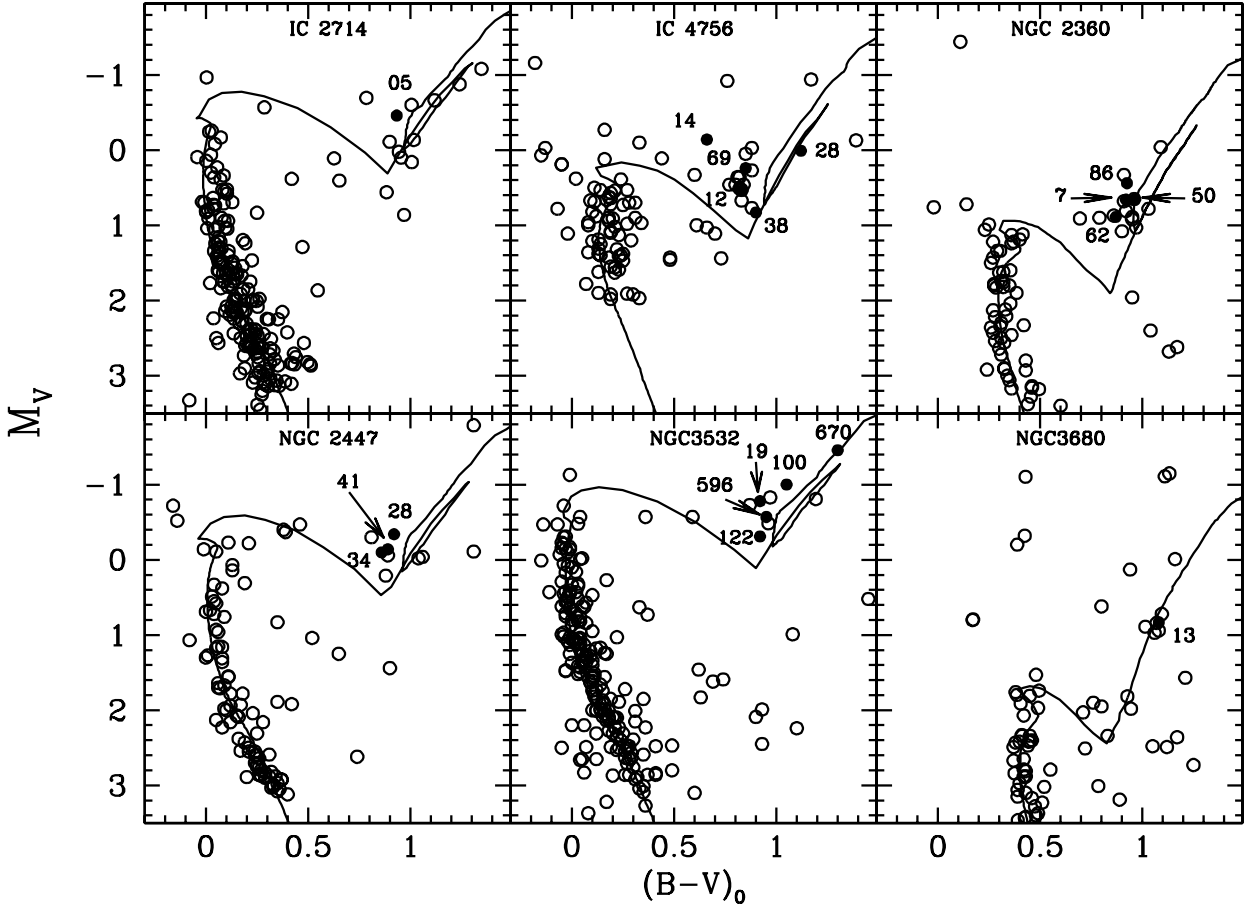}
\caption{The fitting of the color-magnitude diagrams of the clusters 
IC 2714, IC 4756, NGC 2360, NGC 2447, NGC 3532, and NGC 3680 with the 
isochrones by \citet{Sch92}, used to determine the turn-off mass of the 
clusters (except for NGC 2360 and NGC 2447, see text). The observed 
stars are shown as full circles and are identified by their numbers. The 
parameters adopted for the fittings are the ones listed in Table\@ \ref{tab:opc}.}
\label{fig:iso}
\end{figure*}

\begin{figure*}
\centering
\includegraphics[width=13cm,angle=-90]{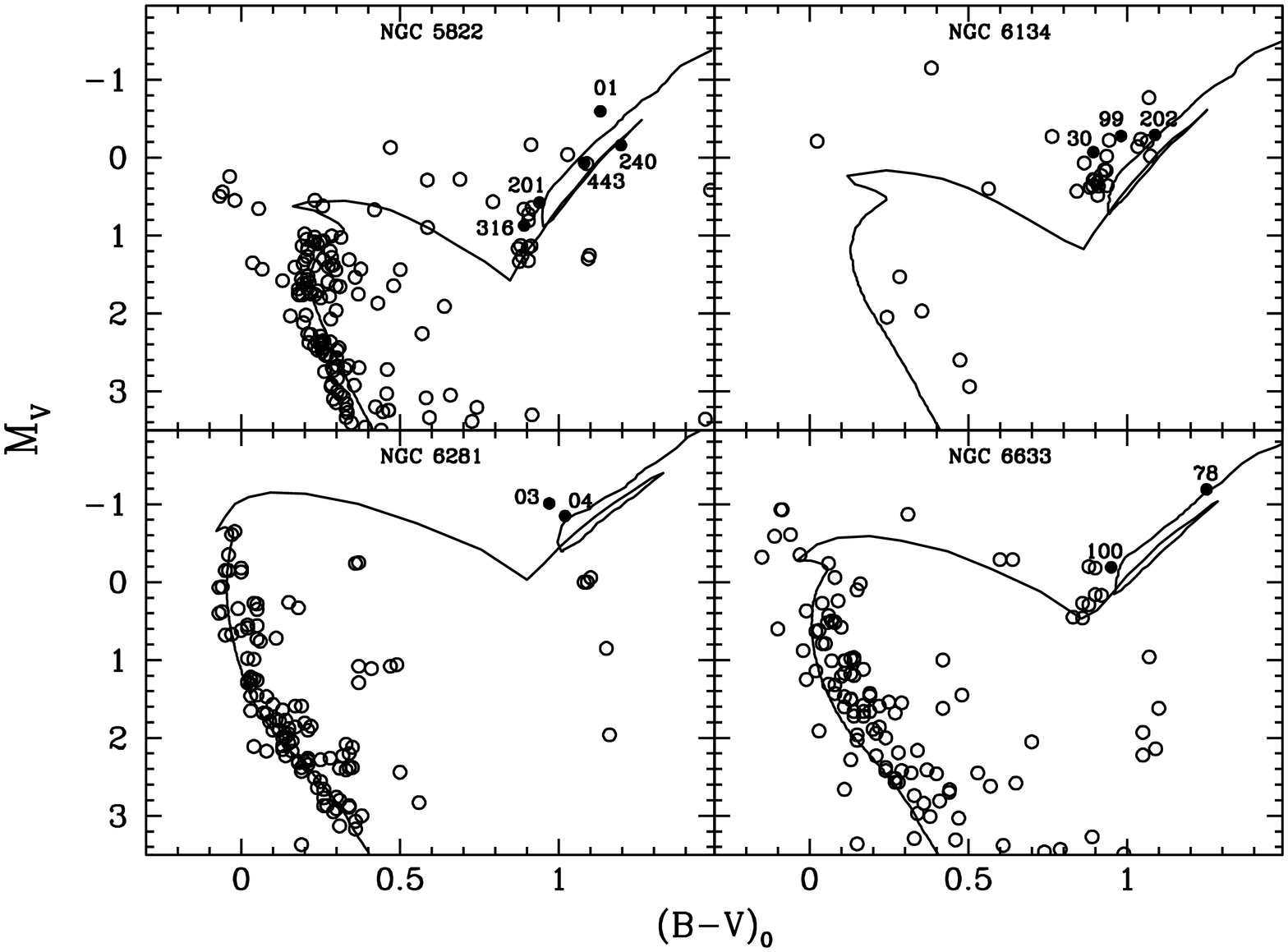}
\caption{The fitting of the color-magnitude diagrams of the clusters 
NGC 5822, NGC 6134, NGC 6281, and NGC 6633 with the 
isochrones by \citet{Sch92}, used to determine the turn-off mass of the 
clusters (except for NGC 6134, see text). The observed 
stars are shown as full circles and identified by their numbers. The 
parameters adopted for the fittings are those listed in Table\@ \ref{tab:opc}.}
\label{fig:iso2}
\end{figure*}

 The data of the open clusters included in the sample are listed in
 Table\@ \ref{tab:opc}. The [Fe/H]\footnote{[A/B] = log [N(A)/N(B)]$_{\rm \star}$ $-$ log
 [N(A)/N(B)]$_{\rm\odot}$} value is the average of all values obtained for 
individual stars in this work. The other adopted parameters, (m$-$M), E(B$-$V), 
age, and distance, of NGC 2360 and NGC 2447, are the same as those listed by 
\citet{Ham00}. For NGC 6134, we adopt the parameters determined by \citet{BFKA99}. 
For the other clusters, new parameters were determined in this work 
using UBV photometry obtained from the WEBDA database, and
 Geneva isochrones \citep{Sch92} with metallicity Z = 0.020. Two values of the 
turn-off mass were determined using the corresponding 
\citet{Sch92} isochrone. The blue turn-off was defined to be the bluest limit 
of the isochrone, excluding the very short-lived phase
 just after core-H exhaustion, as indicated by point B in Fig.\@ 1 of
 \citet{MM91}. The red turn-off is defined to be the reddest point
 just before the short blueward excursion, as indicated by point R in Fig.\@ 1 of
 \citet{MM91}. We also list the mass at the clump as given by
 the isochrones, where the clump is defined as the point of lowest
 luminosity after He ignition. Finally, the Galactocentric distance of 
the cluster given by \citet{Chen03} is given. The color-magnitude diagrams 
and isochrones of all clusters are shown in Figs.\@ \ref{fig:iso} and 
\ref{fig:iso2}, where our sample stars are identified as black dots.

%


\section{Atmospheric parameters}\label{sec:par}

\subsection{Equivalent widths and atomic data}

The atomic-line list adopted in this work is the same as the one 
used by \citet{Ham00} with the addition of a few \ion{Fe}{ii} 
lines. For elements other than Fe, the same oscillator 
strengths ($gfs$) as used by \citet{Ham00} were adopted. 
For the \ion{Fe}{ii} lines, the $gfs$ are those renormalized 
by \citet{MeBa09}, and also used in \citet{S06}. The 
\ion{Fe}{i} $gfs$ are preferentially taken from the 
critical compilation of data by \citet{FuWi06} complemented 
by values from the NIST web database \citep{NIST}. The solar iron 
abundance adopted for the calculations is the one recommended by \citet{GS98}, A(Fe) = 7.50.

For the stars in common with \citet{Ham00}, 
new equivalent widths (Ws) were measured only for the additional  
\ion{Fe}{ii} lines, while the original measurements 
for the remaining lines were adopted. The new 
Ws were measured by fitting Gaussian profiles to the lines using 
IRAF\footnote{IRAF is distributed by the National Optical 
Astronomy Observatory, which is operated by the Association of 
Universities for Research in Astronomy, Inc., under cooperative 
agreement with the National Science Foundation of the USA.}. The same 
normalization of the continuum as used in previous measurements 
was adopted, to ensure consistency between them. We refer the reader to the original 
work for further details of the line list and the data reduction. 

\begin{table*}
\caption{Physical data of the open clusters adopted from the literature or 
calculated in this work (see text).} \label{tab:opc}
\centering
\begin{tabular}{ccccccccccc}
\noalign{\smallskip}
\hline\hline
\noalign{\smallskip}
Cluster & (m$-$M) & E(B-V) & E(b-y) & Age & Distance & [Fe/H] & M$_{\rm
  blue\ TO}$ & M$_{\rm red\ TO}$ & M$_{\rm clump}$ & R$_{\rm GC.}$ \\
 & (mag.) & (mag.) & (mag.) & (log yrs.) & (pc) & &  (M$_{\odot}$) &
 (M$_{\odot}$) & (M$_{\odot}$) & (kpc) \\
\hline
IC 2714 & 11.50 & 0.33 & -- & 8.60 & 1246 & +0.12 & 2.55 & 2.85 &
2.91 & 8.34 \\
IC 4756 &  9.00 & 0.20 & -- & 8.85 &  474 & +0.04 & 2.08 & 2.31 &
2.37 & 7.23 \\
NGC 2360 & 10.40 & 0.07 & -- & 9.06 & 1086 & +0.04 & 1.78 & 1.98 &
2.02 & 6.32 \\
NGC 2447 & 10.25 & 0.04 & -- & 8.65 & 1057 & $-$0.01 & 2.44 & 2.74 &
2.79 & 6.51 \\
NGC 3532 & 8.50 & 0.04 & -- & 8.55 &  473 & +0.04 & 2.67 & 2.96 &
3.03 & 7.87 \\
NGC 3680 & 10.07 & 0.07 & -- & 9.25 & 935 & +0.04 & 1.46 & 1.70 &
1.74 & 7.45 \\
NGC 5822 & 9.65 & 0.14 & -- & 8.95 & 697 & +0.04 & 1.89 & 2.14 &
2.19 & 8.10 \\
NGC 6134 & -- & -- & 0.263 & 8.85 & 1410 & +0.12 & 2.08 & 2.31 & 2.37 & 7.52 \\
NGC 6281 & 8.95 & 0.13 & --  & 8.50 & 512 & +0.05 & 2.78 & 3.09 & 3.18 & 8.47 \\
NGC 6633 & 8.50 & 0.18 & --  & 8.65 & 388 & +0.08 & 2.44 & 2.74 & 2.79 & 8.42 \\
\noalign{\smallskip}
\hline
\end{tabular}
\end{table*}

For the new data, the fitting of the continuum and the measurement of 
the Ws were conducted using the PeakFit software. New 
\ion{Fe}{ii} lines were then added to the line list after these measurements, 
and in this case, new Ws of the \ion{Fe}{ii} lines were also determined 
by fitting Gaussian profiles with IRAF. The same continuum normalization 
was adopted in both cases. A comparison between the \ion{Fe}{ii} Ws 
measured with Peak Fit and IRAF showed excellent agreement. 
Lines with equivalent widths smaller than 10 m{\AA} and larger than 
150 m{\AA} were not used in our analysis. The new equivalent widths
measured in this work are listed in the Appendix, 
Tables \ref{tab:LE} to \ref{tab:LE4}.

\subsection{Determination of the atmospheric parameters}

\begin{table*}
\caption{The atmospheric parameters of the sample stars. The values for [\ion{Fe}{i}/H] and [\ion{Fe}{ii}/H] are
  followed by the standard deviation and the number of lines on which
  the abundance is based. The spectroscopic values are the ones adopted throughout this work.} 
\label{tab:par} \centering
\begin{tabular}{lccccccccc}
\noalign{\smallskip}
\hline\hline
\noalign{\smallskip}
Star & T$_{\rm eff}$ (K) & T$_{\rm eff}$ (K) & log g    & log g   &  $\xi$ (km s$^{-1}$) & [\ion{Fe}{i}/H] $\pm$ $\sigma$\,(\#)& [\ion{Fe}{ii}/H] $\pm$ $\sigma$\,(\#) \\
     & spectr.           &  phot.            &  spectr. &  evolut.  &                      &                                     &                                                \\
\hline
\object{IC 2714\_5}    & 5070 & 5020 & 2.70 & 2.49 & 1.50 & +0.12$\pm$0.09 (38) & +0.12$\pm$0.04 (10) \\
\object{IC 4756\_12}   & 5030 & 5189 & 2.75 & 2.79 & 1.37 & $-$0.01$\pm$0.09 (39) & $-$0.01$\pm$0.06 (13) \\
\object{IC 4756\_14}   & 4720 & 5627 & 2.47 & 2.36 & 1.57 & +0.03$\pm$0.14 (41) & +0.03$\pm$0.08 (11) \\
\object{IC 4756\_28}   & 4620 & 4548 & 2.42 & 2.36 & 1.41 & +0.07$\pm$0.12 (39) & +0.07$\pm$0.09 (12) \\
\object{IC 4756\_38}   & 5075 & 5056 & 3.00 & 2.92 & 1.21 & +0.05$\pm$0.09 (41) & +0.05$\pm$0.07 (13) \\
\object{IC 4756\_69}   & 5130 & 5158 & 3.00 & 2.71 & 1.31 & +0.08$\pm$0.08 (42) & +0.08$\pm$0.06 (11) \\
\object{NGC 2360\_7}   & 5115 & 5016 & 3.00 & 2.82 & 1.21 & +0.11$\pm$0.11 (38) & +0.10$\pm$0.05 (10) \\ 
\object{NGC 2360\_50}  & 5015 & 4897 & 2.90 & 2.77 & 1.37 & $-$0.03$\pm$0.05 (27) & $-$0.02$\pm$0.11 (10) \\
\object{NGC 2360\_62}  & 5105 & 5153 & 3.15 & 2.88 & 0.91 & +0.12$\pm$0.08 (32) & +0.10$\pm$0.12 (13) \\
\object{NGC 2360\_86}  & 4960 & 4906 & 2.65 & 2.62 & 1.18 & $-$0.06$\pm$0.15 (40) & $-$0.07$\pm$0.03 (08) \\
\object{NGC 2447\_28}  & 5060 & 5054 & 2.70 & 2.50 & 1.46 & $-$0.01$\pm$0.14 (38) &  0.00$\pm$0.08 (10) \\
\object{NGC 2447\_34}  & 5120 & 5121 & 2.90 & 2.63 & 1.44 & $-$0.01$\pm$0.12 (38) & $-$0.01$\pm$0.09 (11) \\
\object{NGC 2447\_41}  & 5055 & 5028 & 2.80 & 2.57 & 1.37 & $-$0.02$\pm$0.11 (37) & $-$0.02$\pm$0.09 (11) \\
\object{NGC 3532\_19}  & 4995 & 5033 & 2.65 & 2.36 & 1.52 & +0.11$\pm$0.11 (41) & +0.09$\pm$0.05 (12) \\
\object{NGC 3532\_100} & 4745 & 4731 & 2.15 & 2.13 & 1.66 & +0.01$\pm$0.12 (37) & +0.02$\pm$0.05 (11) \\
\object{NGC 3532\_122} & 5045 & 5042 & 2.60 & 2.56 & 1.54 & $-$0.02$\pm$0.11 (39) & $-$0.02$\pm$0.11 (11) \\
\object{NGC 3532\_596} & 5020 & 4943 & 2.50 & 2.44 & 1.58 & +0.04$\pm$0.11 (41) & +0.04$\pm$0.09 (12) \\
\object{NGC 3532\_670} & 4355 & 4316 & 1.80 & 1.70 & 1.52 & +0.08$\pm$0.11 (28) & +0.08$\pm$0.13 (11) \\
\object{NGC 3680\_13}  & 4660 & 4684 & 2.60 & 2.58 & 1.30 & +0.04$\pm$0.10 (38) & +0.06$\pm$0.12 (13) \\
\object{NGC 5822\_1}   & 4470 & 4559 & 2.00 & 2.00 & 1.38 & +0.03$\pm$0.10 (32) & +0.03$\pm$0.09 (11) \\
\object{NGC 5822\_201} & 5035 & 5035 & 2.85 & 2.78 & 1.32 & +0.05$\pm$0.10 (44) & +0.06$\pm$0.06 (12) \\
\object{NGC 5822\_240} & 4425 & 4467 & 1.95 & 2.12 & 1.34 & +0.02$\pm$0.11 (32) & +0.03$\pm$0.12 (11) \\
\object{NGC 5822\_316} & 5110 & 5125 & 3.05 & 2.92 & 1.28 & +0.16$\pm$0.10 (43) & +0.16$\pm$0.03 (10) \\
\object{NGC 5822\_443} & 4610 & 4648 & 2.10 & 2.34 & 1.53 & $-$0.06$\pm$0.11 (38) & $-$0.06$\pm$0.08 (12) \\
\object{NGC 6134\_30}  & 4980 & 5138 & 2.95 & 2.99 & 1.23 & +0.21$\pm$0.11 (41) & +0.21$\pm$0.08 (10) \\
\object{NGC 6134\_99}  & 4785 & 4898 & 2.55 & 2.81 & 1.39 & +0.10$\pm$0.10 (37) & +0.10$\pm$0.10 (12) \\
\object{NGC 6134\_202} & 4555 & 4677 & 2.25 & 2.67 & 1.34 & +0.04$\pm$0.10 (34) & +0.06$\pm$0.12 (09) \\
\object{NGC 6281\_3}   & 4915 & 4860 & 2.30 & 2.24 & 1.64 & +0.01$\pm$0.09 (38) & +0.01$\pm$0.07 (13) \\
\object{NGC 6281\_4}   & 5015 & 4855 & 2.50 & 2.35 & 1.70 & +0.09$\pm$0.07 (33) & +0.09$\pm$0.04 (10) \\
\object{NGC 6633\_78}  & 4370 & 4383 & 1.80 & 1.79 & 1.51 & +0.04$\pm$0.10 (31) & +0.03$\pm$0.15 (12) \\
\object{NGC 6633\_100} & 5015 & 5016 & 2.85 & 2.56 & 1.44 & +0.11$\pm$0.11 (42) & +0.11$\pm$0.08 (12) \\
\noalign{\smallskip}
\hline
\end{tabular}
\end{table*}

\begin{figure}
\begin{centering}
\includegraphics[width=7cm]{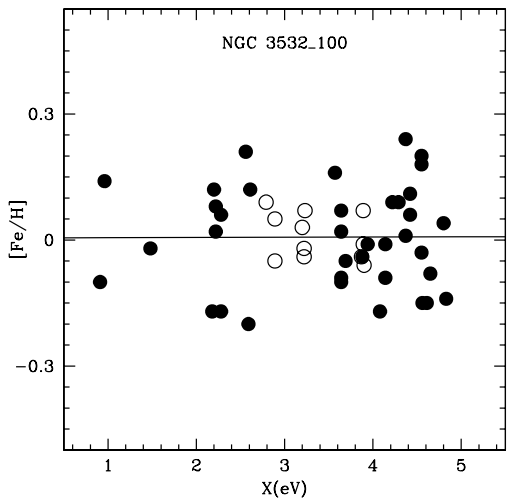}
\caption{Iron abundance of both \ion{Fe}{i} (full circles) and
  \ion{Fe}{ii} lines (open circles) versus the line excitation
  potential for the star NGC 3532\_100. The solid line is a linear fit
  to the \ion{Fe}{i} lines indicating that the excitation equilibrium
  was fulfilled. The ionization equilibrium was also obtained by
  setting the \ion{Fe}{i} and \ion{Fe}{ii} abundances to be equal,
  determining the surface gravity.}
\label{fig:chi}
\end{centering}
\end{figure}

\begin{figure}
\begin{centering}
\includegraphics[width=7cm]{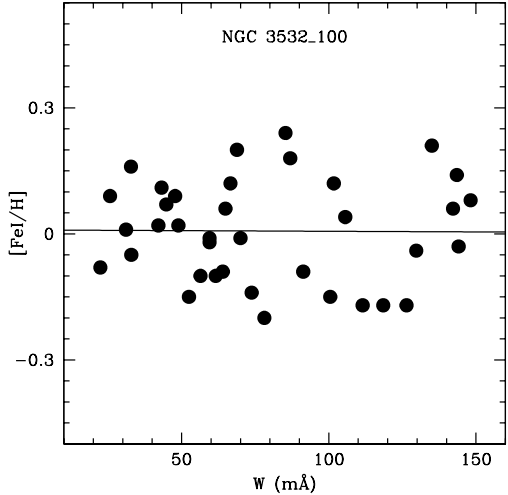}
\caption{Iron abundance versus W for the Fe I lines of
  NGC 3532\_100. This plot was used to determine the microturbulence
  velocity by requiring a null correlation between [\ion{Fe}{i}/H] and
  the Ws.}
\label{fig:le}
\end{centering}
\end{figure}

 Atmospheric parameters of the sample stars were determined using the
 standard spectroscopic approach. The effective temperature (T$_{\rm
 eff}$) was calculated by assuming the excitation equilibrium of the
 \ion{Fe}{i} lines (Fig.\@ \ref{fig:chi}), i.e., requiring a null
 correlation between the iron abundance and the lower level excitation potential
 ($\chi$). The surface gravity was found by assuming the ionization
 equilibrium of Fe, requiring both \ion{Fe}{i} and \ion{Fe}{ii} lines
 to have the same mean abundance (Fig.\@ \ref{fig:chi}). The
 microturbulence velocity ($\xi$) was found by requiring the
 \ion{Fe}{i} abundance to have a null correlation with the equivalent
 widths (Fig.\@ \ref{fig:le}). When these parameters are
 simultaneously constrained, the value of the metallicity,
 [Fe/H], is also determined. The parameters thus
 obtained are listed in Table \ref{tab:par}.

When following this procedure, each time we converge to a set of parameters constrained 
by these criteria, the line list is checked for lines that indicate an
 abundance that departs by more than 2 $\sigma$ from the average 
 value. These lines are then excluded and a new set of parameters is
 calculated. The entire procedure is repeated until the abundance given
 by all lines agree to within 2 $\sigma$. Thus, we believe to be excluding 
 lines that are strongly affected by uncertain $gfs$ or defective equivalent
 widths, and expect the mean abundance given by the remaining set of
 lines to be more reliable.

 For these calculations, we adopted the grids of model atmospheres
 computed with the ATLAS9 code \citep{ipCK03}, without
 overshooting. The ATLAS9  models assume local thermodynamic
 equilibrium (LTE), plane-parallel geometry, and hydrostatic
 equilibrium. 

\subsection{Photometric temperatures}

 We calculated a photometric estimate of the effective temperature 
using the (B-V)$_{0}$ color as a way of testing our spectroscopically determined effective 
temperatures. The Johnson (B-V) color was obtained from the WEBDA database 
and is listed in Table \ref{tab:log}. The color excess listed in Table \ref{tab:opc} 
was used to correct the observed color. Temperatures were calculated using the 
calibrations by \citet{AAM99} and \citet{HBS00}. The average of these two estimates 
is listed in Table \ref{tab:par}.

Temperatures calculated using the two photometric 
calibrations are in good agreement. The mean 
difference between them is 42 K. There is also good agreement 
between our spectroscopically determined T$_{\rm eff}$ 
and the average photometric value. The differences vary between 1 and 160 K, 
excluding the star IC 4756 14. The mean difference in the 
temperatures is 56 K. As detailed below, this is close to the uncertainty that we estimate for 
the spectroscopic temperature.

For star IC 4756 14, the difference between the photometric and 
spectroscopic temperatures is $\sim$ 900 K, probably due to an incorrect 
(B-V) color. The cluster IC 4756 is 
affected by differential reddening \citep{Sc78,Sm83}. We note that 
using a higher temperature we would infer a far higher 
metallicity, incompatible with the other cluster members. Therefore, 
the spectroscopic estimate is to be preferred.

This comparison shows that the spectroscopic method provides 
a reliable temperature scale. We therefore adopt these 
temperatures throughout the paper.

\subsection{Evolutionary gravities}

To test the spectroscopically determined gravities, we 
also determined log g values using the stellar masses obtained 
from the isochrone fittings. For this test, the stellar mass 
was considered to be equal to the stellar mass of the clump 
at the given isochrone. Since there is little variation in mass between 
the red turn-off and the clump, this choice should introduce no 
important effect.

The evolutionary gravities were calculated using the classical 
equation, log g$_{\star}$ = log g$_{\odot}$  + log
 (M$_{\star}$/M$_{\odot}$) + 4 log (T$_{\rm eff}$$_{\star}$/T$_{\rm
 eff}$$_{\odot}$) $-$ log (L$_{\star}$/L$_{\odot}$), where log
 (L$_{\star}$/L$_{\odot}$) = $-$0.4 (M$_{\rm bol}$$_{\star}$ -
 M$_{\rm Bol}$$_{\odot}$). Luminosities were calculated with the 
parameters listed in Table \ref{tab:opc} and bolometric 
corrections calculated with the relations by \citet{AAM99}. For 
the Sun, we adopted T$_{\rm eff}$ = 5777 K, log g = 4.44 dex, and 
M$_{\rm bol}$ = 4.75 mag. The gravities calculated in this way are given 
in Table \ref{tab:par}.

The evolutionary log g is in good agreement with the 
spectroscopically determined log g. The differences 
vary between 0.01 and 0.42 dex. The mean difference between the 
gravities is on the order of 0.14 dex. As discussed below, this 
is close to the uncertainty that we estimate for 
the spectroscopic gravity. Most of the spectroscopic values are systematically higher 
than the evolutionary ones. The disagreement between these 
two methods is well known in the literature and is also found 
for field giants when parallaxes are used to derive log g 
\citep{AlP99,DaS06}. The precise reasons remain unknown, although 
departures from LTE are usually blamed. We consider it to be possible that our
 gravity values are systematically overestimated by an amount close 
to its associated uncertainty. This has little effect on the 
metallicities, since these are mainly derived from the 
gravity-insensitive \ion{Fe}{i} lines.

This comparison again shows that the use of the 
spectroscopic method establishes a reliable gravity scale. 
We therefore adopt these gravities throughout the paper.

\subsection{Uncertainties in the atmospheric parameters}
                  
 The uncertainties in the atmospheric parameters were calculated for a
 representative star, IC 4756\_14, whose atmospheric parameters are close 
to the median value for the entire sample. The 1$\sigma$ uncertainties in 
T$_{\rm eff}$ and $\xi$ were determined by the
 uncertainties in the linear fits used to constrain these
 parameters. The uncertainties were given by the variation in these parameters 
necessary to match the angular coefficient value of the linear fit to 
the value of its own uncertainty, in
the diagram of \ion{Fe}{i} abundance versus the line excitation potential for
 T$_{\rm eff}$, and in the \ion{Fe}{i} abundance versus W diagram for
 $\xi$. The 1$\sigma$ uncertainty in the surface gravity was found by changing
 the gravity value until the difference between the mean abundances
 from \ion{Fe}{i} and \ion{Fe}{ii} equals the larger of the standard
 deviation values.

The atmospheric parameters are not trully independent of each other 
and thus, for example, an error in the effective temperature may also 
introduce an error in the gravity. We thus also evaluated the influence of 
the uncertainty in each parameter on the remaining ones. This cross-terms 
were combined with the uncertainties calculated above to produce the 
total uncertainties caused by the analysis method, listed in the second column 
 of Table \ref{tab:uncer}.

Uncertainties in the measurement of the equivalent widths of the 
\ion{Fe}{i} lines, caused by the S/N and the accuracy of the continuum definition, can also 
affect the calculation of the atmospheric parameters. After conducting some 
tests, we determined that the equivalent widths are affected by at most an 
uncertainty of $\sigma$ = $\pm$ 3.0 m\AA. This uncertainty propagates 
into the atmospheric parameters at a level listed in the third column of Table \ref{tab:uncer}.

In addition, we recalculated the parameters of three stars using a 
2.5 $\sigma$ clipping factor, to test the influence of this choice on the 
derived atmospheric parameters. Some of 
the lines excluded before with 2 $\sigma$ were still excluded when using 2.5 
$\sigma$. After increasing the clipping factor, only a few extra lines were taken into account (from 3 to 9 
\ion{Fe}{i} lines), resulting mostly in similar parameters. The average differences 
in the atmospheric parameters calculated by adopting each of the two clipping factors are 
listed in the fourth column of Table \ref{tab:uncer}. This was considered 
another source of uncertainty inherent to the method, and thus added to the other ones 
to calculate the final total uncertainty in the parameters. 

The microturbulence velocity and the temperature were found to be mostly insensitive to 
the choice clipping factor, while log g was found to vary by a 
significant amount. This result shows that the addition of a few extra lines does not 
significantly affect the excitation equilibrium or the method used to constrain $\xi$. On the 
other hand, these extra lines do affect the mean abundances of \ion{Fe}{i} 
and \ion{Fe}{ii} and thus force a change in log g to maintain the ionization 
equilibrium.

As a typical error in metallicity, we adopt the standard deviation in the 
\ion{Fe}{i} values of star IC 4756 14, $\sigma_{[Fe/H]}$ = $\pm$0.14. This is 
one of the largest standard deviations, as seen in Table \ref{tab:par}, and 
can thus be seen as a rather conservative choice. The effect of uncertainties 
related to the cross-terms ($\pm$0.04), both the S/N and continuum ($\pm$0.05), and the sigma-clipping 
procedure ($\pm$0.03) have an almost negligible effect on the final 
total uncertainty in the metallicity (which would increase to $\pm$0.16).

Other systematic effects might be present, because of the use of 
1D stellar atmospheres, and the neglect of departures from the LTE, among others effects. These, 
however, are likely to affect all the sample stars in much the same way. Since we 
are mostly interested in a relative comparison between the stars, these effects 
do not significantly affect either the analysis or the conclusions. They might be important, however, when 
the results obtained here are compared to those obtained by other works 
in the literature.

\subsection{Comparison with previous results}

\subsubsection{The sample of \citet{Ham00}}

 Although we use almost the same data as \citet{Ham00}, the methods that we
 employ to determine the atmospheric parameters are different. \citet{Ham00}
 followed an iterative procedure, where the T$_{\rm eff}$ is given by
 minimizing the spread in the abundances of the iron-peak
 elements, and the surface gravity is calculated for this temperature
 and the masses given by the isochrones (see more details in the
 original paper). For reference, we list the parameters found by
 \citet{Ham00} in Table \ref{tab:ham}.

\begin{table}
\caption{Uncertainties in the adopted atmospheric parameters.}
\label{tab:uncer}
\centering
\begin{tabular}{ccccc}
\noalign{\smallskip}
\hline\hline
\noalign{\smallskip}
Parameter     & $\sigma$   &  $\sigma$     & $\sigma$ & $\sigma$ \\
              &  method    &  S/N \& cont. & sigma-clip. & total \\
\hline
T$_{\rm eff}$ (K)      & $\pm$ 55   &  $\pm$ 10   & $\pm$ 20 & $\pm$ 60 \\
log g       (dex)    & $\pm$ 0.20  & $\pm$ 0.07 & $\pm$ 0.15 & $\pm$ 0.26 \\
$\xi$   (km s$^{-1}$) & $\pm$ 0.05  & $\pm$ 0.01 & $\pm$ 0.06 & $\pm$ 0.08 \\
\noalign{\smallskip}
\hline
\end{tabular}
\end{table}
\begin{table*}
\caption{Atmospheric parameters derived by \citet{Ham00} for the
  stars in common with our sample.}
\label{tab:ham}
\centering
\begin{tabular}{cccccc}
\noalign{\smallskip}
\hline\hline
\noalign{\smallskip}
Star & T$_{\rm{eff}}$ & log g & $\xi$ & [\ion{Fe}{i}/H] $\pm$
$\sigma$\,(\#) & [\ion{Fe}{ii}/H] $\pm$ $\sigma$\,(\#) \\
\hline
NGC 2360\_7 & 5230 & 2.89 & 1.57 & +0.15$\pm$0.17 (57) &
+0.15$\pm$0.10 (04)\\
NGC 2360\_50 & 5170 & 2.86 & 1.69 & +0.01$\pm$0.17 (55) & +0.11$\pm$0.20 (04)\\
NGC 2360\_62 & 5180 & 2.94 & 1.44 & +0.08$\pm$0.17 (52) & +0.16$\pm$0.17 (04)\\
NGC 2360\_86 & 5130 & 2.73 & 1.52 & +0.04$\pm$0.16 (55) & +0.04$\pm$0.13 (04)\\
NGC 2447\_28 & 5140 & 2.56 & 1.75 & $-$0.01$\pm$0.21 (55) & +0.06$\pm$0.16 (04)\\
NGC 2447\_34 & 5250 & 2.70 & 1.77 & +0.05$\pm$0.19 (56) & +0.03$\pm$0.13 (04)\\
NGC 2447\_41 & 5200 & 2.65 & 1.70 & +0.05$\pm$0.19 (57) & +0.04$\pm$0.16 (04)\\
\noalign{\smallskip}
\hline
\end{tabular}
\end{table*}

 Our temperatures are systematically lower than those found by
 \citet{Ham00}. The difference varies from 75 K to 170 K, with an
 average of $\sim$ 125 K, larger than the estimated uncertainty. We believe
 that this systematic difference is related to a combination of the
 method and the final linelist. In particular, we note that, although
 we employ the same equivalent widths, our $gf$ list is
 different. Our final \ion{Fe}{i} mean
 abundance is also based on fewer lines. The first reason for that is the exclusion from 
the analysis of lines in the saturated part of the curve of growth, W $\geq$ 150 m\AA. 
A second reason is the $\sigma$-clipping procedure explained above. It seems that 
the lines we systematically exclude, in both the 2 or 2.5 $\sigma$ clipping, are the 
most significant causes for the different temperatures that we derive.

 On the other hand, with the exception of one star, the gravities
 derived in this work are systematically higher than those adopted
 by \citet{Ham00}. The differences vary from 0.04 to 0.21
 dex, with an average of 0.13 dex. The average difference is very 
close to the 1 $\sigma$ uncertainty estimated for log g.
 While we rely on the ionization
 equilibrium, \citet{Ham00} calculated the log g using the mass
 given by the isochrones and the known cluster distances. As discussed 
before, the spectroscopic method is known to result in systematically higher 
gravities. In addition, we note that this difference in log g 
does not seem to be related to the difference in temperature. 
An increase in the temperature of our stars would be followed 
by an increase in our log g values, which would only increase 
the discrepancies. The derived microturbulence velocity values behave 
in a similar way to the temperatures, probably for the same reason. We note that the range
 of our values is in excellent agreement with that found, for
 example, by \citet{DaS06} for field giants with a similar range of gravities 
and temperatures.

 In spite of the different parameters, the metallicities ([Fe/H]) of
 the stars show good agreement with those of \citet{Ham00}.
 The average difference is +0.05 dex, a value well within the uncertainties. 
We believe this result to be evidence that reliable abundances can be derived
 regardless of small-scale differences introduced by the different methods
 of atmospheric-parameter calculation.

 \citet{G89} appears to be the only other work that included another member of NGC
 2360, star 12. It was found to have [Fe/H] = +0.20, a value that
 seems too high when compared with most of the metallicities obtained by us and
 by \citet{Ham00}. For NGC 2447, the only previous high-resolution analysis
 seems to be the one by \citet{Ham00}.

\subsubsection{The new sample}

 In this section, we compare our results with literature data. We
 restrict the discussion to other analyses based on high-resolution
 spectra. In cases where the sample stars were targets of other analyses, the
 entire set of parameters is compared. A comparison of cluster
 metallicity is presented in cases where different cluster members were
 analyzed. To the best of our knowledge, this is the first time that atmospheric
 parameters and abundances have been derived for members of IC 2714 and NGC
 6281 with high-resolution spectroscopy.

 Members of IC 4756 were analyzed by \citet{G89}, \citet{L94}, and
 \citet{Jac07}. We  have no star in common with \citet{L94} and
 only IC 4756\_69 in common with both \citet{G89} and \citet{Jac07}. 
While \citet{G89} found T$_{\rm eff}$ = 5200 K,
 log g = 3.20, $\xi$ = 2.00 km s$^{-1}$, and [Fe/H] = 0.00, in
 reasonable agreement with our parameters, except for the
 microturbulence, \citet{Jac07} derived T$_{\rm eff}$ = 5000 K, log g
 = 2.20, $\xi$ = 1.50 km s$^{-1}$, and [Fe/H] = $-$0.15. While T$_{\rm
 eff}$ and $\xi$ have close values, log g and [Fe/H] are 
 clearly discrepant. The reasons for this large discrepancy are
 probably related to the small number of \ion{Fe}{ii} lines used by
 \citet{Jac07}, only three. Our mean cluster metallicity, [Fe/H]
 = +0.04, is in good agreement with the ones found by
 \citeauthor{L94}, [Fe/H] = $-$0.03,  and \citet{G89}, [Fe/H] = +0.04,
 while the mean metallicity derived by \citet{Jac07}, [Fe/H] =
 $-$0.15, disagrees with ours and those of other authors. 

 Three of the stars from NGC 3532, stars 19, 596, and 670, were also
 analyzed by \citet{L94}. The parameters found by \citet{L94} are
 T$_{\rm eff}$ = 5000 K, log g = 2.36, $\xi$ = 2.00 km s$^{-1}$, [Fe/H]
 = +0.13 for star 19, T$_{\rm eff}$ = 5000 K, log g = 2.25, $\xi$ =
 2.00 km s$^{-1}$, [Fe/H] = +0.08 for star 596, and  T$_{\rm eff}$ =
 4500 K, log g = 2.00, $\xi$ = 2.40 km s$^{-1}$, [Fe/H] = +0.09 for
 star 670. Temperatures and particularly the metallicities show 
good agreement, while our log g differ by 0.20 to 0.35 dex and the
 microturbulence velocities are systematically lower. The reason for
 this difference is unclear since the methods used in estimating log g and
 $\xi$ were the same and the derived temperatures are similar. Our mean cluster
 metallicity is [Fe/H] = +0.04 in good agreement with the mean
 [Fe/H] = +0.07 found by \citet{L94}. 

 Two high-resolution analysis of NGC 3680 members are reported in the
 literature, \citet{PRP01} and \citet{PP08}. The first work found an
 average of [Fe/H] = $-$0.27, although they considered the most accurate estimate
 to be [Fe/H] = $-$0.17 probably due to systematic effects. The latter
 work found [Fe/H] = $-$0.04, for two main-sequence stars. Our value
 for the only star that we analyze that is not included in these works, NGC
 3680\_13, is [Fe/H] = +0.04, which agrees with \citet{PP08} within
 the uncertainties.

 Among our sample stars in NGC 5822, only star 01 had been analyzed before
 with high-resolution spectroscopy by \citet{L94}. The parameters
 adopted by \citet{L94} are T$_{\rm eff}$ = 4800 K, log g = 2.50,
 $\xi$ = 2.50 km s$^{-1}$, and [Fe/H] = +0.13, which are clearly
 different from ours. An increase of 300 K in our
 temperature would still cause disagreement between the parameters including a
 very high metallicity of [Fe/H] = +0.33. We favor our lower value of 
temperature, which would be in excellent agreement with the two
 photometric estimates, 4500 and 4450 K also listed but not adopted by
 \citet{L94}.

 None of the stars that we studied in NGC 6134 had been previously analyzed with
 high-resolution spectroscopy. \citet{CBGT04} analyzed 6 stars in 
this cluster and found a mean metallicity of [Fe/H] = +0.15 in very
 good agreement with our mean metallicity, [Fe/H] = +0.12.

 Of the two stars that we analyze in NGC 6633, star 100 was analyzed by
 both \citet{G89} and \citet{VaFi05}. \citet{G89} derived T$_{\rm
 eff}$ = 4800 K, log g = 2.70, $\xi$ = 1.80 km s$^{-1}$, and [Fe/H] =
 0.00, while \citet{VaFi05} derived T$_{\rm eff}$ = 5245 K, log g
 = 3.11, and [Fe/H] = +0.35. Our results are in closer agreement with
 those of \citet{G89}, although our microturbulence is smaller
 and the metallicity is higher.


\section{Abundances}\label{sec:abu}

\subsection{Abundances using equivalent widths}

 Abundances were calculated for the elements Mg, Si, Ca, Sc, Ti, V,
 Cr, Co, and Ni using equivalent widths. Lines weaker than
 10 m\AA, largely affected by the S/N and uncertainties in the
 continuum normalization, and stronger than 150 m\AA, on the saturated
 part of the curve of growth, were not used. The resulting
 abundances are listed in Table\@ \ref{tab:able}. We adopt the 
 solar abundances reported in \citet{GS98}.
 
 In this analysis, the hyperfine structure (HFS) of the lines of elements 
 such as Mg, Sc, V, Mn, and Co was not taken into account. Thus, caution is necessary 
 when comparing these abundances with those obtained in other analyses. A 
 reliable comparison is possible only within the sample analyzed here.
 
 Table\@ \ref{tab:able} includes the standard deviation of the mean values 
 when possible. In these cases, it is possible to note that all abundances 
 agree with the solar ones to within 2$\sigma$. The only exception is the Cr 
 abundances given by the \ion{Cr}{ii} lines in some stars. The abundance given by the 
ionized Cr lines is always higher than the one given by the neutral species. The 
same is also seen in the case of V. As discussed before, it is possible that our 
spectroscopic log g values are slightly overestimated. This would also 
overestimate the abundances given by the ionized lines. In the cases of Cr and 
V, the abundance given by the neutral species is more reliable and should be 
preferred.

Star IC 4756\_69 was found by \citet{MM90} to be a spectroscopic binary. This system 
has a long period, 1994 days, but a very circularized orbit of eccentricity 
0.0043 \citep{MALM07}. Assuming a mass of 2.37 M$_{\odot}$ for the primary, a 
minimum mass of 0.59 M$_{\odot}$ is estimated for the secondary, which is consistent 
with a possible white-dwarf nature.

The possible white-dwarf nature of the secondary, the extremely circularized 
orbit, and the abundance anomalies are all consistent with there having been a past 
mass-transfer event in the system. The orbital elements, in particular, are consistent 
with those observed for barium stars by \citet{Jor98}. Barium stars are also 
enriched in carbon and s-process elements \cite[see][and references therein]{Sm07}. 
Star 69, however, was not found to be enriched in carbon and s-process elements 
compared to the other stars in the same cluster. This implies that the mass-transfer 
event occurred before the companion star reached the thermal pulses in the AGB 
phase and enriched itself with the products of s-process nucleosynthesis.

\begin{table*}
\caption{Mean elemental abundances for the program stars, followed
by the standard deviation, when applicable, and the number of
lines on which the abundance is based.}
\centering
\label{tab:able}
\begin{tabular}{ccccccc}
\noalign{\smallskip}
\hline\hline
\noalign{\smallskip}
[X/Fe] & 2714\_5 & 4756\_12 & 4756\_14 & 4756\_28 & 4756\_38 & 4756\_69 \\
\hline
Mg  & $-$0.02 (01)            & $-$0.02 (01)            & $-$0.05 (01)            &  $-$0.06 (01)           & $-$0.05 (01)            & $-$0.07 (01) \\ 
Si  & +0.03 $\pm$ 0.12 (07)   & +0.04 $\pm$ 0.09 (07)   & +0.10 $\pm$ 0.07 (07)   & +0.10 $\pm$ 0.09 (07)   & +0.01 $\pm$ 0.06 (06)   & +0.02 $\pm$ 0.04 (06) \\
Ca  & +0.04 $\pm$ 0.04 (06)   & +0.07 $\pm$ 0.10 (07)   & +0.03 $\pm$ 0.05 (05)   & $-$0.05 $\pm$ 0.04 (05) & +0.04 $\pm$ 0.09 (07)   & +0.04 $\pm$ 0.09 (07) \\
Sc  & $-$0.05 (02)            & +0.01 (02)              & +0.19 (02)              & +0.02 (02)              & +0.03 (02)              & +0.07 (02) \\
Ti  & $-$0.03 $\pm$ 0.09 (08) & $-$0.04 $\pm$ 0.11 (09) & $-$0.03 $\pm$ 0.16 (09) & $-$0.06 $\pm$ 0.05 (07) & $-$0.04 $\pm$ 0.10 (09) & $-$0.06 $\pm$ 0.04 (07) \\
\ion{V}{i}  & +0.04 $\pm $0.11 (08)   & +0.03 $\pm$ 0.14 (08)   & +0.09 $\pm$ 0.12 (08)   & +0.01 $\pm$ 0.15 (08)   & $-$0.06 $\pm$ 0.08 (07) & $-$0.05 $\pm$ 0.07 (06) \\
\ion{V}{ii} & +0.08 (02)              & +0.19 (02)              & +0.39 (02)              & +0.22 (02)              & +0.16 (02)              & +0.19 (02) \\
\ion{Cr}{i} & $-$0.01 $\pm$ 0.16 (13) & +0.03 $\pm$ 0.08 (15)   & +0.11 $\pm$ 0.10 (13)   & $-$0.04 $\pm$ 0.18 (15) & +0.05 $\pm$ 0.08 (15)   & +0.03 $\pm$ 0.09 (14) \\
\ion{Cr}{ii} & +0.06 $\pm$ 0.11 (04)  & +0.16 $\pm$ 0.10 (04)   & +0.27 $\pm$ 0.08 (04)   & +0.20 $\pm$ 0.12 (04)   & +0.14 $\pm$ 0.06 (04)   & +0.21 $\pm$ 0.11 (04) \\
Co  & +0.05 $\pm$ 0.13 (08)   & +0.05 $\pm$ 0.11 (08)   & +0.12 $\pm$ 0.15 (08)   & +0.05 $\pm$ 0.19 (08)   & +0.02 $\pm$ 0.14 (08)   & +0.06 $\pm$ 0.14 (08) \\
Ni  & +0.01 $\pm$ 0.06 (09)   & $-$0.02 $\pm$ 0.05 (10) & +0.03 $\pm$ 0.09 (09)   & $-$0.01 $\pm$ 0.05 (07) & $-$0.02 $\pm$ 0.05 (11) & $-$0.01 $\pm$ 0.05 (10) \\
\noalign{\smallskip}
\hline\hline
\noalign{\smallskip}
[X/Fe] & 3532\_19 & 3532\_100 & 3532\_122 & 3532\_596 & 3532\_670 & 3680\_13 \\
\hline
Mg   & $-$0.07 (01)            & $-$0.03 (01)            & $-$0.03 (01)            &  0.00 (01)              &     --                  & $-$0.23 (01) \\
Si   & +0.06 $\pm$ 0.06 (07)   & +0.14 $\pm$ 0.09 (07)   & +0.08 $\pm$ 0.16 (07)   & +0.09 $\pm$ 0.11 (07)   & +0.14 $\pm$ 0.11 (06)   & +0.10 $\pm$ 0.09 (07) \\
Ca   & +0.10 $\pm$ 0.12 (07)   & +0.02 $\pm$ 0.06 (05)   & +0.15 $\pm$ 0.12 (07)   & +0.13 $\pm$ 0.11 (07)   & +0.04 $\pm$ 0.04 (03)   & +0.01 $\pm$ 0.04 (05) \\
Sc   & $-$0.04 (02)            & $-$0.07 (02)            & $-$0.08 (02)            & +0.04 (02)              & +0.12 (02)              & +0.08 (02) \\
Ti   & $-$0.01 $\pm$ 0.08 (08) & $-$0.04 $\pm$ 0.08 (07) & $-$0.03 $\pm$ 0.13 (09) & $-$0.02 $\pm$ 0.10 (09) & +0.01 $\pm$ 0.09 (06)   & $-$0.01 $\pm$ 0.04 (07) \\
\ion{V}{i}  & $-$0.01 $\pm$ 0.09 (07) & $-$0.04 $\pm$ 0.13 (08) & +0.05 $\pm$ 0.10 (07)   & +0.02 $\pm$ 0.14 (08)   & +0.16 $\pm$ 0.16 (07)   & +0.09 $\pm$ 0.15 (08) \\
\ion{V}{ii}  & +0.16 (02)              & +0.14 (02)              & +0.16 (01)              & +0.14 (02)              & +0.36 (02)              & +0.24 (02) \\
\ion{Cr}{i}  &  0.00 $\pm$ 0.10 (13)   & $-$0.04 $\pm$ 0.11 (13) & +0.06 $\pm$ 0.17 (15)   & +0.03 $\pm$ 0.06 (11)   & $-$0.01 $\pm$ 0.13 (12) &  0.00 $\pm$ 0.13 (13) \\
\ion{Cr}{ii} & +0.17 $\pm$ 0.08 (04)   & +0.12 $\pm$ 0.06 (04)   & +0.06 $\pm$ 0.11 (04)   & +0.19 $\pm$ 0.05 (04)   & +0.28 $\pm$ 0.21 (04)   & +0.18 $\pm$ 0.06 (04) \\
Co   & +0.03 $\pm$ 0.13 (08)   & +0.01 $\pm$ 0.14 (08)   & +0.12 $\pm$ 0.14 (08)   & +0.10 $\pm$ 0.18 (08)   & +0.13 $\pm$ 0.25 (07)   & +0.14 $\pm$ 0.19 (08) \\
Ni   & $-$0.01 $\pm$ 0.07 (09) &  0.00 $\pm$ 0.07 (09)   & $-$0.01 $\pm$ 0.08 (10) & $-$0.01 $\pm$ 0.09 (09) & +0.16 $\pm$ 0.15 (09)   & +0.02 $\pm$ 0.08 (09) \\
\noalign{\smallskip}
\hline
\end{tabular}
\end{table*}
\setcounter{table}{5}
\begin{table*}
\caption{continued.}
\centering
\begin{tabular}{ccccccc}
\noalign{\smallskip}
\hline\hline
\noalign{\smallskip}
[X/Fe] & 5822\_1 & 5822\_201 & 5822\_240 & 5822\_316 & 5822\_443 & 6134\_30 \\
\hline
Mg   &    --                   & $-$0.01 (01)          &     --                  & $-$0.02 (01)            & +0.03 (01)              &  0.00 (01) \\
Si   & +0.13 $\pm$ 0.12 (07)   & +0.06 $\pm$ 0.07 (07) & +0.11 $\pm$ 0.09 (06)   & $-$0.04 $\pm$ 0.04 (06) & +0.17 $\pm$ 0.08 (07)   & +0.03 $\pm$ 0.10 (07) \\
Ca   & $-$0.02 $\pm$ 0.06 (03) & +0.03 $\pm$ 0.15 (07) &  0.00 $\pm$ 0.08 (03)   &  0.00  $\pm$ 0.06 (06)  & +0.06 $\pm$ 0.14 (06)   & +0.06 $\pm$ 0.14 (06) \\
Sc   & +0.03 (02)              &  0.00 (02)            & +0.03 (02)              & $-$0.17 (02)            & +0.06 (02)              & +0.10 (02) \\
Ti   & $-$0.02 $\pm$ 0.08 (07) & +0.04 $\pm$ 0.04 (07) & $-$0.01 $\pm$ 0.08 (07) & $-$0.02 $\pm$ 0.10 (09) & $-$0.04 $\pm$ 0.06 (07) & $-$0.04 $\pm$ 0.06 (08) \\
\ion{V}{i}  & +0.11 $\pm$ 0.16 (08)   & +0.05 $\pm$ 0.14 (08) & +0.12 $\pm$ 0.15 (07)   &  0.00 $\pm$ 0.08 (07)   & +0.02 $\pm$ 0.13 (08)   & $-$0.04 $\pm$ 0.05 (07) \\
\ion{V}{ii}  & +0.28 (02)              & +0.11 (02)            & +0.27 (02)              & +0.13 (02)              & +0.26 (02)              & +0.17 (02) \\
\ion{Cr}{i}  & $-$0.03 $\pm$ 0.17 (14) & +0.06 $\pm$ 0.10 (15) & +0.02 $\pm$ 0.07 (10)   & +0.03 $\pm$ 0.10 (15)   & +0.04 $\pm$ 0.10 (13)   & +0.04 $\pm$ 0.09 (10) \\
\ion{Cr}{ii} & +0.11 $\pm$ 0.15 (04)   & +0.16 $\pm$ 0.09 (04) & +0.16 $\pm$ 0.12 (04)   & +0.19 $\pm$ 0.12 (04)   & +0.19 $\pm$ 0.06 (04)   & +0.20 $\pm$ 0.13 (04) \\
Co   & +0.06 $\pm$ 0.20 (08)   & +0.10 $\pm$ 0.15 (08) & +0.12 $\pm$ 0.20 (08)   & +0.03 $\pm$ 0.15 (07)   & +0.08 $\pm$ 0.15 (08)   & +0.11 $\pm$ 0.16 (07) \\
Ni   & +0.05 $\pm$ 0.09 (09)   & +0.01 $\pm$ 0.06 (09) & +0.08 $\pm$ 0.13 (09)   & $-$0.03 $\pm$ 0.05 (09) & +0.05 $\pm$ 0.05 (08)   & +0.03 $\pm$ 0.10 (10) \\
\noalign{\smallskip}
\noalign{\smallskip}
\hline\hline
\noalign{\smallskip}
[X/Fe] & 6134\_99 & 6134\_202 & 6281\_3 & 6281\_4 & 6633\_78 & 6633\_100 \\
\hline
Mg   & $-$0.02 (01)            & +0.01 (01)              & +0.04 (01)              & +0.02 (01)              &    --                   & $-$0.06 (01) \\
Si   & +0.09 $\pm$ 0.08 (07)   & +0.16 $\pm$ 0.11 (07)   & +0.13 $\pm$ 0.07 (07)   & +0.10 $\pm$ 0.07 (07)   & +0.10 $\pm$ 0.09 (06)   & +0.04 $\pm$ 0.08 (07) \\
Ca   & +0.05 $\pm$ 0.08 (06)   & +0.09 $\pm$ 0.16 (05)   & +0.10 $\pm$ 0.11 (06)   & +0.11 $\pm$ 0.10 (06)   & $-$0.03 $\pm$ 0.08 (03) & $-$0.01 $\pm$ 0.05 (06) \\
Sc   & +0.11 (02)              & +0.02 (02)              & $-$0.02 (02)            & $-$0.01 (02)            & +0.12 (02)              & +0.11 (02) \\
Ti   &  0.00 $\pm$ 0.07 (07)   & $-$0.06 $\pm$ 0.07 (07) & +0.01 $\pm$ 0.10 (07)   & $-$0.03 $\pm$ 0.12 (09) & +0.05 $\pm$ 0.07 (07)   & $-$0.03 $\pm$ 0.10 (09) \\
\ion{V}{i}   & +0.10 $\pm$ 0.13 (08)   & +0.05 $\pm$ 0.16 (08)   & $-$0.02 $\pm$ 0.08 (07) & +0.04 $\pm$ 0.12 (08)   & +0.16 $\pm$ 0.16 (07)   & $-$0.04 $\pm$ 0.07 (07) \\
\ion{V}{ii}  & +0.28 (02)              & +0.20 (02)              & +0.15 (02)              & +0.12 (02)              & +0.40 (02)              & +0.30 (02) \\
\ion{Cr}{i} & +0.09 $\pm$ 0.11 (13)   & +0.06 $\pm$ 0.14 (13)   & +0.01 $\pm$ 0.08 (13)   & +0.02 $\pm$ 0.10 (13)   &  0.00 $\pm$ 0.15 (12)   & +0.03 $\pm$ 0.11 (13) \\
\ion{Cr}{ii} & +0.22 $\pm$ 0.10 (04)   &  0.00 $\pm$ 0.13 (04)   & +0.15 $\pm$ 0.06 (04)   & +0.14 $\pm$ 0.06 (04)   & +0.27 $\pm$ 0.12 (04)   & +0.25 $\pm$ 0.09 (04) \\
Co   & +0.08 $\pm$ 0.17 (08)   & +0.14 $\pm$ 0.14 (07)   & +0.07 $\pm$ 0.18 (09)   & +0.02 $\pm$ 0.14 (08)   & +0.17 $\pm$ 0.23 (08)   & +0.07 $\pm$ 0.16 (08) \\
Ni   & +0.05 $\pm$ 0.08 (09)   & +0.07 $\pm$ 0.08 (09)   & +0.01 $\pm$ 0.06 (08)   & +0.06 $\pm$ 0.06 (08)   & +0.09 $\pm$ 0.14 (09)   & $-$0.03 $\pm$ 0.08 (10) \\
\noalign{\smallskip}
\hline
\end{tabular}
\end{table*}

\subsection{Abundances using spectrum synthesis}

 The $^{12}$C/$^{13}$C ratio and the abundances of C, N, O, and Na were
 derived using spectrum synthesis. The codes for calculating synthetic
 spectra are described by \citet{Bar03} and \citet{CB05}. We adopted
 the solar abundances for C, N, and Na as recommended by \citet{GS98},
 A(C) = 8.52, A(N) = 7.92, and A(Na) = 6.33. For O, we adopted the
 abundance suitable to the 1D atmospheric models recommended by
 \citet{ALA01}, A(O) = 8.77.

\begin{figure}
\begin{centering}
\includegraphics[width=7cm]{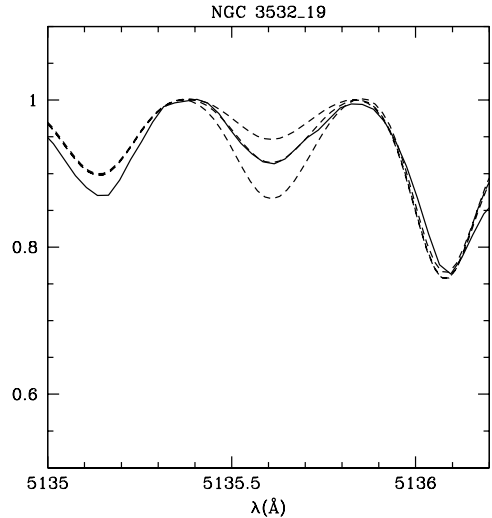}
\caption{Fit to the C$_2$ $\lambda$ 5135.62 \AA\ feature in
  NGC 3532\_19. The observed spectrum is shown as a solid line. Synthetic
  spectra with [C/Fe] = $-$0.11, $-$0.21, and $-$0.31 are shown as
  dashed lines.}
\label{fig:car}
\end{centering}
\end{figure}

 The carbon abundance was calculated using the C$_{2}$(0,1) bandhead of
the Swan system at
 $\lambda$ 5135 \AA\@. The data of the C$_2$
 molecule are those adopted by \citet{Bar85}, i.e., dissociation potential
 D$_0$(C$_2$) = 6.21 eV and electronic-vibrational oscillator strength
 $f_{00}$ = 0.0184. We consider the C abundance derived from this
 feature to be the sum $^{12}$C+$^{13}$C. An example of the fit is shown
 in Fig.\@ \ref{fig:car}.
 
\begin{figure}
\begin{centering}
\includegraphics[width=7cm]{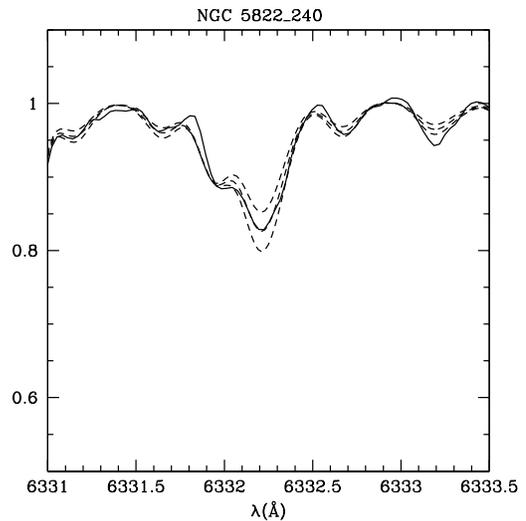}
\caption{Fit to the CN $\lambda$ 6332.18 \AA\ feature in
  NGC 5822\_240. The observed spectrum is shown as a solid
  line. Synthetic spectra with [N/Fe] = +0.23, +0.33, and +0.43 are
  shown as dashed lines.}
\label{fig:nit}
\end{centering}
\end{figure}

 The nitrogen abundance was derived using the CN(5,1)
 $\lambda$ 6332.18 \AA\ bandhead of the A$^2\Pi$-X$^2\Sigma$ red
 system. The parameters for the atomic and molecular lines are the
 same as those used by \citet{S06} and \citet{Mil92}, i.e., dissociation potential
 D$_0$(CN) = 7.65 eV and electronic oscillator strength $f_{el}$ =
 6.76$ \times 10^{-3}$. An example of the fit is shown in Fig.\@
 \ref{fig:nit}.

\begin{figure}
\begin{centering}
\includegraphics[width=7cm]{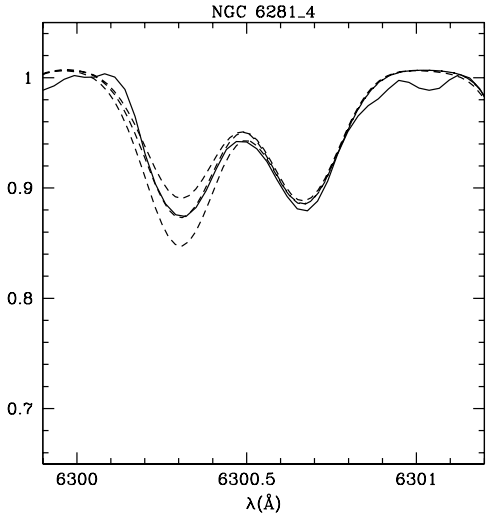}
\caption{Fit to the forbidden [\ion{O}{I}] feature in $\lambda$ 6300
  \AA\ in NGC 6281\_4. The observed spectrum is shown as a solid
  line. Synthetic spectra with [O/Fe] = $-$0.18, $-$0.08, and +0.02 are
  shown as dashed lines.}
\label{fig:oxi}
\end{centering}
\end{figure}

 The oxygen abundance was calculated from the [OI] 6300.311 \AA\@
 forbidden line. The forbidden line is blended with a weak
 \ion{Ni}{i} line at $\lambda$ 6300.34 \AA, which is included in the
 synthesis with parameters recommended by \citet{ALA01}. It also has a
 nearby \ion{Sc}{ii} line at $\lambda$ 6300.70 \AA\, for which we
 adopted the hyperfine structure by \citet{SBS89}. An example of the
 fit is shown in Fig.\@ \ref{fig:oxi}.

\begin{figure}
\begin{centering}
\includegraphics[width=7cm]{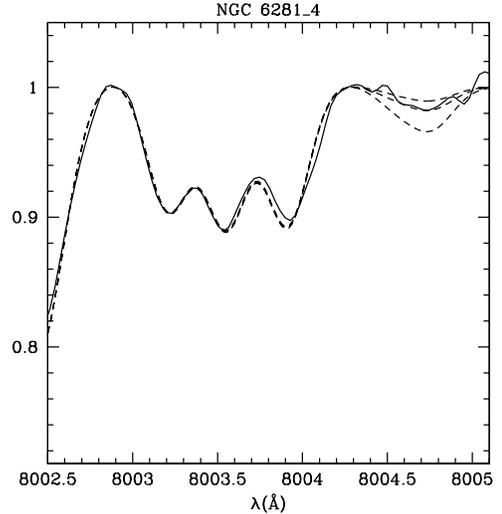}
\caption{Fit to the $^{12}$CN and $^{13}$CN features in the region of
  8005 \AA\ in NGC 6281\_4. The observed spectrum is shown as a solid
  line. Synthetic spectra with $^{12}$CN/$^{13}$CN = 06, 12, and 20 are
  shown as dashed lines.}
\label{fig:c13}
\end{centering}
\end{figure}

 The $^{12}$C/$^{13}$C ratio was derived by fitting the $^{12}$CN and
 $^{13}$CN lines in the $\lambda\lambda$ 8000-8006 \AA\ region. The
 molecular data are the same as those adopted by \citet{DDB95} and described
 in \citet{Bar92}. An example of the fit to this region is shown in
 Fig.\@ \ref{fig:c13}. As a check, we determined the $^{12}$C/$^{13}$C
 ratio in the spectrum of \object{Arcturus}, adopting the atmospheric
 parameters and CNO abundances by \citet{Mel03}. The value that we obtained
 is $^{12}$C/$^{13}$C = 6, in excellent agreement with other
 literature results (see for example \citeauthor{CBW98}
 \citeyear{CBW98} and references therein).

\begin{figure}
\begin{centering}
\includegraphics[width=7cm]{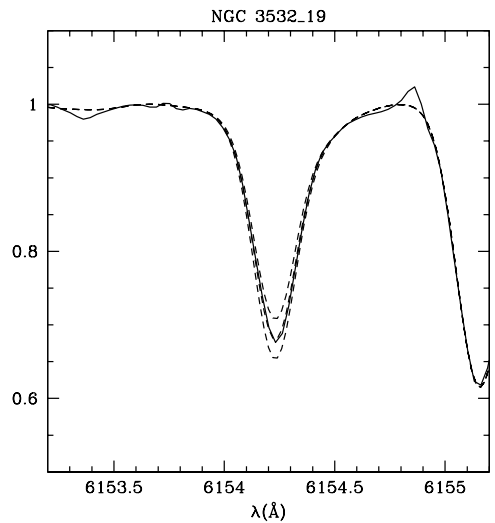}
\caption{Fit to the \ion{Na}{I} line at 6154 \AA\ in NGC 3532\_19. The
  observed spectrum is shown as a solid line. Synthetic spectra with
  [Na/Fe] = $-$0.10, 0.00, and +0.10 are shown as dashed lines.}
\label{fig:na}
\end{centering}
\end{figure}

 The Na abundance was derived by fitting the \ion{Na}{i} lines at
 $\lambda$ 6154.23 \AA\ and 6160.753 \AA. We adopted the line parameters
 determined by \citet{Bar06} by fitting the solar spectrum, i.e., log $gf$ = $-$1.56, C$_{6}$ = 0.90
 $\times 10^{-31}$ and log $gf$ = $-$1.26, C$_{6}$ = 0.30 $\times
 10^{-31}$, respectively. An example of the fit to the line $\lambda$
 6154 \AA\@ is shown in Fig.\@ \ref{fig:na}.
 
 Several works have estimated the influence of departures from the LTE
 on sodium abundances. In general, it is found that the lines at
$\lambda$ 6154/6160 \AA\@ are less affected and that the
 effects are stronger for metal-poor stars
 \citep{BBG98,GCEG99,MSS00,Tak03,SGZ04}. For these two lines,
 \citet{BBG98} estimated a correction between 0.00 and $-$0.04 dex for
 the Sun, in good agreement with the $-$0.04 dex, for $\lambda$ 6154
 \AA, and $-$0.06 dex, for $\lambda$ 6160 \AA, estimated by
 \citet{Tak03}, who tabulated an extensive grid of NLTE corrections
 for a large range of atmospheric parameters. From these same grids, an
 average correction of $-$0.07 dex to $\lambda$ 6154 \AA\@ and $-$0.11
 dex to $\lambda$ 6160 \AA\@ would correspond to the range in
 parameters of our sample, or $-$0.03 and $-$0.05 in the comparison
 with the Sun for the lines $\lambda$ 6154 \AA\@ and 6160 \AA, 
respectively. In Table \ref{tab:abun}, we list the mean [Na/Fe]
 already corrected for NLTE effects. 

 \subsection{Uncertainties in the abundances}

 An important source of uncertainties in the abundances are the
 uncertainties in the determination of the atmospheric parameters. These were 
estimated by changing each atmospheric parameter by its
 uncertainty, keeping the others to their adopted values,
 and recalculating the abundances. In this way, we measured the effect of
 the parameter uncertainty on the abundance. The results are listed in
 Table \ref{tab:sigma}. The total uncertainty was calculated by quadratically
 adding the individual uncertainties.

 The carbon isotopic ratio is rather robust, as can be seen from Table\@
 \ref{tab:sigma}, since both isotopic species are expected to react in
 similar ways to the change in the parameters. It is well known 
that only the uncertainty in the microturbulence
 velocity is usually important. In our case, however, the uncertainties 
that we estimate for these parameters are too small to introduce significant 
changes in the carbon isotopic ratio. For this ratio, the most
 important sources of uncertainties are the photon noise and the
 placement of the continuum. 

 The uncertainty due to photon noise was estimated as follows. The
 star IC 4756\_14 was again used as a template, since its parameters
 lie close to the median defined by the sample. Using its atmospheric
 parameters and abundances, two synthetic spectra were calculated, one
 with $^{12}$C/$^{13}$C = 10 and the other with $^{12}$C/$^{13}$C =
 20. Then, a Gaussian noise level equivalent to signal-to-noise ratios of
 100, 200, and 350 (the range of S/N of the observed spectra) were
 introduced into each spectrum, producing six synthetic spectra. The
 $^{12}$C/$^{13}$C ratio was then measured for each spectrum
 by searching for lower and upper values that would be considered
 reasonable fits. The difference between the original ratio and the
 upper and lower limits determined in this way therefore provided a measurement of the
 uncertainty introduced by the photon noise and is listed in 
Table\@ \ref{tab:phnoi}. 

\begin{table*}
\caption{Abundances of Na, C, N, and O, and the $^{12}$C/$^{13}$C
  ratio. We list the NLTE sodium abundances already corrected, as
  discussed in the text. The [N/C] ratio and the sums C+N, O+N, C+O, and C+N+O are also given.}
\centering 
\label{tab:abun}
\begin{tabular}{ccccccccccc}
\noalign{\smallskip}
\hline\hline
\noalign{\smallskip}
 Star & [Na/Fe] & [C/Fe] & [N/Fe]  & [O/Fe] & $^{12}$C/$^{13}$C & [N/C] & C+N & O+N & C+O & C+N+O \\
\hline
IC 2714\_5   &  0.00 & $-$0.17 &  0.51 &    --  &  -- & 0.68 &  8.81 &    --  &    --  &    -- \\
IC 4756\_12  &  0.05 & $-$0.14 &  0.50 &  $-$0.03 &  11 & 0.64 &  8.69 &  8.89 &  9.01 &  8.90  \\
IC 4756\_14  & $-$0.02 &  $-$0.14 &  0.45 &  $-$0.01 &  17 & 0.59 &  8.71 &  8.94 &  9.05 &  8.94  \\
IC 4756\_28  & $-$0.05 &  $-$0.15 &  0.32 &  $-$0.02 &  15 & 0.47 &  8.68 &  8.97 &  9.06 &  8.94  \\
IC 4756\_38  & $-$0.01 &  $-$0.18 &  0.34 &  0.00 &  10 & 0.52 &  8.65 &  8.96 &  9.05 &  8.94   \\
IC 4756\_69  &  0.15 &  $-$0.60 &  0.55 &  $-$0.17 &  05 & 1.15 &  8.66 &  8.76 &  8.97 &  8.92  \\
NGC 2360\_7   & $-$0.02 &    --  &    --  &    --  &  -- &   --  &    --  &    --  &    --  &  -- \\
NGC 2360\_50  &  0.02 &  $-$0.18 &  0.40 &    --  &  -- & 0.58 &  8.6 &    --  &    --  &    -- \\
NGC 2360\_62  & $-$0.03 &  $-$0.24 &  0.22 &    --  &  -- & 0.46 &  8.64 &    --  &    --  &    -- \\
NGC 2360\_86  & $-$0.03 &  $-$0.18 &  0.45 &  $-$0.10 &  -- & 0.63 &  8.60 &  8.78 &  8.90 &  8.79  \\
NGC 2447\_28  &  0.05 &  $-$0.18 &  0.58 &  $-$0.17 &  -- & 0.76 &  8.72 &  8.78 &  8.96 &  8.84 \\
NGC 2447\_34  &  0.03 &  $-$0.18 &  0.48 &  $-$0.13 &  -- & 0.66 &  8.66 &  8.81 &  8.95 &  8.83  \\
NGC 2447\_41  &  0.06 &  $-$0.15 &    --  &  $-$0.12 & -- &    --  &    --  &  8.81 &    --  &  -- \\
NGC 3532\_19  & $-$0.02 &  $-$0.25 &  0.34 &  $-$0.21 &  12 & 0.59 &  8.68 &  8.85 &  8.97 &  8.85  \\
NGC 3532\_100 &  0.05 &  $-$0.20 &  0.47 &  $-$0.13 &  10 & 0.67 &  8.67 &  8.82 &  8.96 &  8.84 \\
NGC 3532\_122 &  0.08 &  $-$0.15 &  0.48 &    --  &  -- & 0.63 &  8.67 &    --  &    --  &    -- \\
NGC 3532\_596 &  0.06 &  $-$0.22 &  0.40 &  $-$0.22 &  -- & 0.62 &  8.65 &  8.78 &  8.92 &  8.79 \\
NGC 3532\_670 &  0.08 &  $-$0.15 &  0.37 &  $-$0.08 &  20 & 0.52 &  8.71 &  8.94 &  9.04 &  8.92  \\
NGC 3680\_13  & $-$0.07 &    --  &    --  &    --  &    -- & --  &    --  &    --  &    --  &  -- \\
NGC 5822\_1   & $-$0.05 &  $-$0.15 &  0.43 &    --  &  13 & 0.58 &  8.69 &    --  &    --  &  -- \\
NGC 5822\_201 &  0.03 &  $-$0.19 &  0.47 &    --  &  13 & 0.66 &  8.71 &    --  &    --  &  --  \\
NGC 5822\_240 & $-$0.02 &  $-$0.10 &  0.33 &    --  &  17 & 0.43 &  8.66 &    --  &    --  &    --  \\
NGC 5822\_316 & $-$0.07 &  $-$0.21 &  0.39 &    --  &  -- & 0.60 &  8.77 &    --  &    --  &    -- \\
NGC 5822\_443 & $-$0.01 &  $-$0.13 &  0.50 &    --  &  10 & 0.63 &  8.65 &    --  &    --  &    --  \\
NGC 6134\_30  & $-$0.03 &  $-$0.21 &  0.42 &    --  &  12 & 0.63 &  8.84 &    --  &    --  &    --  \\
NGC 6134\_99  &  0.07 &    --  &    --  &    --  & -- &    --  &    --  &    --  &    --  &    --  \\
NGC 6134\_202 & $-$0.02 &  $-$0.09 &  0.36 &    --  &  13 & 0.45 &  8.70 &    --  &    --  &    --  \\
NGC 6281\_3   &  0.09 &  $-$0.24 &  0.55 &  $-$0.15 &  12 & 0.79 &  8.70 &  8.79 &  8.97 &  8.86  \\
NGC 6281\_4   &  0.02 &  $-$0.22 &  0.40 &  $-$0.08 &  12 & 0.62 &  8.70 &  8.93 &  9.04 &  8.93  \\
NGC 6633\_78  & $-$0.01 &  $-$0.17 &  0.38 &  $-$0.11 &  18 & 0.55 &  8.67 &  8.87 &  8.98 &  8.86  \\
NGC 6633\_100 & $-$0.03 &  $-$0.19 &  0.45 &  $-$0.06 &  21 & 0.64 &  8.76 &  8.97 &  9.09 &  8.98  \\
\noalign{\smallskip}
\hline
\end{tabular}
\end{table*}

 The uncertainty caused by the continuum placement was, again, estimated
 using the star IC 4756\_14. The approach in this case was
 straightforward. The normalization of the continuum was repeated in
 the region of the $^{13}$CN line, adopting different but yet
 reasonable points to define the pseudo-continuum. The carbon isotopic ratio
 was then redetermined in each case. The uncertainty related to the variations in the 
the continuum was estimated to be $\sigma$ = $\pm$2.

\begin{table}
\caption{The uncertainties in the abundances introduced by the
  uncertainties in the atmospheric parameters.}
\centering 
\label{tab:sigma}
\begin{tabular}{c c c c c c}
\noalign{\smallskip}
\hline\hline
\noalign{\smallskip}
 Elem. & $\sigma_{\rm Teff}$ & $\sigma_{\rm log g}$ & $\sigma_{\xi}$ &
 $\sigma_{\rm [Fe/H]}$ & $\sigma_{\rm total}$ \\
\hline
Na & $\pm$0.04 & $\pm$0.01 & $\pm$0.02 & $\pm$0.02 & $\pm$0.05 \\
C  & $\pm$0.01 & $\pm$0.02 & $\pm$0.01 & $\pm$0.05 & $\pm$0.06 \\
N  & $\pm$0.03 & $\pm$0.05 & $\pm$0.01 & $\pm$0.03 & $\pm$0.07 \\
O  & $\pm$0.02 & $\pm$0.08 & $\pm$0.01 & $\pm$0.07 & $\pm$0.11 \\
$^{12}$C/$^{13}$C & $\pm$0.00  & $\pm$0.00 &  $\pm$0.00 & $\pm$0.00 & $\pm$0.00 \\
\noalign{\smallskip}
\hline
\end{tabular}
\end{table}
\begin{table}
\caption{Uncertainty in the $^{12}$C/$^{13}$C ratio caused by the photon
  noise.}
\centering 
\label{tab:phnoi}
\begin{tabular}{c c c}
\noalign{\smallskip}
\hline\hline
\noalign{\smallskip}
 $^{12}$C/$^{13}$C & S/N & $\sigma$ \\
\hline
10 & 100 & $^{-1}_{+2}$ \\
10 & 200 & $^{-1}_{+1}$ \\
10 & 350 & $^{-0}_{+1}$ \\
20 & 100 & $^{-3}_{+5}$ \\
20 & 200 & $^{-2}_{+4}$ \\
20 & 350 & $^{-1}_{+2}$ \\
\noalign{\smallskip}
\hline
\end{tabular}
\end{table}
%


\section{Evolutionary state of the sample stars}\label{sec:evo}

We used the isochrone fitting shown in Figs. \ref{fig:iso} and 
\ref{fig:iso2} to derive the evolutionary status of each of the sample 
stars. The luminosity, evolutionary state, membership, and multiplicity 
of each of the sample stars are listed in Table\@ \ref{tab:mas}. 

\begin{table}
\caption{Luminosities, evolutionary stage, membership, and binarity of the sample stars.}
\centering 
\label{tab:mas}
\begin{tabular}{cccccc}
\noalign{\smallskip}
\hline\hline
\noalign{\smallskip}
 Star & log(L/L$_{\odot}$) & Evol. Stage & Member & Binary\\
\hline
IC 2714\_5    & 2.18 & clump  & m & -- \\
IC 4756\_12   & 1.78 & clump  & m & no \\
IC 4756\_14   & 2.10 & clump  & m & no \\
IC 4756\_28   & 2.06 & RGB  & nm/m & no \\
IC 4756\_38   & 1.66 & clump & m & no \\
              &      & or RGB  &   &    \\
IC 4756\_69   & 1.89 & clump  & m & yes \\
NGC 2360\_7   & 1.71 & clump  & m & no \\
NGC 2360\_50  & 1.73 & clump  & m & no \\
NGC 2360\_62  & 1.64 & clump  & m & yes \\
NGC 2360\_86  & 1.85 & clump  & m & no \\
              &      & or early-AGB &   &    \\
NGC 2447\_28  & 2.16 & clump  & m & no \\
NGC 2447\_34  & 2.04 & clump  & m & no \\
NGC 2447\_41  & 2.09 & clump  & m & no \\
NGC 3532\_19  & 2.31 & clump  & m & prob. \\
NGC 3532\_100 & 2.44 & early-AGB & m & no \\
NGC 3532\_122 & 2.12 & clump  & nm/m & prob. \\
              &      & or RGB &      &      \\
NGC 3532\_596 & 2.23 &  clump  & m & no \\
NGC 3532\_670 & 2.72 & early-AGB & m & prob. \\
             &      & or RGB tip &  &  \\
NGC 3680\_13  & 1.73 & bump RGB  & m & no \\
              &      & or clump &    &   \\
NGC 5822\_1   & 2.33 & early-AGB & m & no \\
NGC 5822\_201 & 1.76 & clump  & m & yes \\
NGC 5822\_240 & 2.19 & RGB  & m & no \\
NGC 5822\_316 & 1.64 & clump  & m & no \\
NGC 5822\_443 & 2.04 & RGB & m & prob.\\
              &      & or early-AGB &   &      \\
NGC 6134\_30  & 1.56 & clump  & m & yes \\
              &      & or early-AGB &  & \\
NGC 6134\_99  & 1.68 & clump  & m & no \\
              &      & or early-AGB & & \\
NGC 6134\_202 & 1.73 & RGB & m & no \\
              &      & or early-AGB & & \\
NGC 6281\_3   & 2.42 & clump  & m & -- \\
NGC 6281\_4   & 2.34 & clump  & m & -- \\
NGC 6633\_78  & 2.61 & early-AGB & m & no \\
NGC 6633\_100 & 2.08 & clump  & m & no \\
\noalign{\smallskip}
\hline
\end{tabular}
\end{table}

\subsection*{IC 2714}

For the relatively poorly studied southern Galactic cluster IC 2714, we 
adopted the numbering system of \citet{B60}. Only the star IC 2714\_5 was 
included in our sample. According to \citet{CMPM94}, IC 2714\_5 is a 
confirmed member of the cluster based on kinematic and photometric 
criteria. As already noted by them, its position in the CMD (Fig.\@ 
\ref{fig:iso}) indicates that the star is a clump giant in the core 
helium-burning phase.

\subsection*{IC 4756}

Five stars of the Galactic cluster IC 4756 were included in our sample, 
IC 4756\_12, 14, 28, 38, and 69, where we adopt the numbering system of 
\citet{K43}. Based on proper motions, \citet{HSS75} concluded that stars 
38 and 69 have high probabilities of being members, while star 28 is a 
non-member. However, the values of its radial velocity measured in this work and 
by \citet{MM90} implies that this star is a member of the 
cluster. All five sample stars were considered to be members by 
\citet{MM90}. The radial velocity monitoring by \citet{MM90} also 
indicated that stars 12, 14, 28, and 38 are most probably single stars, 
while star 69 was shown to be a binary system with a period of 2000 days.

The color-magnitude diagram (Fig.\@ \ref{fig:iso}) of this cluster should be interpreted 
with caution because it is affected by differential reddening 
\citep{Sc78,Sm83}. On the basis of their positions in the CMD, stars 12, 14, and 69 are 
possible clump giants, while star 28 seems to be a first-ascent red giant. 
Star 38 is either at the base of the RGB or in the clump.

\subsection*{NGC 2360}

The four stars analyzed by \citet{Ham00}, NGC 2360\_7, 50, 62, and 86, 
were included in our sample, where the numbering system of \citet{Beck76} 
was adopted. All four stars were considered to be cluster members by 
\citet{MM90}. They also found that stars 7, 50, and 86 are most probably 
single stars while, star 62 is a spectroscopic binary. Stars 7 and 86 
were found by \citet{BSW00} to have a 72\% probability of membership 
based on Hipparcos \citep{ESA} proper motions. The positions of the stars 
in the CMD (Fig.\@ \ref{fig:iso}) seem to indicate that stars 07, 50, 
and 62 are clump giants, while star 86 might be a clump or an early-AGB star. 

\subsection*{NGC 2447}

The three stars of this cluster analyzed by \citet{Ham00} were included 
in our sample, NGC 2447\_28, 34, and 41, where the numbering system 
from \citet{Beck76} was also adopted. \citet{MM89} found that the three stars 
were cluster members with no evidence of binarity. The position of the 
stars in the CMD (Fig.\@ \ref{fig:iso}) seems to indicate that they are 
clump giants \citep{CPLP05}.

\subsection*{NGC 3532}

Five stars in this cluster were also included in our sample, stars NGC 
3532\_19, 100, 122, 596, and 670, where we adopt the numbering system 
of \citet{FS80}. According to \citet{GL02}, stars 19, 122, and 670 
show probable variations in the radial velocities and might be binaries. 
Stars 100 and 596 do not exhibit radial-velocity variability and are possible 
single stars. Kinematically, 
\citet{GL02} consider all five stars to be cluster members, although 
they flag star 670 with a doubtful membership because of its large 
distance from the cluster center. The photometric criteria by 
\citet{CL88} for assigning membership indicate that all stars except star 122 
are cluster members.
 
As noted by \citet{CL88}, the positions of the stars in the 
CMD (Fig.\@ \ref{fig:iso}) seem to indicate that stars 19 and 
596 are possible clump giants in the core helium-burning phase. 
Star 122 might be either a clump or an RGB star. Star 
100 seems to be slightly more evolved, possibly on the early AGB. Star 
670 is the most luminous star of our sample. It is above the 
luminosity expected for the end of the core helium-burning and is thus 
probably an AGB star. We note however that \citet{CL88} classified 
this same star as a tip RGB object, a possibility that we cannot exclude.

\subsection*{NGC 3680}

Only one star of this cluster was included in our analysis, star NGC 
3680\_13, adopting the numbering system of \citet{E69}. Based on 
proper motions, \citet{KP95} concluded that star 13 is a member of the 
cluster. \citet{MANM95} found no evidence of binarity and thus we 
consider star 13 to be a single star. \citet{CL83} classified the 
object as a core He-burning clump star. The isochrone that we used from 
\citet{Sch92} does not extend beyond the He flash. Based only in the 
CMD (Fig.\@ \ref{fig:iso}), we identify NGC 3680\_13 
instead as a bump RGB star. However, we do not exclude the possibility 
of it being a clump giant. 

\subsection*{NGC 5822}

Five stars of this open cluster were included in our analysis, stars 
NGC 5822\_1, 201, 240, 316, and 443, where we adopt the numbering 
system of \citet{Boz74}. According to \citet{MM90}, all five stars 
seem to be true members of the cluster. In this same work, star 201 was 
found to be a spectroscopic binary and all other sample stars are most 
probably single stars. According to the CMD (Fig.\@ \ref{fig:iso2}), we 
classified stars 240 as a first ascent giant, stars 201 and 316 
as clump stars, and star 01 as an early-AGB star. Star 443 might 
be either a RGB or an early-AGB star.

\subsection*{NGC 6134}

Three stars in this cluster were included in our analysis, NGC 
6134\_30, 99, and 202, according to the numbering system of 
\citet{L72}. According to \citet{CM92}, all three stars are members of 
the cluster, stars 99 and 202 are single stars and star 30 is a 
spectroscopic binary. According to their positions in the CMD 
(Fig.\@ \ref{fig:iso2}), stars 30 and 99 are probable clump or 
early-AGB stars, while star 202 might be a RGB or an early-AGB star.

\subsection*{NGC 6281}

Two stars in this open cluster were included in our sample, NGC 
6281\_3 and 4, according to the numbering system of \citet{FF74}. The 
photometric criteria of \citet{CLM89} and the proper motions of 
\citet{DLA01} indicate that both stars are cluster members. In the CMD 
(Fig.\@ \ref{fig:iso2}), both stars seem to be clump giants.

\subsection*{NGC 6633}

Two stars of this open cluster are included in our sample, NGC 
6633\_78 and 100, where we adopt the numbering system of 
\citet{K43}. The proper motion study by \citet{VKP58} indicated that star 
100 is a member. The radial velocity measurements of 
\citet{MM89} indicated that both stars are single stars and true 
members. According to their positions in the CMD (Fig.\@ \ref{fig:iso2}), 
star 100 is a clump giant and star 78 might be an early-AGB star.

\section{Discussion}\label{sec:dis}

\begin{figure}
\begin{centering}
\includegraphics[width=7cm]{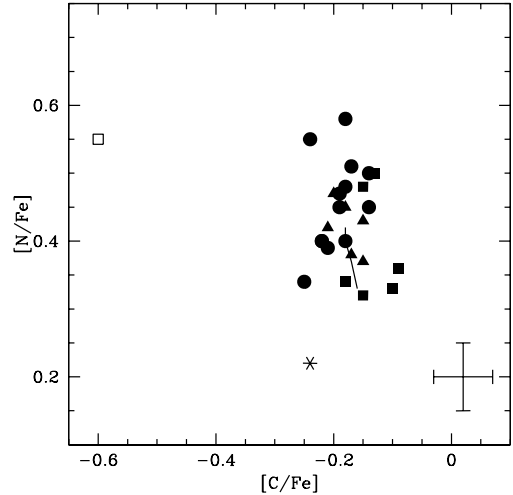}
\caption{Nitrogen abundances, [N/Fe], as a function of the carbon abundances, [C/Fe], for 
the sample stars. Star IC 4756\_69 is shown as an open square, star NGC 2360\_62 as 
a starred symbol. Possible RGB stars, including the ones with dubious classification, 
are shown as full squares, clump giants as full circles, and possible early-AGBs as 
full triangles. The solid line connects the values expected for stars between 1.7 and 
4.0 M$_{\odot}$ in the models of \citet{Sch92}. A typical error bar is shown.}
\label{fig:cfenfe}
\end{centering}
\end{figure}

\subsection{Carbon, nitrogen, and oxygen}

When the material is processed by the CNO-cycle, the relative abundances 
of C, N, and O change, but the sum C+N+O should remain constant. In 
Table \ref{tab:abun}, we list the sums C+N, O+N, C+O, and C+N+O for the 
sample stars. The sum of C+N+O in our sample varies slightly from 
8.79 to 8.98, with an average of 8.88 $\pm$ 0.06. This value is close 
to the solar one, 9.00, and agrees with the almost solar average 
metallicity of our sample, [Fe/H] = +0.05 $\pm$ 0.06. The agreement 
between stars in a given cluster is also excellent, confirming that 
the observed mixing effects on CNO abundances are the result of the 
CNO-cycle.

In Fig.\@ \ref{fig:cfenfe}, we plot the nitrogen abundances of the sample 
stars as a function of the carbon abundances. The isolated open square with 
high N and low C is star IC 4756\_69, a binary star with a low-mass 
companion, probably a white dwarf. For this object, the observed CNO abundances 
(in reality, an upper limit for carbon and a lower limit for nitrogen) are 
compatible with those observed in more massive giants \citep{S06} and 
are probably caused by a mass-transfer event in the system. We thus 
excluded IC 4756\_69 from the following plots and discussion. The star 
with the lowest N content is NGC 2360\_62 (starred symbol in Fig.\@ 
\ref{fig:cfenfe}), which is also a binary star. However, the reason for its low 
nitrogen abundance is unclear. 

\begin{figure}
\begin{centering}
\includegraphics[width=7cm]{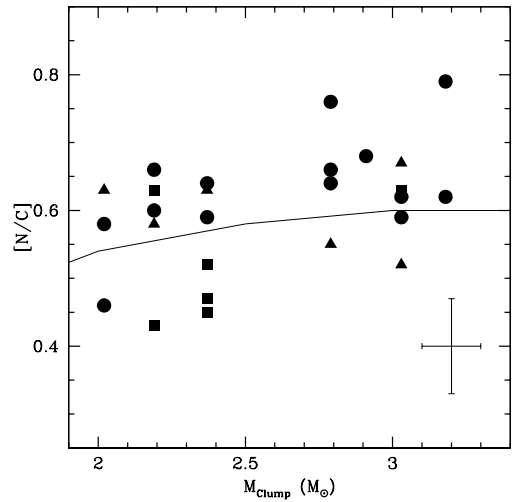}
\caption{[N/C] ratio as a function of the mass of the clump stars. Possible RGB 
stars are shown as full squares, clump giants as full circles, and possible 
early-AGBs as full triangles.The solid line represents the predicted [N/C] 
as a function of initial stellar mass given by the models of \citet{Sch92}.}
\label{fig:ncmcl}
\end{centering}
\end{figure}

Excluding these 2 stars, one can note that there is no correlation between 
C and N; a correlation coefficient of $\rho$ = $-$0.14 is found. The 
remaining stars have an average [C/Fe] =  $-$0.17 $\pm$ 0.04 and an average 
[N/Fe] = +0.43 $\pm$ 0.07. The low rms values imply that the stars have very 
similar abundances. To discuss mixing, however, it is more appropriate to 
use the [N/C] ratio rather than the N or C abundances. This ratio is given 
in Table \ref{tab:abun}. The sample has an average of [N/C] = +0.61 $\pm$ 
0.08, in excellent agreement with the ratio predicted after the first 
dredge-up by the models of \citet{Sch92} for a star with 2.5 M$_{\odot}$, 
[N/C] = +0.58 \citep{Ch94}.

In Fig.\@ \ref{fig:ncmcl}, we plot the [N/C] ratio as a function of the 
stellar mass at the clump of each cluster. The clump mass must be very close 
to the mass of the sample stars in the individual clusters. Only small 
differences are expected, as can be seen by comparing the mass at 
the red turn-off and the mass at the clump, as listed in Table \ref{tab:opc}. 
In the same figure, we show as a solid line the [N/C] predicted by the models 
of \citet{Sch92} as a function of mass after the first dredge-up. A small 
increase in [N/C] with stellar mass is predicted, in agreement with the 
observed behavior. A small difference in the average abundances on each 
side of the mass gap, between 2.4 and 2.8 M$_{\odot}$, is suggested by the 
figure. The stars of lower mass have, on average [N/C] = +0.57 $\pm$ 0.08, 
while the stars of higher mass have [N/C] = +0.64 $\pm$ 0.08. Given the 
observational uncertainties, it is however difficult to judge whether 
this difference is real.

The stars in all figures of this section are plotted with different 
symbols according to their possible evolutionary stages. The group of 
stars with higher mass in Fig.\@ \ref{fig:ncmcl} ($>$ 2.5 M$_{\odot}$) 
mostly consist of more evolved clump (circles) and early-AGB stars 
(triangles). On the other hand, most less evolved RGB stars 
(squares) have masses lower than 2.5 M$_{\odot}$. These less evolved stars 
tend to have lower [N/C] ratios than clump and early-AGBs of 
the same mass.

The four RGB stars in the lower mass range with smaller [N/C] 
are IC 4756\_28, IC 4756\_38, NGC 5822\_240, and NGC 6134\_202. The 
circle with low [N/C] is star NGC 2360\_62, the binary with an 
anomalous low N abundance. This group of four stars has an average of [N/C] = +0.47 
$\pm$ 0.04. The remaining, and possibly more evolved, stars with masses 
lower than 2.4 M$_{\odot}$ have [N/C] = +0.63 $\pm$ 0.03. In case the 
difference in the abundances is real, it might be related to the 
evolutionary status of the stars.

There are three other stars in the sample that could potentially 
be RGB stars based on the CMDs, NGC 3532\_122, NGC 3680\_13, and NGC 5822\_443. 
We do not have the [N/C] ratio for NGC 3680\_13. Both of the other stars have [N/C] = 
+0.63. The higher [N/C] value of these last two stars might argue 
they are not RGB stars but more evolved giants.

We note that the models by \citet{Sch92} may overestimate the effect of the 
first dredge-up for stars with $\sim $2.2 M$_{\odot}$. For a 1.5 M$_{\odot}$ star, 
\citet{Sch92} predict [N/C] = +0.44, closer to the average value of the RGB stars discussed above. 
Observations must thus be compared to predictions of models 
that include up-to-date initial abundances and solar 
mixtures. Work is in progress in that direction.

\begin{figure}
\begin{centering}
\includegraphics[width=7cm]{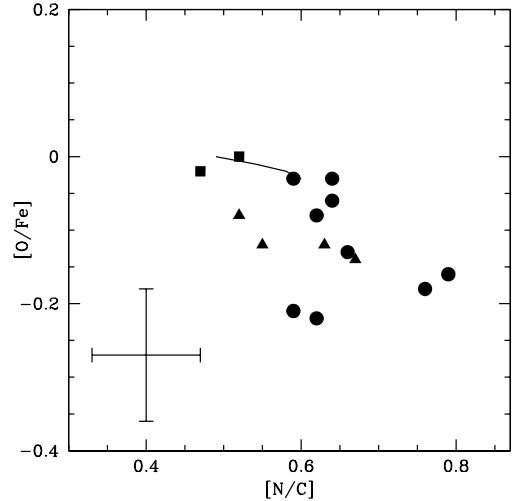}
\caption{The oxygen abundance of the sample stars as a function of the 
[N/C] ratio. Star IC 4756\_69 is not shown. Symbols are as in 
Fig.\@ \ref{fig:ncmcl}. The solid line connects the values expected for stars between 1.7 and 
4.0 M$_{\odot}$ in the models of \citet{Sch92}. A typical error bar is shown.}
\label{fig:ncoxy}
\end{centering}
\end{figure}

In Fig.\@ \ref{fig:ncoxy}, we show the oxygen abundances of the sample 
stars as a function of the [N/C] ratio. A weak anti-correlation is evident 
with $\rho$ = $-$0.53. This correlation was not detected by \citet{S06} 
for more massive stars, M/M$_{\odot}$ $\geq$ 4. However, the correlation 
was seen by \citet{Luck06} for these more massive stars. Given the uncertainties, 
one needs to be cautious in these interpretations. Further analyses with 
larger samples are necessary before a conclusion can be made.

\begin{figure}
\begin{centering}
\includegraphics[width=7cm]{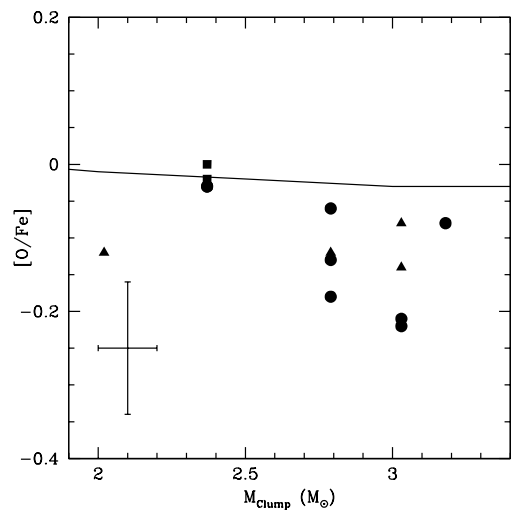}
\caption{The oxygen abundance of the sample stars as a function of the stellar mass 
at the clump. Symbols are as in Fig.\@ \ref{fig:ncmcl}. The solid line represents 
the predicted [O/Fe] as a function of initial stellar mass given by the models 
of \citet{Sch92}.}
\label{fig:mclofe}
\end{centering}
\end{figure}

In Fig.\@ \ref{fig:mclofe}, we show the [O/Fe] ratio as a function of the 
stellar mass. The solid line represents the oxygen abundance in the 
models of \citet{Sch92} after the first dredge-up. Only a small change in 
[O/Fe] is expected. Our results suggest a slightly more pronounced dependence 
with the mass. However, it is interesting that the stars on 
the theoretical curve belong to the same cluster, IC 4756. This includes 
two of the RGB stars, IC 4756\_28 and 38, but also the clump giants IC 4756\_12 
and 14. If the effect that we observe is real, it seems to be an effect related to 
the stellar mass and not to an extra-mixing affecting giants with different 
evolutionary stages. 

\subsection{Sodium}

Sodium abundances have been calculated for giants in a number of open clusters. 
Very different results have however been reported in the literature. Some works 
measured sodium overabundances of as high as [Na/Fe] = +0.60 \citep{Jac07}, some reported 
only a mild overabundance of [Na/Fe] = +0.20 \citep{Ham00}, and others reported 
solar or almost solar abundances \citep{Ses07}. Among the factors responsible for 
this discrepancy was the adoption of different $gfs$ for the Na lines. For 
example, the values of $gfs$ that we adopt were derived with respect to the Sun for the lines 
$\lambda$ 6154 \AA\@ and 6160 \AA\ and by using a solar sodium abundance of A(Na) = 
6.33 \citep{GS98}. These values are the same as those reported in the NIST database. 
They are 0.18 and 0.24 dex higher, respectively, than the values adopted by \citet{Jac07}, which 
had been derived with respect to Arcturus. Another factor is the adoption of 
different NLTE corrections, or even no correction. As discussed before, we 
adopted corrections in our analysis that are based on the work by \citet{Tak03}. 

\begin{figure}
\begin{centering}
\includegraphics[width=7cm]{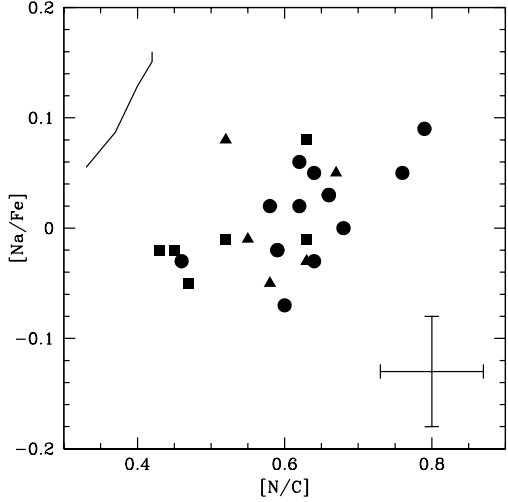}
\caption{Sodium abundance, [Na/Fe], as a function of the [N/C] ratio for the sample 
stars. Symbols are as in the previous figures. The solid line connects the values 
for [Na/Fe] expected for stars between 1.8 and 4.0 M$_{\odot}$ in the models of 
\citet{Mow99} to the [N/C] values of \citet{Sch92}. A typical error bar is shown.}
\label{fig:ncnafe}
\end{centering}
\end{figure}

In Fig.\ \ref{fig:ncnafe}, we plot the sodium abundances derived in this work 
as a function of [N/C]. We obtain a correlation coefficient 
of $\rho$ = 0.49, which indicates if anything only a weak correlation. The 
average sodium abundance of the sample is [Na/Fe] = 0.01 $\pm$ 0.05, for 
the complete interval of [N/C].

\begin{figure}
\begin{centering}
\includegraphics[width=7cm]{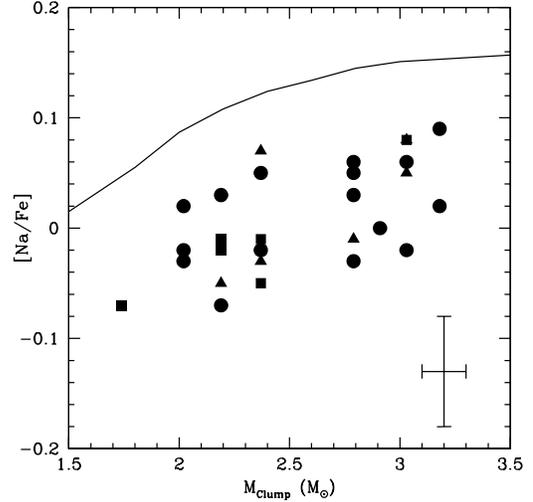}
\caption{Sodium abundances, [Na/Fe], as a function of the mass of the clump stars. A typical 
error bar is shown. The solid line represents the predicted [Na/Fe] as a function of 
initial stellar mass, given by the models of \citet{Mow99}.}
\label{fig:namcl}
\end{centering}
\end{figure}

The sodium abundances are displayed in Fig.\ \ref{fig:namcl} 
as a function of  stellar mass, together with the first dredge-up 
predictions from a standard evolution code \citep{Mow99,Ham00}. 
Both the observed and predicted abundances show an increase of  
about 0.10 dex over the mass range covered by our sample. A 
$2 \times 2$ contingency table was compiled with the limits 
M$_{\rm clump}$ = 2.5 and [Na/Fe] = 0.00. The two-tailed Fisher test 
yields a probability $P = 0.011$ that the samples are drawn from the 
same parent population, which demonstrates that the correlation is significant. The  
agreement would be perfect were it not for a slight offset of about  
0.08 dex between the observed and predicted data. Since we do not  
expect the sodium surface abundance to decrease during the evolution  
of a star by means of the first and second dredge-ups, we may argue that  
our observed values are, on average, slightly underestimated by about  
0.08 dex. While this remains within the margins of error, the most significant  
result of our data is the relative increase in the Na abundances  
as a function of stellar mass by an amount that agrees with model  
predictions.

In summary, our results do not support the claim by 
\citet[and references therein]{Jac07} of sodium overabundances 
as high as +0.60 in giants of open clusters, in agreement with 
the conclusions of \citet{R06} and \citet{Ses07,Ses08}. 
Our results instead are consistent with a slight dependence, 
as predicted by standard stellar evolution models, of the sodium 
abundance at the surface of red giants on stellar mass.

\subsection{$^{12}$C/$^{13}$C}

\citet{G89} and \citet{GB91} analyzed stars from 20 open clusters and showed 
that the $^{12}$C/$^{13}$C ratio in giants of clusters with turn-off masses higher 
than $\sim$ 2.2M$_{\odot}$ agree with the expected behavior from standard 
first dredge-up models, $^{12}$C/$^{13}$C $\sim$ 20 \citep{Sch92} (see also 
Fig.~6 of \citeauthor{Ch94} \citeyear{Ch94}). They also found, on the other hand, 
that giants in clusters with turn-off masses lower than $\sim$ 2.2M$_{\odot}$ 
have a decreasing carbon isotopic ratio with decreasing turn-off mass. 

Although \citet{GB91} found no evidence in M67 for different carbon isotopic 
ratios between red giants and clump giants, \citet{T00} analyzing a larger 
sample of M67 clump giants found a small difference between the isotopic 
ratios of these two groups. A possible additional mixing after the He-core 
flash was suggested. A similar difference between clump and red giants, 
however, was not found by \citet{T05} in NGC 7789.

\begin{figure}
\begin{centering}
\includegraphics[width=7cm]{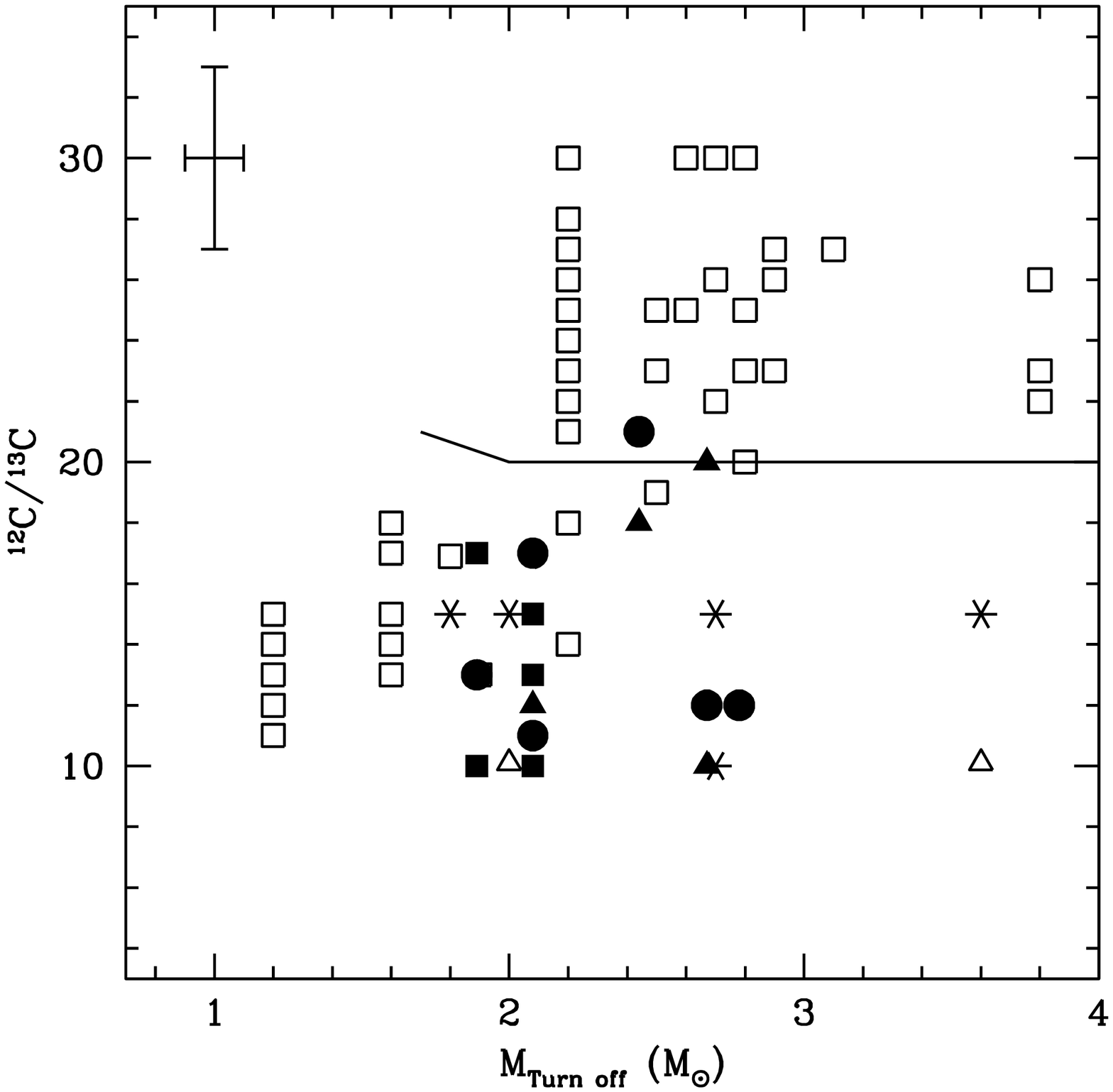}
\caption{The carbon isotopic ratio, $^{12}$C/$^{13}$C, as a function of 
turn-off mass. Stars analyzed in this work are plotted as in the previous 
figures: the stars analyzed by \citet{G89} are shown as open squares, and the stars analyzed by 
\citet{L94} are shown as starred symbols, or as an open triangle when only a lower 
limit was available. The theoretical value of $^{12}$C/$^{13}$C as a function of 
the initial stellar mass from the models of \citet{Sch92} is shown as a solid 
line.}
\label{fig:c12c13mto}
\end{centering}
\end{figure}

In Fig.\@ \ref{fig:c12c13mto}, we plot the carbon isotopic ratios of our sample as a 
function of the turn-off mass. The isotopic ratios of the cluster stars 
analyzed by \citet{G89} and \citet{L94} are also shown. In this figure, we decided 
to plot turn-off mass and not the clump mass, which is a more reliable indicator of 
the true stellar mass, to facilitate the comparison with previous results. 
This figure shows that part of our determinations are in good 
agreement with the relation between $^{12}$C/$^{13}$C and turn-off 
mass found by \citet{G89}. However, it is also clear that some 
stars have lower $^{12}$C/$^{13}$C ratios than previously found. 
In the case of the cluster stars with M$_{\rm TO}$ $\sim$ 2.0 M$_{\odot}$, 
we find additional stars with low carbon isotopic ratios. The mean 
ratio for stars in this region is $^{12}$C/$^{13}$C = 13.1 $\pm$ 2.6. 
The second group of low $^{12}$C/$^{13}$C stars, however, have higher 
masses, which cannot be explained by the observational scatter. Four stars 
in clusters with  M$_{\rm TO}$ $\geq$ 2.6 M$_{\odot}$ have an average 
of $^{12}$C/$^{13}$C = 11.5 $\pm$ 1.0. These stars, NGC 3532\_19, NGC 
3532\_100, NGC 6281\_3, and NGC 6281\_4, deviate considerably from 
other stars in the same mass range. This behavior has not been clearly detected or reported before. 
This result shows that intermediate-mass stars may also experience an 
extra-mixing episode, which produces a drop of the $^{12}$C/$^{13}$C ratio.

In low-mass RGB stars, the low carbon isotopic ratio is attributed to 
thermohaline mixing operating immediately after the RGB bump \citep{CZ07}. 
For intermediate-mass stars, thermohaline mixing is not expected to occur 
during the RGB because these stars ignite central helium-burning before 
reaching the bump. However, as suggested by \citet{ChB00}, a similar 
mechanism might operate in higher mass stars during the early-AGB 
phase (see also \citeauthor{CaLan08} \citeyear{CaLan08}). This effect 
might explain the observed low $^{12}$C/$^{13}$C ratio in these stars. 

However, when we consider the stellar evolutionary status that we have determined, 
some difficulties remain. Indeed of the four stars with M/M$_{\odot}$ $\geq$ 
2.0 and low $^{12}$C/$^{13}$C, only NGC 3532\_100 was classified as an early-AGB, 
the remaining three stars being classified as clump stars. Furthermore, the 
two stars classified as AGBs with M/M$_{\odot}$ $\geq$ 2.0 have high $^{12}$C/$^{13}$C 
(NGC 3532\_670 and NGC 6633\_78). This could imply that not all 
of these stars will develop a low $^{12}$C/$^{13}$C ratio. As discussed by \citet{CZ07b}, 
thermohaline mixing could indeed be inhibited by fossil magnetic fields in stars 
that are descendants of Ap stars. On the other hand, one should also keep in mind 
that a reliable classification of these stars is not an easy task and this could affect our findings.

The $^{12}$C/$^{13}$C ratio of the four probable RGB stars was determined 
and found to be low for three of them, IC 4756\_28, 38, and NGC 5822\_443. In 
the CMD two of the stars seem to be post-bump giants, which explains their low ratio. 
However, IC 4756\_38 was classified as a pre-bump giant or alternatively as a 
clump star in spite of its low [N/C]. We note that 
the cluster IC 4756 is affected by differential reddening, which complicates 
the classification of its member stars.

\section{Conclusions}\label{sec:con}

We have reported the observational results of a homogeneous abundance 
analysis of elements affected by evolutionary mixing in giants 
of open clusters. Abundances of C, N, O, Na and the $^{12}$C/$^{13}$C 
ratio are derived from high resolution, high S/N spectra using 
spectrum synthesis. Our sample consists of 31 objects, including 
red giants, clump giants, and early-AGB stars, in 10 open clusters.

The average [N/C] ratio of the sample, [N/C] = +0.61, is in 
very good agreement with the values predicted by the \citet{Sch92} 
models after the first dredge-up. However, we identify a 
group of first ascent red giants with average [N/C] = +0.43 
$\pm$ 0.04, which is lower than for the more evolved stars, with [N/C] = +0.63 
$\pm$ 0.03. This result might indicate a real difference 
in mixing between red giants and clump or early-AGB giants, 
in contrast to that expected in the standard stellar models, 
but in agreement with models including thermohaline convection 
(Charbonnel \& Zahn 2007a; Charbonnel et al. in preparation).

A weak trend with mass, in the sense of smaller [O/Fe] for higher 
mass stars, is also suggested. This must however be interpreted with 
caution because of the observational error bars.

The sodium abundances derived in our sample are between 
the values [Na/Fe]=-0.08 and 0.10 dex. We thus do not measure any high sodium  
overabundance, i.e., values of up to +0.60 dex, as previously reported 
in the literature. However, the strength of our analysis is the 
range of stellar masses covered by our stars, inbetween 1.8 and 3.2 M$_{\odot}$, 
in which stellar evolution models predict an increase of 0.10 dex in 
the surface sodium enhancement as a function of stellar mass. Our 
results agree with both the continuous increase in [Na/Fe] 
predicted as a function of stellar mass and the predicted amplitude of this 
increase of 0.10 between 1.8 and 3.2 M$_{\odot}$.

The well-known correlation between $^{12}$C/$^{13}$C and 
mass, for stars with M/M$_{\odot}$ $\leq$ 2.5, is also seen in 
our results. However, we also discuss 
for the first time a group of slightly more massive 
stars with low $^{12}$C/$^{13}$C. Since these more massive stars 
do not go through the bump while on the RGB, the extra-mixing 
events that modify their surface abundances do not occur at 
the same evolutionary phase as for stars with M/M$_{\odot}$ 
$\leq$ 2.5. However, as discussed by \citet{ChB00}, an extra-mixing 
event might take place in these stars during the early-AGB and 
could be related to thermohaline convection \citep[see][]{CZ07,CaLan08}

A detailed comparison of these observational results with the predictions of 
evolutionary models that include the effects of the thermohaline mixing, will be 
presented in a forthcoming paper.

\begin{acknowledgements}
We are indebted to Dr. Daniel Erspamer, former PhD student at Institut
d'Astronomie de l'Universit de Lausanne, for having carried out the
observations. This research has made use of the WEBDA database, operated at the
Institute for Astronomy of the University of Vienna and of the Simbad
database operated at CDS, Strasbourg, France. R.S. acknowledges 
a FAPESP PhD fellowship (04/13667-4). R.G., P.N., and C.C. thank the 
Swiss National Science Foundation for its support. 
\end{acknowledgements}

\bibliographystyle{aa}
\bibliography{rsmiljanic}



\Online

\appendix

\section{Equivalent widths}

\begin{longtable}{ccccccccccc}
\caption{\label{tab:LE} Equivalent widths of the lines used in the abundance analysis 
of the stars IC 2714\_5, IC 4756\_12, 14, 28, 38, 69, and NGC 3532\_19. 
Lines with equivalent widths smaller than 10 m$\AA$ and larger than 150 m$\AA$ were 
not used.} \\ 
\hline\hline
$\lambda$ (\AA) & Elem. & $\chi$ (eV) & log gf & 2714\_5 & 4756\_12 & 4756\_14 & 
4756\_28 & 4756\_38 & 4756\_69 & 3532\_19 \\
\hline  
\endfirsthead
\caption{continued.} \\
\hline\hline
$\lambda$ (\AA) & Elem. & $\chi$ (eV) & log gf & 2714\_5 & 4756\_12 & 4756\_14 & 
4756\_28 & 4756\_38 & 4756\_69 & 3532\_19 \\
\hline
\endhead
\hline
\endfoot
6154.22 & NA1 & 2.10 & $-$1.560 &  85.4 &  73.3 &  85.5 &  96.0 &  68.6 &  80.0 &  89.8 \\
6160.75 & NA1 & 2.10 & $-$1.260 & 103.0 &  93.5 & 114.4 & 114.9 &  91.9 & 100.9 & 105.8 \\
5528.42 & MG1 & 4.34 & $-$0.470 & 234.9 & 225.1 & 246.6 & 257.3 & 228.4 & 212.6 & 235.4 \\
5711.09 & MG1 & 4.34 & $-$1.750 & 137.6 & 127.7 & 145.8 & 145.6 & 125.9 & 126.0 & 137.1 \\
5772.15 & SI1 & 5.06 & $-$1.790 &  91.2 &  76.9 &  79.0 &  78.0 &   --  &  --   &  82.4 \\
6125.03 & SI1 & 5.59 & $-$1.660 &  45.8 &  43.3 &  44.0 &  55.5 &  44.4 &  45.5 &  55.3 \\
6131.58 & SI1 & 5.59 & $-$1.840 &  45.9 &  28.1 &  38.4 &  37.2 &  35.7 &  34.0 &  42.2 \\
6131.86 & SI1 & 5.59 & $-$1.770 &  37.1 &  32.0 &  36.6 &  40.4 &  33.3 &  38.1 &  43.4 \\
6142.53 & SI1 & 5.59 & $-$1.580 &  52.5 &  43.7 &  44.0 &  45.1 &  44.8 &  48.7 &  49.7 \\
6145.08 & SI1 & 5.59 & $-$1.500 &  54.8 &  45.9 &  47.8 &  49.2 &  45.4 &  48.4 &  58.3 \\
6155.14 & SI1 & 5.59 & $-$0.890 &  96.6 &  93.6 &  96.8 &  96.9 &  92.5 &  94.7 & 105.8 \\
5867.57 & CA1 & 2.92 & $-$1.760 &  50.0 &  46.7 &  62.1 &  63.5 &  45.8 &  45.5 &  52.9 \\
6122.23 & CA1 & 1.88 & $-$0.180 & 215.8 & 197.2 & 228.2 & 235.4 & 199.8 & 198.1 & 213.4 \\
6156.03 & CA1 & 2.51 & $-$2.580 &  30.5 &  18.9 &  33.8 &  38.7 &  18.6 &  18.6 &  43.1 \\
6161.29 & CA1 & 2.51 & $-$1.370 &  --   & 103.3 &  --   &  --   &  98.5 & 100.5 & 120.5 \\
6166.44 & CA1 & 2.51 & $-$1.270 & 105.7 &  97.7 & 119.3 & 119.8 &  95.1 &  97.5 & 110.5 \\
6169.04 & CA1 & 2.51 & $-$0.800 & 129.2 & 123.9 & 144.7 & 146.6 & 120.3 & 122.2 & 134.1 \\
6169.56 & CA1 & 2.51 & $-$0.580 & 147.6 & 137.1 & 157.7 & 160.1 & 135.3 & 138.1 & 149.1 \\
6493.78 & CA1 & 2.51 & $-$0.280 & 172.2 & 155.6 & 178.1 & 178.9 & 154.9 & 162.1 & 169.3 \\
6499.65 & CA1 & 2.51 & $-$0.970 & 128.4 & 121.0 & 140.1 & 141.6 & 116.0 & 117.7 & 134.0 \\
5318.34 & SC2 & 1.35 & $-$1.890 &  45.6 &  39.6 &  53.1 &  48.7 &  38.8 &  42.3 &  50.0 \\
5334.22 & SC2 & 1.49 & $-$2.200 &  23.6 &  20.0 &  32.8 &  28.6 &  17.9 &  20.8 &  22.4 \\
5145.47 & TI1 & 1.45 & $-$0.590 &  87.8 &  78.5 & 110.0 & 113.0 &  77.9 &  76.3 &  93.8 \\
5295.78 & TI1 & 1.06 & $-$1.790 &  48.9 &  42.7 &  78.5 &  79.4 &  43.3 &  --   &  53.5 \\
5299.98 & TI1 & 1.05 & $-$1.750 &  42.8 &  39.9 &  64.7 &  --   &  41.7 &  42.0 &  49.2 \\
5338.33 & TI1 & 0.82 & $-$2.100 &  --  &  32.2 &  49.2 &   --   &  31.9 &  34.7 &  38.3 \\
5351.07 & TI1 & 2.77 & $-$0.210 &  24.8 &  30.9 &  48.5 &  46.7 &  28.2 &  27.3 &  37.6 \\
5766.33 & TI1 & 3.28 &  0.220 &  28.0 &  25.9 &  36.9 &  45.3 &  29.1 &   --  &  32.6 \\
6121.01 & TI1 & 1.87 & $-$1.480 &  22.3 &  15.7 &  33.7 &  42.7 &  13.9 &  15.6 &  23.3 \\
6126.22 & TI1 & 1.06 & $-$1.480 &  76.5 &  62.6 &  98.2 & 105.5 &  61.7 &  62.8 &  79.8 \\
6497.68 & TI1 & 1.44 & $-$2.070 &  22.9 &  19.9 &  42.6 &  48.6 &  18.5 &  13.0 &  26.0 \\
5846.27 & V1  & 3.12 &  0.700 &  15.7 &  16.6 &  24.2 &  26.1 &  12.7 &   9.9 &  18.6 \\
6002.65 & V1  & 1.05 & $-$1.720 &  18.4 &  16.2 &  36.7 &  42.7 &   --  &   --  &  --  \\
6039.69 & V1  & 1.06 & $-$0.740 &  51.0 &  44.6 &  80.9 &  84.7 &  40.6 &  38.7 &  56.6 \\
6111.65 & V1  & 1.04 & $-$0.720 &  51.7 &  43.0 &  87.1 &  97.5 &  41.2 &  42.4 &  55.7 \\
6119.53 & V1  & 1.06 & $-$0.560 &  73.5 &  64.5 &  94.4 &  97.0 &  57.4 &  58.6 &  74.1 \\
6135.37 & V1  & 1.05 & $-$0.910 &  52.4 &  41.2 &  81.0 &  88.6 &  38.6 &  40.9 &  55.0 \\
6150.15 & V1  & 0.30 & $-$1.680 &  56.9 &  47.8 &  98.6 & 110.5 &  45.0 &  43.9 &  67.1 \\
6504.19 & V1  & 1.18 & $-$0.830 &  41.6 &  36.9 &  61.5 &  68.0 &  36.3 &  33.7 &  48.1 \\
5303.22 & V2  & 2.27 & $-$2.040 &  26.4 &  23.0 &  37.6 &  31.3 &  19.0 &  26.5 &  31.0 \\
6028.28 & V2  & 2.48 & $-$1.990 &  15.8 &  15.0 &  18.9 &  16.4 &  13.6 &  12.4 &  17.3 \\
5122.12 & CR1 & 1.03 & $-$3.240 &  58.0 &  49.5 &  83.9 &  95.1 &  49.9 &  48.5 &  65.6 \\
5296.69 & CR1 & 0.98 & $-$1.510 & 152.5 & 139.3 & 177.2 & 180.4 & 136.0 & 141.8 & 155.1 \\
5300.75 & CR1 & 0.98 & $-$2.230 & 119.2 & 104.2 & 134.3 & 131.5 &  98.0 & 102.2 & 113.7 \\
5304.18 & CR1 & 3.45 & $-$0.770 &  33.7 &  29.6 &  48.1 &  50.1 &  31.0 &  33.7 &  37.1 \\
5312.88 & CR1 & 3.43 & $-$0.690 &  53.6 &  35.4 &  54.1 &  55.8 &  40.4 &  37.4 &  47.5 \\
5318.78 & CR1 & 3.43 & $-$0.800 &  30.4 &  31.9 &  52.8 &  51.2 &  32.6 &  32.3 &  37.3 \\
5329.12 & CR1 & 2.90 & $-$0.140 &  99.7 & 102.2 & 128.1 & 130.1 & 100.9 & 105.5 & 109.0 \\
5340.44 & CR1 & 3.42 & $-$0.840 &  --   &  33.2 &  50.6 &  50.4 &  31.9 &  35.7 &  38.0 \\
5348.32 & CR1 & 1.00 & $-$1.370 & 151.9 & 140.8 & 174.6 & 180.2 & 138.8 & 142.8 & 154.5 \\
5783.07 & CR1 & 3.31 & $-$0.400 &  69.6 &  63.2 &  76.6 &  77.5 &  59.8 &  60.3 &  63.6 \\
5783.87 & CR1 & 3.31 & $-$0.560 &  --   &  --   &   --  &   --  &   --  &  --   &  --   \\
5787.99 & CR1 & 3.31 & $-$0.260 &  81.4 &  75.2 &  91.2 &  87.3 &  71.7 &  --   &  82.7 \\
5788.39 & CR1 & 3.00 & $-$1.720 &  21.3 &  21.9 &  38.1 &  40.8 &  18.4 &  16.5 &  25.0 \\
5844.61 & CR1 & 3.00 & $-$1.820 &  24.9 &  18.4 &  30.6 &  36.5 &  17.4 &  12.6 &  25.0 \\
5863.96 & CR1 & 3.11 & $-$1.970 &  --   &  --   &  22.5 &  25.9 &   --  &   --  &  --  \\
6135.78 & CR1 & 4.80 &  0.550 &  34.9 &  30.1 &  38.9 &  43.9 &  29.7 &  30.5 &  40.6 \\
6501.21 & CR1 & 0.98 & $-$3.730 &  25.4 &  23.7 &  44.3 &  49.5 &  22.8 &  21.1 &  35.3 \\
6630.02 & CR1 & 1.03 & $-$3.240 &  40.3 &  --   &   --  &  73.9 &   --  &  --   &  --  \\
5305.87 & CR2 & 3.81 & $-$2.240 &  50.1 &  47.1 &   --  &  45.0 &  41.3 &  51.4 &  56.5 \\
5310.70 & CR2 & 4.05 & $-$2.410 &  31.7 &  26.7 &  27.5 &  25.4 &  25.8 &  28.1 &  35.8 \\
5313.59 & CR2 & 4.06 & $-$1.840 &  66.8 &  60.2 &  57.6 &  61.0 &  53.3 &  62.8 &  67.7 \\
5334.88 & CR2 & 4.05 & $-$1.750 &  58.9 &  54.0 &  62.1 &  58.5 &  54.4 &  56.8 &  62.7 \\
5133.69 & FE1 & 4.18 &  0.140 & 179.8 & 169.5 & 189.0 & 198.7 & 171.6 & 168.3 & 178.6 \\
5141.75 & FE1 & 2.42 & $-$2.240 & 139.7 & 129.2 & 159.7 & 146.1 & 128.3 & 125.1 & 144.0 \\
5143.73 & FE1 & 2.20 & $-$3.690 &  78.4 &  71.2 & 104.9 &  --   &  69.0 &  64.3 &  86.7 \\
5293.97 & FE1 & 4.14 & $-$1.840 &  62.4 &  58.8 &  75.6 &  71.1 &  55.7 &  59.7 &  67.6 \\
5294.55 & FE1 & 3.64 & $-$2.810 &  40.1 &  34.1 &  56.7 &  54.1 &  34.4 &  36.2 &  41.8 \\
5295.32 & FE1 & 4.42 & $-$1.670 &  56.1 &  42.5 &  66.8 &  61.3 &  49.1 &  53.1 &  59.0 \\
5307.36 & FE1 & 1.61 & $-$2.978 & 154.4 & 139.5 & 170.7 & 165.4 & 133.3 & 139.6 & 155.2 \\
5315.07 & FE1 & 4.37 & $-$1.550 &  --   &  --   &  74.0 &  75.9 &  61.3 &  64.4 &  77.8 \\
5320.05 & FE1 & 3.64 & $-$2.490 &  50.2 &  42.4 &  59.5 &  60.0 &  43.9 &  44.8 &  52.5 \\
5321.11 & FE1 & 4.44 & $-$1.090 &  --   &  --   &  79.7 &  --   &  --   &  --   &  --   \\
5322.05 & FE1 & 2.28 & $-$2.800 & 112.0 &  --   & 126.8 & 124.4 &  --   & 101.5 & 115.3 \\
5326.79 & FE1 & 4.42 & $-$2.090 &  36.9 &  33.5 &  --   &  49.6 &  34.8 &  35.0 &  37.9 \\
5339.94 & FE1 & 3.27 & $-$0.720 & 172.4 & 162.1 & 182.9 & 183.3 & 161.9 & 161.3 & 170.8 \\
5358.10 & FE1 & 3.29 & $-$3.400 &  --   &  --   &  --   &  --   &  --   &  --   &  --   \\
5367.47 & FE1 & 4.42 &  0.443 & 153.5 &   --  & 152.5 & 152.4 &  -- &  --  & 151.3 \\
5369.97 & FE1 & 4.37 &  0.536 &  --   &   --  & 168.3 & 170.8 &  --   &  --   &  --   \\
5568.81 & FE1 & 3.64 & $-$2.950 &  33.6 &  28.4 &  41.5 &  48.9 &  28.0 &  30.6 &  37.3 \\
5759.27 & FE1 & 4.65 & $-$2.070 &  18.9 &  16.5 &  22.6 &  --   &  17.9 &  16.6 &  19.7 \\
5760.35 & FE1 & 3.64 & $-$2.490 &  54.2 &  45.8 &  60.4 &  64.7 &  44.7 &  46.2 &  56.9 \\
5775.09 & FE1 & 0.05 & $-$1.298 &   --  &   --  &  --   &  --   &  --   &  --   &   --  \\
5778.47 & FE1 & 2.59 & $-$3.430 &  70.2 &  61.4 &  81.3 &  83.2 &  55.9 &  60.3 &  65.7 \\
5784.69 & FE1 & 3.40 & $-$2.532 &  --   &  62.0 &  77.5 &  75.3 &  55.1 &  60.4 &  65.5 \\
5838.42 & FE1 & 3.94 & $-$2.290 &  49.2 &  42.4 &  60.0 &  58.0 &  41.6 &  42.2 &  52.9 \\
5849.70 & FE1 & 3.69 & $-$2.990 &  24.2 &  29.3 &  43.4 &  44.5 &  29.7 &  24.8 &  27.0 \\
5852.19 & FE1 & 4.55 & $-$1.300 &  75.6 &  66.8 &  89.0 &  --   &  63.8 &  66.8 &  78.5 \\
5853.18 & FE1 & 1.48 & $-$5.270 &  39.0 &  33.8 &  59.7 &  62.6 &  32.8 &  35.2 &  46.9 \\
5855.09 & FE1 & 4.61 & $-$1.478 &   --  &  40.8 &  50.4 &  54.2 &  40.5 &  41.8 &  46.0 \\
5856.08 & FE1 & 4.29 & $-$1.328 &  --   &   --  &  --   &  --   &  --   &  --   &  --   \\
5858.77 & FE1 & 4.22 & $-$2.260 &  37.6 &  32.7 &  42.6 &  44.9 &  30.6 &  31.1 &  38.5 \\
5859.61 & FE1 & 4.53 & $-$0.600 &  --   &   --  &  --   &  --   &  --   &  --   &  --   \\
5862.36 & FE1 & 4.53 & $-$0.250 &  --   &  --   &  --   &  --   &  --   &  --   &  --   \\
6003.03 & FE1 & 3.88 & $-$1.110 & 118.2 & 109.6 & 123.1 & 124.4 & 105.8 & 107.4 & 121.6 \\
6007.96 & FE1 & 4.63 & $-$0.750 &  --   &  --   &   --  & --    &   --  &   --  &  --   \\
6008.58 & FE1 & 3.87 & $-$1.100 &  --   &  --   &   --  & --    &   --  &   --  &  --   \\
6015.25 & FE1 & 2.22 & $-$4.680 &  28.5 &  23.4 &  38.1 &  48.4 &  23.4 &  21.0 &  32.3 \\
6019.36 & FE1 & 3.57 & $-$3.360 &  22.7 &  19.3 &  30.5 &  34.6 &  19.5 &  16.7 &  24.6 \\
6024.07 & FE1 & 4.55 & $-$0.110 & 138.4 & 126.6 & 134.1 & 137.9 & 123.5 & 126.6 & 137.0 \\
6027.06 & FE1 & 4.08 & $-$1.089 & 107.7 &  95.1 & 106.7 & 107.5 &  90.4 &  94.8 & 104.9 \\
6034.04 & FE1 & 4.31 & $-$2.420 &  30.9 &   --  &  29.6 &  37.1 &  19.5 &  20.9 &  35.2 \\
6035.34 & FE1 & 4.29 & $-$2.590 &  17.3 &  15.5 &  23.6 &  27.0 &  17.7 &  17.8 &  20.7 \\
6054.10 & FE1 & 4.37 & $-$2.300 &  24.3 &  18.7 &  29.0 &  29.0 &  18.7 &  18.6 &  26.2 \\
6120.25 & FE1 & 0.91 & $-$5.970 &  39.0 &  31.6 &  59.1 &  65.9 &  28.6 &  27.9 &  40.0 \\
6151.62 & FE1 & 2.18 & $-$3.299 & 101.7 &  91.4 & 109.7 & 116.6 &  86.0 &  90.5 & 105.7 \\
6157.73 & FE1 & 4.08 & $-$1.220 & 106.4 &  98.6 & 121.4 & 121.4 &  94.8 &  97.7 & 119.3 \\
6165.37 & FE1 & 4.14 & $-$1.474 &  78.3 &  69.6 &  87.1 &  85.5 &  67.7 &  69.3 &  82.4 \\
6173.34 & FE1 & 2.22 & $-$2.880 & 131.8 & 116.9 & 143.3 & 143.0 & 109.0 & 111.7 & 132.4 \\
6475.63 & FE1 & 2.56 & $-$2.940 & 114.3 & 104.4 & 133.9 & 130.3 &  97.9 & 100.2 & 118.1 \\
6481.87 & FE1 & 2.28 & $-$2.984 & 121.9 & 108.2 & 134.4 & 137.2 & 105.3 & 103.4 & 125.1 \\
6483.94 & FE1 & 1.48 & $-$5.650 &   --  &  --   &  --   &  --   &  --   &   --  &  --   \\
6495.74 & FE1 & 4.83 & $-$0.920 &  84.1 &  61.8 &  80.5 &  78.8 &  60.0 &  65.5 &  69.8 \\
6496.47 & FE1 & 4.80 & $-$0.610 &  90.9 &  92.7 & 110.3 & 104.4 &  87.8 &  88.7 &  97.8 \\
6498.95 & FE1 & 0.96 & $-$4.687 & 113.3 & 106.3 & 140.2 & 142.3 &  99.8 & 102.6 & 125.8 \\
6627.56 & FE1 & 4.55 & $-$1.680 &  56.6 &  46.5 &  60.4 &  64.4 &  47.8 &  48.7 &  60.7 \\
6633.42 & FE1 & 4.83 & $-$1.490 &  --   &  --   &  62.7 &  60.6 &  --   &  --   &  --   \\
6633.76 & FE1 & 4.56 & $-$0.799 &  89.6 &  88.3 & 100.0 &  98.3 &  88.4 &  90.3 &  98.5 \\
6646.98 & FE1 & 2.61 & $-$3.990 &  48.5 &  36.7 &  63.5 &  69.0 &  36.9 &  37.6 &  54.9 \\
6648.08 & FE1 & 1.01 & $-$5.918 &   --  &  --   &  --   &   --  &  --   &  --   &  --   \\
5132.67 & FE2 & 2.79 & $-$4.110 &  60.8 &  49.5 &  55.2 &  50.4 &  50.9 &  47.9 &  61.6 \\
5256.94 & FE2 & 2.89 & $-$4.050 &  56.9 &  45.4 &  49.2 &  49.1 &  41.5 &  46.9 &  55.2 \\
5264.81 & FE2 & 3.23 & $-$3.200 &   --  &  73.0 &  69.7 &  69.5 &  67.0 &  --   &   --  \\
5325.56 & FE2 & 3.22 & $-$3.160 &  83.0 &  69.9 &  72.6 &  64.5 &  66.3 &  74.3 &  81.1 \\
5414.08 & FE2 & 3.22 & $-$3.650 &  60.2 &  50.5 &  47.1 &  42.5 &  45.8 &  49.0 &  55.1 \\
5425.26 & FE2 & 3.20 & $-$3.220 &  77.8 &  66.3 &  --   &  59.9 &  61.7 &  64.8 &  73.5 \\
6084.10 & FE2 & 3.20 & $-$3.760 &  53.3 &  44.8 &  45.0 &  39.4 &  38.9 &  43.4 &  50.8 \\
6113.33 & FE2 & 3.22 & $-$4.110 &  --   &  32.9 &  34.8 &  --   &  29.8 &  --   &  37.5 \\
6129.70 & FE2 & 3.20 & $-$4.600 &  --   &  --   &  --   &  --   &  --   & --    &  --   \\
6149.24 & FE2 & 3.87 & $-$2.700 &  --   &  54.9 &  53.1 &  50.8 &  51.1 &  57.8 &  66.5 \\
6247.56 & FE2 & 3.89 & $-$2.310 &  89.1 &  76.2 &   --  &  65.9 &  72.2 &  79.0 &  87.6 \\
6369.46 & FE2 & 2.89 & $-$4.150 &  52.2 &  42.5 &  40.8 &  40.5 &  38.7 &  44.3 &  50.6 \\
6416.93 & FE2 & 3.89 & $-$2.720 &  67.5 &  60.6 &  57.8 &  58.3 &  56.3 &  63.7 &  70.4 \\
6456.39 & FE2 & 3.90 & $-$2.060 & 107.3 &  89.5 &  81.4 &  78.8 &  85.2 &  92.3 & 100.4 \\
5301.04 & CO1 & 1.70 & $-$2.080 &  67.0 &  64.0 &  93.3 &  97.4 &  62.0 &  65.3 &  74.8 \\
5325.28 & CO1 & 4.21 & $-$0.100 &  26.3 &  23.1 &  31.5 &  31.0 &  24.5 &  24.2 &  27.3 \\
5342.70 & CO1 & 4.00 &  0.550 &  56.4 &  50.8 &  64.2 &  61.2 &  47.3 &  51.3 &  56.9 \\
5352.05 & CO1 & 3.56 & $-$0.020 &  62.7 &  52.3 &  72.9 &  70.7 &  50.8 &  53.1 &  60.8 \\
5359.20 & CO1 & 4.13 &  0.010 &  22.2 &  19.4 &  23.3 &  23.3 &  17.1 &  18.4 &  23.0 \\
5369.59 & CO1 & 1.73 & $-$1.730 &  --   &  --   &   --  &   --  &   --  &   --  &  --  \\
6117.00 & CO1 & 1.78 & $-$2.570 &  34.8 &  26.6 &  50.1 &  58.1 &  24.4 &  25.6 &  36.3 \\
6490.34 & CO1 & 2.03 & $-$2.580 &  31.2 &  22.0 &  32.5 &  41.4 &  19.3 &  22.3 &  30.1 \\
6632.47 & CO1 & 2.27 & $-$2.060 &  38.7 &  31.2 &  55.5 &  58.6 &  30.3 &  29.7 &  44.1 \\
5137.08 & NI1 & 1.67 & $-$1.630 & 152.2 & 144.2 & 165.2 & 165.0 & 141.1 & 147.5 &  --   \\
5593.74 & NI1 & 3.90 & $-$0.930 &  71.3 &  63.9 &  72.7 &  73.9 &  63.8 &  61.1 &  72.2 \\
5760.83 & NI1 & 4.09 & $-$0.850 &  71.1 &  56.4 &  71.2 &  71.2 &  51.8 &  55.7 &  63.7 \\
5847.01 & NI1 & 1.67 & $-$3.480 &  73.6 &  63.9 &  91.8 &   --  &  58.4 &  63.7 &  77.7 \\
6007.31 & NI1 & 1.67 & $-$3.400 &  73.2 &  64.7 &  82.3 &  86.8 &  61.0 &  61.0 &  76.2 \\
6053.68 & NI1 & 4.22 & $-$1.110 &  48.0 &  35.4 &  47.7 &  48.1 &  35.6 &  36.8 &  51.0 \\
6111.06 & NI1 & 4.07 & $-$0.900 &  64.3 &  55.4 &  64.8 &  66.7 &  54.1 &  56.0 &  63.2 \\
6128.99 & NI1 & 1.67 & $-$3.400 &  76.4 &  63.7 &  89.3 &  95.4 &  59.5 &  63.2 &  77.1 \\
6130.13 & NI1 & 4.25 & $-$1.030 &  42.4 &  33.3 &  38.5 &  45.8 &  34.0 &  --  &  43.9 \\
6635.15 & NI1 & 4.40 & $-$0.830 &  48.0 &  45.7 &  51.6 &  50.5 &  41.4 &  40.9 &  56.9 \\
6643.64 & NI1 & 1.67 & $-$1.980 & 164.8 & 150.1 & 177.2 & 180.8 & 144.0 & 149.1 & 165.1 \\
5119.12 & Y2  & 0.99 & $-$1.370 &  52.6 &  44.1 &  60.5 &  63.9 &  46.1 &  44.0 &  63.5 \\
5289.82 & Y2  & 1.03 & $-$1.870 &  22.8 &  21.8 &  37.0 &  31.0 &  17.5 &  24.7 &  30.8 \\
5330.58 & CE2 & 0.87 & $-$0.280 &  28.0 &  22.7 &  39.6 &  35.9 &  20.7 &  25.7 &  31.3 \\
6043.39 & CE2 & 1.21 & $-$0.340 &  15.4 &  13.0 &  18.3 &  21.0 &   8.2 &   6.3 &  18.2 \\
6645.11 & EU2 & 1.37 &  0.170 &  41.1 &  31.3 &  36.5 &  38.6 &  25.4 &  33.8 &  42.2 \\
\hline
\end{longtable}
\begin{longtable}{cccccccccc}
\caption{\label{tab:LE2} Equivalent widths of the lines used in the abundance analysis
of the stars NGC 3532\_100, 122, 596, 670, NGC 3680\_13, NGC 5822\_1, 201, and 240. 
Lines with equivalent widths smaller than 10 m$\AA$ and larger than 150 m$\AA$ were 
not used.} \\ 
\hline\hline
$\lambda$ (\AA) & Element & 3532\_100 & 3532\_122 & 3532\_596 & 
3532\_670 & 3680\_13 & 5822\_1 & 5822\_201 & 5822\_240 \\
\hline  
\endfirsthead
\caption{continued.} \\
\hline\hline
$\lambda$ (\AA) & Element  & 3532\_100 & 3532\_122 & 3532\_596 & 
3532\_670 & 3680\_13 & 5822\_1 & 5822\_201 & 5822\_240 \\
\hline
\endhead
\hline
\endfoot
6154.22 & NA1 & 108.8 &  81.1 &  87.7 & 133.8 &  91.5 & 111.5 &  78.5 & 113.2 \\ 
6160.75 & NA1 & 120.2 &  99.2 & 106.5 & 148.9 & 109.5 & 125.9 &  97.3 & 126.9 \\ 
5528.42 & MG1 & 250.4 & 234.8 & 235.1 & 277.5 & 250.1 & 254.5 & 226.3 & 267.9 \\ 
5711.09 & MG1 & 146.3 & 129.0 & 136.5 & 162.6 & 129.3 & 152.9 & 131.4 & 151.5 \\ 
5772.15 & SI1 &  89.2 &  82.2 &  88.0 &  83.6 &  74.5 &  79.5 &  77.4 &  77.8 \\ 
6125.03 & SI1 &  60.0 &  48.8 &  54.0 &   --  &  51.5 &  59.8 &  49.0 &  61.1 \\ 
6131.58 & SI1 &  45.7 &  48.1 &  45.5 &  41.9 &  34.2 &  35.9 &  35.1 &  35.3 \\ 
6131.86 & SI1 &  43.1 &  25.1 &  36.0 &  46.6 &  33.9 &  41.2 &  37.8 &  41.3 \\ 
6142.53 & SI1 &  50.0 &  43.7 &  49.5 &  45.4 &  43.1 &  43.5 &  46.1 &  43.0 \\ 
6145.08 & SI1 &  60.2 &  51.4 &  54.3 &  54.0 &  47.2 &  47.7 &  49.2 &  47.5 \\ 
6155.14 & SI1 & 106.6 &  93.0 & 102.0 & 105.8 &  94.5 &  95.2 &  95.8 &  94.2 \\ 
5867.57 & CA1 &  53.3 &  50.8 &  51.9 &  80.6 &  58.6 &  69.4 &  46.4 &  71.8 \\ 
6122.23 & CA1 & 228.0 & 222.8 & 207.1 & 284.6 & 232.8 & 248.3 & 207.2 & 257.7 \\ 
6156.03 & CA1 &  33.1 &  35.5 &  37.7 &  66.0 &  38.5 &  45.4 &  23.6 &  47.1 \\ 
6161.29 & CA1 &  --   & 110.2 & 116.1 & 169.9 &  --   & 147.4 & 106.4 &  --   \\ 
6166.44 & CA1 & 120.3 & 101.0 & 107.6 & 145.5 & 118.2 & 127.6 & 100.0 & 130.1 \\ 
6169.04 & CA1 & 145.7 & 122.7 & 129.3 & 174.2 & 143.2 & 153.9 & 126.0 & 157.1 \\ 
6169.56 & CA1 & 159.9 & 146.3 & 147.9 & 188.9 & 159.4 & 168.3 & 121.5 & 170.5 \\ 
6493.78 & CA1 & 181.2 & 170.9 & 166.7 & 207.4 & 172.5 & 190.2 & 158.5 & 191.1 \\ 
6499.65 & CA1 & 148.1 & 127.0 & 132.1 & 177.0 & 138.6 & 154.9 & 124.5 & 154.6 \\ 
5318.34 & SC2 &  51.2 &  43.7 &  55.5 &  66.9 &  43.4 &  54.1 &  41.4 &  53.7 \\ 
5334.22 & SC2 &  26.4 &  16.2 &  25.6 &  48.5 &  27.8 &  36.1 &  18.8 &  36.0 \\ 
5145.47 & TI1 & 107.3 &  86.5 &  90.6 & 154.5 & 108.2 & 126.3 &  85.9 & 127.9 \\ 
5295.78 & TI1 &  67.6 &  46.4 &  55.0 & 111.2 &  76.9 &  91.7 &  53.2 &  95.2 \\ 
5299.98 & TI1 &  --   &  31.7 &  47.2 &   --  &   --  &  --   &   --  &   --  \\ 
5338.33 & TI1 &   --  &  33.7 &  38.6 &   --  &   --  &  --   &   --  &   --  \\ 
5351.07 & TI1 &  42.4 &  34.0 &  38.3 &  68.8 &  44.1 &  51.5 &  31.7 &  56.7 \\ 
5766.33 & TI1 &  34.2 &  25.7 &  23.2 &  55.4 &  39.8 &  46.1 &  28.0 &  46.1 \\ 
6121.01 & TI1 &  31.3 &  19.2 &  17.7 &  70.7 &  38.4 &  55.0 &  20.4 &  57.0 \\ 
6126.22 & TI1 &  96.8 &  64.8 &  70.8 & 145.3 &  99.1 & 120.9 &  70.5 & 123.7 \\ 
6497.68 & TI1 &  40.1 &  11.0 &  20.5 &  85.8 &  44.5 &  65.1 &  20.4 &  66.4 \\ 
5846.27 & V1  &  22.3 &  12.2 &  16.0 &  54.2 &  25.4 &  35.0 &  19.4 &  34.1 \\ 
6002.65 & V1  &  33.2 &   --  &  18.8 &  74.6 &  35.4 &  52.9 &  18.8 &  56.6 \\ 
6039.69 & V1  &  74.6 &  45.8 &  48.6 & 121.5 &  82.4 &  99.0 &  48.8 & 104.3 \\ 
6111.65 & V1  &  74.1 &  45.6 &  47.0 & 145.6 &  95.9 & 118.2 &  49.1 & 125.4 \\ 
6119.53 & V1  &  88.3 &  69.0 &  72.9 & 135.6 &  93.0 & 115.2 &  64.2 & 123.0 \\ 
6135.37 & V1  &  72.2 &  50.3 &  47.6 & 128.5 &  84.7 & 105.3 &  47.8 & 109.8 \\ 
6150.15 & V1  &  86.0 &  56.6 &  59.8 & 162.8 & 105.7 & 130.8 &  54.4 &   --  \\ 
6504.19 & V1  &  63.1 &  42.2 &  40.2 & 105.0 &  64.7 &  84.5 &  40.4 &  88.3 \\ 
5303.22 & V2  &  32.3 &  23.1 &  34.8 &  50.7 &  25.5 &  39.4 &  23.3 &  38.8 \\ 
6028.28 & V2  &  18.9 &   9.1 &  13.9 &  25.2 &  15.9 &  20.0 &  11.6 &  19.0 \\ 
5122.12 & CR1 &  80.1 &  61.1 &  55.6 &  --   &  --   &   --  &  59.6 &  --  \\ 
5296.69 & CR1 & 174.4 & 149.0 & 155.6 & 231.1 & 170.3 & 191.8 & 144.4 & 200.7 \\ 
5300.75 & CR1 & 129.4 & 108.3 & 116.6 & 164.4 & 127.4 & 144.1 & 105.0 & 151.3 \\ 
5304.18 & CR1 &  41.4 &  35.8 &  39.4 &  74.3 &  46.5 &  48.5 &  33.0 &  60.8 \\ 
5312.88 & CR1 &  52.0 &  46.1 &  46.4 &  73.2 &  52.4 &  59.9 &  43.2 &  64.0 \\ 
5318.78 & CR1 &  42.6 &  24.4 &  38.6 &  80.2 &  49.8 &  58.6 &  34.3 &  64.6 \\ 
5329.12 & CR1 & 119.8 & 115.6 &   --  & 174.6 & 128.5 & 140.2 & 109.7 &  --   \\ 
5340.44 & CR1 &  43.7 &  20.1 &  39.0 &  64.8 &  51.1 &  53.6 &  34.7 &  56.3 \\ 
5348.32 & CR1 & 172.3 & 137.9 & 153.4 & 223.8 & 169.7 & 191.1 & 145.8 & 198.8 \\ 
5783.07 & CR1 &  72.9 &  62.4 &  66.3 &  97.1 &  75.0 &  86.3 &  62.3 &  88.2 \\ 
5783.87 & CR1 &  --   &   --  &   --  &  --   &  --   &   --  &   --  &  --   \\ 
5787.99 & CR1 &  86.1 &  76.2 &  82.0 & 104.5 &  86.5 &  92.2 &  75.7 &  97.9 \\ 
5788.39 & CR1 &  30.6 &  25.7 &   --  &  61.6 &  34.3 &  49.6 &  23.8 &  47.3 \\ 
5844.61 & CR1 &  32.3 &  18.7 &  17.7 &  50.4 &  29.8 &  39.5 &  20.1 &  39.9 \\ 
5863.96 & CR1 &   --  &   --  &   --  &  33.2 &  24.2 &  30.6 &   --  &  29.8 \\ 
6135.78 & CR1 &  40.7 &  31.9 &  32.7 &  54.0 &  38.7 &  47.0 &  34.2 &  46.9 \\ 
6501.21 & CR1 &  46.3 &  22.4 &  28.6 &  86.8 &  49.2 &  65.9 &  26.6 &   --  \\ 
6630.02 & CR1 &   --  &   --  &   --  & 116.4 &   --  &  91.6 &   --  &   --  \\ 
5305.87 & CR2 &  57.5 &  44.4 &  58.8 &  51.0 &  40.2 &  47.4 &  47.7 &  45.2 \\ 
5310.70 & CR2 &  31.6 &  27.3 &  38.0 &  32.5 &  24.2 &  22.4 &  27.9 &  28.2 \\ 
5313.59 & CR2 &  64.3 &  62.6 &  68.7 &  78.4 &  51.2 &  59.3 &  59.5 &  58.1 \\ 
5334.88 & CR2 &  63.0 &  54.8 &  67.7 &  59.6 &  53.1 &  51.6 &  54.2 &  48.6 \\ 
5133.69 & FE1 & 201.0 & 187.0 & 188.2 & 230.0 & 202.3 & 206.2 & 179.3 & 210.9 \\
5141.75 & FE1 & 161.4 & 136.9 & 144.6 & 194.8 & 154.2 & 167.6 & 134.0 & 171.3 \\
5143.73 & FE1 & 101.7 &  75.1 &  84.5 &  --   &  --   &  --   &  76.4 &  --   \\
5293.97 & FE1 &  70.0 &  56.9 &  68.1 &  85.1 &  71.4 &  76.7 &  60.3 &  78.7 \\
5294.55 & FE1 &  48.9 &  37.4 &  44.4 &  74.2 &  51.0 &  58.8 &  39.6 &  59.5 \\
5295.32 & FE1 &  64.9 &  51.4 &  62.2 &  75.5 &  59.0 &  64.7 &  55.2 &  64.4 \\
5307.36 & FE1 & 172.5 & 144.7 & 158.5 & 199.7 & 160.7 & 178.7 & 139.6 & 180.2 \\
5315.07 & FE1 &  85.3 &  62.7 &  --   &   --  &  72.3 &  --   &  67.0 &   --  \\
5320.05 & FE1 &  61.6 &  40.5 &  47.0 &  74.2 &  58.9 &  62.5 &  41.4 &  63.4 \\
5321.11 & FE1 &  --   &   --  &  73.9 &  94.7 &  --   &   --  &  68.4 &  81.2 \\
5322.05 & FE1 & 126.4 &   --  & 119.3 & 155.7 & 114.4 & 132.5 & 104.1 & 134.0 \\
5326.79 & FE1 &  43.2 &  36.3 &  39.5 &  --   &  --   &  51.3 &  35.0 &  --   \\
5339.94 & FE1 & 168.0 & 185.2 & 173.6 & 207.6 & 181.7 & 190.0 & 166.1 & 199.2 \\
5358.10 & FE1 &  --   &  --   &  --   &  --   &  --   &  --   &  --   &  --   \\
5367.47 & FE1 & 158.2 & 153.2 & --    & 163.5 & 150.2 & 154.9 &  --   & 156.3 \\
5369.97 & FE1 & 177.5 &  --   &  --   & 191.8 &  --   & 177.3 &  --   & 179.8 \\
5568.81 & FE1 &  44.8 &  31.0 &  35.3 &  57.0 &  43.8 &  53.8 &  31.1 &  48.6 \\
5759.27 & FE1 &  22.4 &  16.3 &  17.1 &  31.3 &  22.9 &  26.1 &  18.1 &  25.2 \\
5760.35 & FE1 &  64.0 &  55.4 &  64.3 &  74.1 &  64.2 &  68.3 &  48.0 &  67.7 \\
5775.09 & FE1 &  --   &  --   &  --   &  --   &  --   &  --   &  --   &   --  \\
5778.47 & FE1 &  78.1 &  54.5 &  63.3 &  99.6 &  75.7 &  88.6 &  62.4 &  88.3 \\
5784.69 & FE1 &   --  &  54.3 &  68.4 &  90.2 &  74.9 &  83.4 &  61.6 &  87.3 \\
5838.42 & FE1 &  59.5 &  --   &  53.0 &  74.8 &  53.6 &  62.1 &  46.3 &  62.7 \\
5849.70 & FE1 &  32.9 &  22.2 &  23.8 &  47.1 &  34.9 &  38.8 &  24.1 &  40.4 \\
5852.19 & FE1 &  86.9 &  68.3 &  74.2 &  --   &   --  &  --   &  73.5 &  --   \\
5853.18 & FE1 &  59.5 &  35.8 &  42.1 &  89.2 &  60.1 &  74.1 &  38.9 &  75.8 \\
5855.09 & FE1 &  52.5 &  39.2 &  44.2 &  60.0 &  46.0 &  54.0 &  45.4 &  54.8 \\
5856.08 & FE1 &   --  &   --  &  --   &  --   &   --  &  --   &  66.3 &  --   \\
5858.77 & FE1 &  47.8 &  33.5 &  38.6 &  60.9 &  39.5 &  46.8 &  32.7 &  47.7 \\
5859.61 & FE1 &   --  &   --  &  --   &  --   &   --  &  --   &   --  &   --  \\
5862.36 & FE1 &   --  &   --  &  --   &  --   &   --  &  --   &  --   &   --  \\
6003.03 & FE1 & 129.7 & 116.0 & 120.3 & 138.8 & 120.2 & 128.9 & 111.8 & 127.7 \\
6007.96 & FE1 &  --   &  --   &  --   &  --   &  --   &  --   &  --   &  --   \\
6008.58 & FE1 &  --   &  --   &  --   &  --   &  --   &  --   &  --   &  --   \\
6015.25 & FE1 &  42.1 &  21.5 &  24.6 &  67.5 &  37.0 &  52.9 &  23.2 &  54.2 \\
6019.36 & FE1 &  32.8 &  16.1 &  19.4 &  50.4 &  31.6 &  38.2 &  20.4 &  39.9 \\
6024.07 & FE1 & 144.1 & 125.8 & 130.2 & 150.7 & 131.3 & 138.5 & 127.0 & 136.7 \\
6027.06 & FE1 & 111.5 &  96.4 & 106.7 & 120.2 &  98.6 & 110.4 &  94.8 & 109.5 \\
6034.04 & FE1 &  --   &  26.9 &  18.7 &  --   &  31.3 &  41.7 &  20.3 &  40.9 \\
6035.34 & FE1 &  25.7 &  14.9 &  16.6 &  34.8 &  24.4 &  31.0 &  19.3 &  31.0 \\
6054.10 & FE1 &  31.1 &  19.2 &  24.7 &  42.9 &  29.2 &  34.9 &  21.8 &  33.4 \\
6120.25 & FE1 &  56.4 &  33.8 &  37.4 &  96.9 &  59.1 &  78.5 &  35.5 &  79.9 \\
6151.62 & FE1 & 118.5 &  93.4 & 101.7 & 146.5 & 111.5 & 129.0 &  92.8 & 128.0 \\
6157.73 & FE1 &   --  & 110.1 & 116.2 & 157.3 & 114.8 &  --   & 100.9 &  --   \\
6165.37 & FE1 &  91.3 &  73.6 &  81.1 & 104.6 &  83.9 &  88.6 &  71.5 &  90.2 \\
6173.34 & FE1 & 148.2 & 122.3 & 133.0 & 180.2 & 138.1 & 158.0 & 117.1 & 156.4 \\
6475.63 & FE1 & 135.0 & 106.0 & 106.2 & 165.1 & 108.5 &  --   & 102.9 &  --   \\
6481.87 & FE1 & 142.2 & 115.9 & 120.4 & 172.5 & 125.2 & 149.0 & 111.1 & 150.2 \\
6483.94 & FE1 &  --   &  --   &   --  &  --   &   --  &  --   &   --  &   --  \\
6495.74 & FE1 &  73.8 &  74.6 &  72.0 &  95.4 &  74.5 &  --   &  67.2 &  88.5 \\
6496.47 & FE1 & 105.6 &  80.3 &  90.0 & 111.5 &  95.6 & 107.1 &  91.1 & 105.0 \\
6498.95 & FE1 & 143.5 & 115.7 & 120.0 & 202.5 & 131.8 & 164.5 & 106.4 & 160.8 \\
6627.56 & FE1 &  68.8 &  49.6 &  53.3 &  76.9 &  59.3 &  68.4 &  54.4 &  68.6 \\
6633.42 & FE1 &   --  &  34.3 &  --   &  --   &  52.3 &   --  &   --  &  --   \\
6633.76 & FE1 & 100.5 &  --   & 102.0 & 110.8 &  95.6 & 101.3 &  92.9 & 102.4 \\
6646.98 & FE1 &  66.6 &  --   &  49.5 &   --  &  54.7 &  76.8 &  41.4 &  79.9 \\
6648.08 & FE1 &   --  &  --   &   --  &   --  &  --   &   --  &  --   &   --  \\
5132.67 & FE2 &  64.0 &  --   &  63.4 &  59.6 &  46.4 &  53.5 &  52.9 &  50.8 \\
5256.94 & FE2 &  59.1 &  --   &  55.6 &  --   &  49.7 &   --  &  49.7 &   --  \\
5264.81 & FE2 &  82.8 &  81.4 &  84.4 &  69.6 &  62.7 &  70.1 &   --  &  66.3 \\
5325.56 & FE2 &  81.3 &  86.2 &  91.4 &  76.0 &  65.0 &  63.1 &  69.4 &  64.6 \\
5414.08 & FE2 &  56.2 &  52.2 &  57.1 &  48.1 &  39.6 &  48.1 &  49.5 &  43.3 \\
5425.26 & FE2 &   --  &  68.7 &  74.8 &  60.2 &  51.9 &  59.4 &  65.8 &  57.5 \\
6084.10 & FE2 &  54.7 &  50.3 &  55.7 &  49.0 &  38.6 &  45.8 &  43.3 &  43.2 \\
6113.33 & FE2 &   --  &  33.9 &  42.9 &   --  &  24.6 &  30.8 &  33.4 &  30.8 \\
6129.70 & FE2 &  --   &   --  &  --   &   --  &   --  &  --   &  --   &  --   \\
6149.24 & FE2 &  66.3 &  64.7 &  68.2 &  55.9 &  46.6 &  48.8 &  55.8 &  48.9 \\
6247.56 & FE2 &  86.5 &  89.0 &  87.7 &  62.9 &  57.7 &  63.9 &  75.4 &  58.6 \\
6369.46 & FE2 &  48.6 &  41.1 &  --   &  37.8 &  33.2 &  40.4 &  43.0 &  --   \\
6416.93 & FE2 &  69.7 &  59.7 &  68.9 &  60.3 &  53.8 &   --  &  62.5 &  57.5 \\
6456.39 & FE2 &  96.2 & 103.6 & 101.6 &  74.2 &  69.0 &  78.3 &  89.7 &  69.9 \\
5301.04 & CO1 &  89.1 &  59.7 &  71.7 &  --   &  92.9 & 104.1 &  70.0 & 106.3 \\ 
5325.28 & CO1 &  28.6 &  18.4 &  25.4 &  40.7 &  31.8 &  35.1 &  25.8 &  35.8 \\ 
5342.70 & CO1 &  60.2 &  52.7 &  51.6 &  75.3 &  58.0 &  65.7 &  51.6 &  68.4 \\ 
5352.05 & CO1 &  69.4 &  57.6 &  54.6 &  85.0 &  70.9 &  76.3 &  55.2 &  76.3 \\ 
5359.20 & CO1 &  25.5 &   --  &   --  &  30.5 &  25.6 &  27.1 &  18.1 &  27.2 \\ 
5369.59 & CO1 &   --  &  95.5 &  99.4 &   --  &  --   & 131.6 &  --   &  --   \\ 
6117.00 & CO1 &  46.3 &  31.1 &  33.5 &  84.0 &  53.4 &  68.6 &  30.1 &  68.4 \\ 
6490.34 & CO1 &  44.0 &  26.5 &  35.5 &  85.7 &  44.6 &  38.0 &  22.9 &  53.2 \\ 
6632.47 & CO1 &  52.9 &  35.5 &  36.4 &  86.0 &  54.6 &  68.2 &  35.5 &  69.8 \\ 
5137.08 & NI1 & 166.6 & 152.7 & 157.0 & 184.7 & 169.4 & 167.4 & 150.3 & 173.8 \\ 
5593.74 & NI1 &  76.9 &  66.5 &  73.5 &  85.0 &  68.9 &  74.7 &  62.5 &  75.0 \\ 
5760.83 & NI1 &  72.5 &  66.2 &  71.7 &  96.0 &  67.0 &  72.9 &  60.1 &  75.5 \\ 
5847.01 & NI1 &  91.0 &  67.0 &  73.2 & 118.5 &  85.8 & 100.5 &  69.1 & 105.3 \\ 
6007.31 & NI1 &  88.1 &  66.2 &  69.3 & 107.9 &  77.2 &  94.6 &  64.9 &  95.5 \\ 
6053.68 & NI1 &  54.9 &  44.1 &  45.1 &  62.6 &  48.4 &  50.5 &  41.3 &  52.0 \\ 
6111.06 & NI1 &  69.8 &  53.9 &  59.4 &  75.0 &  60.5 &  68.0 &  60.3 &  67.3 \\ 
6128.99 & NI1 &  97.7 &  68.1 &  72.1 & 127.6 &  86.8 & 106.3 &  70.8 & 106.1 \\ 
6130.13 & NI1 &  47.7 &  39.3 &  40.1 &  59.4 &  42.6 &  49.3 &  35.8 &  43.7 \\ 
6635.15 & NI1 &  57.9 &  42.1 &  51.8 &  66.9 &  46.8 &  56.3 &  47.5 &  55.4 \\ 
6643.64 & NI1 & 181.0 & 147.3 & 156.6 & 213.7 & 170.6 & 193.0 & 151.6 & 195.4 \\ 
5119.12 & Y2  &  67.9 &  52.7 &  56.1 &  80.2 &  58.8 &  71.8 &  55.9 &  69.3 \\ 
5289.82 & Y2  &  36.6 &  21.5 &  33.5 &  49.2 &  22.6 &  38.3 &  23.2 &  37.4 \\ 
5330.58 & CE2 &  33.9 &  23.8 &  31.2 &  64.0 &  31.4 &  45.8 &  24.7 &  48.7 \\ 
6043.39 & CE2 &  27.4 &  14.0 &  15.6 &  37.1 &  18.9 &  22.1 &  13.4 &  25.5 \\ 
6645.11 & EU2 &  45.7 &  28.5 &  30.8 &  51.8 &  31.9 &  43.8 &  33.2 &  41.4 \\
\hline
\end{longtable}
\begin{longtable}{ccccccccccc}
\caption{\label{tab:LE3} Equivalent widths of the lines used in the abundance analysis
of the stars NGC 5822\_316, 443, NGC 6134\_30, 99, 202, NGC 6181\_3, 4, NGC 6633\_78, 
and 100. Lines with equivalent widths smaller than 10 m$\AA$ and larger than 150 m$\AA$ 
were not used.} \\ 
\hline\hline
$\lambda$ (\AA) & Elem. & 5822\_316 & 5822\_443 & 6134\_30 & 6134\_99 & 
6134\_202 & 6181\_1 & 6181\_4 & 6633\_78 & 6633\_100 \\
\hline  
\endfirsthead
\caption{continued.} \\
\hline\hline
$\lambda$ (\AA) & Elem. & 5822\_316 & 5822\_443 & 6134\_30 & 6134\_99 & 
6134\_202 & 6181\_1 & 6181\_4 & 6633\_78 & 6633\_100 \\
\hline
\endhead
\hline
\endfoot
6154.22 & NA1 &  73.8 &  95.7 &  94.6 & 101.5 & 105.4 &  91.8 &  86.7 & 125.5 &  80.8 \\
6160.75 & NA1 &  98.9 & 114.6 & 113.2 & 117.0 & 121.4 & 110.0 & 106.1 & 138.3 & 102.1 \\
5528.42 & MG1 & 232.4 & 246.4 & 246.4 & 256.5 & 256.7 & 238.8 & 243.0 & 271.5 & 234.4 \\
5711.09 & MG1 & 136.7 & 146.3 & 145.7 & 145.4 & 147.5 & 142.8 & 144.8 & 160.4 & 136.7 \\
5772.15 & SI1 &   --  &  80.4 &  86.9 &  85.2 &  84.6 &  85.8 &  89.8 &  76.8 &  83.1 \\
6125.03 & SI1 &  47.6 &  54.8 &  53.8 &  49.2 &  55.5 &  58.6 &  60.4 &   --  &  49.6 \\
6131.58 & SI1 &  34.6 &  37.7 &  41.8 &  43.3 &  37.2 &  42.4 &  46.0 &  40.4 &  39.4 \\
6131.86 & SI1 &  39.7 &  43.5 &  39.8 &  43.4 &  40.3 &  44.8 &  42.5 &  42.5 &  38.1 \\
6142.53 & SI1 &  46.6 &  46.3 &  50.8 &  47.7 &  46.6 &  51.0 &  56.3 &  43.3 &  48.0 \\
6145.08 & SI1 &  49.8 &  49.9 &  54.3 &  55.4 &  48.8 &  58.9 &  62.8 &  49.7 &  51.4 \\
6155.14 & SI1 &  97.2 &  97.2 & 106.3 & 103.7 &  97.7 & 105.2 & 109.5 &  99.5 & 101.1 \\
5867.57 & CA1 &  50.7 &  64.3 &  56.9 &  61.8 &  65.5 &  55.2 &  55.6 &  73.8 &  51.7 \\
6122.23 & CA1 & 205.2 & 234.1 & 217.8 & 228.2 & 245.7 & 223.0 & 225.4 & 280.1 & 209.5 \\
6156.03 & CA1 &  24.1 &  34.6 &  26.5 &  34.4 &  41.6 &  26.7 &  32.8 &  53.1 &  24.3 \\
6161.29 & CA1 &  --   & 131.7 & 121.1 & 120.1 & 136.3 & 120.9 & 123.4 & 165.3 &  --   \\
6166.44 & CA1 & 103.0 & 117.9 & 110.7 & 118.3 & 122.6 & 112.2 & 115.8 & 141.1 & 105.6 \\
6169.04 & CA1 & 125.2 & 142.6 & 138.9 & 143.2 & 150.5 & 136.5 & 137.9 & 172.7 & 131.1 \\
6169.56 & CA1 & 140.4 & 158.4 & 153.3 & 160.8 & 164.6 & 154.1 & 158.0 & 182.6 & 147.1 \\
6493.78 & CA1 & 154.7 & 177.7 & 171.2 & 175.9 & 181.1 & 176.4 & 179.7 & 203.3 & 164.6 \\
6499.65 & CA1 & 120.2 & 143.1 & 133.5 & 136.9 & 148.4 & 138.9 & 135.9 & 165.8 & 127.4 \\
5318.34 & SC2 &  42.2 &  51.6 &  47.0 &  52.5 &  47.4 &  52.1 &  55.6 &  66.0 &  49.5 \\
5334.22 & SC2 &  10.5 &  32.3 &  30.5 &  33.5 &  31.7 &  27.3 &  26.2 &  45.9 &  26.9 \\
5145.47 & TI1 &  86.8 & 111.0 &  91.1 & 107.5 & 117.5 &  93.2 &  91.8 & 141.3 &  92.1 \\
5295.78 & TI1 &  50.8 &  77.1 &  59.9 &  72.6 &  78.5 &  56.2 &  58.0 & 108.7 &  57.5 \\
5299.98 & TI1 &  43.3 &   --  &  55.0 &   --  &   --  &   --  &  47.0 &  --   &  47.4 \\
5338.33 & TI1 &  35.3 &   --  &   --  &   --  &   --  &   --  &  37.8 &   --  &  39.3 \\
5351.07 & TI1 &  35.0 &  46.8 &  35.8 &  40.7 &  48.5 &  38.2 &  38.0 &  66.4 &  38.0 \\
5766.33 & TI1 &  27.1 &  35.9 &  36.4 &  34.5 &  41.6 &  29.4 &  29.0 &  54.7 &  27.6 \\
6121.01 & TI1 &  19.4 &  41.2 &  25.9 &  37.7 &  46.9 &  25.5 &  26.8 &  75.3 &  22.8 \\
6126.22 & TI1 &  67.9 & 101.0 &  79.6 &  92.3 & 104.6 &  81.6 &  81.1 & 140.7 &  73.9 \\
6497.68 & TI1 &  22.1 &  45.5 &  27.9 &  42.4 &  52.5 &  30.4 &  21.4 &  80.0 &  23.5 \\
5846.27 & V1  &  16.7 &  28.6 &  18.4 &  32.9 &  27.9 &  17.4 &  18.8 &  45.7 &  14.2 \\
6002.65 & V1  &   --  &  38.3 &   --  &  33.5 &  45.3 &   --  &  19.8 &  72.2 &   --  \\
6039.69 & V1  &  51.5 &  81.2 &  63.0 &  79.1 &  91.0 &  58.6 &  55.1 & 121.4 &  55.6 \\
6111.65 & V1  &  47.8 &  92.9 &  64.3 &  80.0 & 105.1 &  58.5 &  56.2 & 142.8 &  52.1 \\
6119.53 & V1  &  66.8 &  96.0 &  77.4 &  98.7 &  97.4 &  79.6 &  79.6 & 133.1 &  73.1 \\
6135.37 & V1  &  46.3 &  83.6 &  59.2 &  75.0 &  92.8 &  55.9 &  58.1 & 123.4 &  51.5 \\
6150.15 & V1  &  52.2 & 101.2 &  58.0 &  91.9 & 116.8 &  67.5 &  63.9 & 155.8 &  59.1 \\
6504.19 & V1  &  39.3 &  66.2 &  54.0 &  61.2 &  73.6 &  50.0 &  44.5 &  99.2 &  43.7 \\
5303.22 & V2  &  22.4 &  34.0 &  25.8 &  31.6 &  26.9 &  31.7 &  34.2 &  49.0 &  32.4 \\
6028.28 & V2  &  14.5 &  18.5 &  17.9 &  22.1 &  19.5 &  18.8 &  16.4 &  27.0 &  19.3 \\
5122.12 & CR1 &  55.6 &  86.9 &  67.0 &  80.9 &  94.8 &  62.4 &  65.8 &  --   &  59.9 \\
5296.69 & CR1 & 146.2 & 175.3 & 155.8 & 169.7 & 182.7 & 162.5 & 166.3 & 226.4 & 155.2 \\
5300.75 & CR1 & 104.8 & 129.9 & 117.5 & 124.0 & 134.0 & 118.9 & 118.2 & 158.5 & 112.4 \\
5304.18 & CR1 &  37.5 &  47.4 &  42.1 &  48.0 &  51.8 &  36.5 &  39.8 &  68.5 &  42.8 \\
5312.88 & CR1 &  44.3 &  56.0 &  53.2 &  58.0 &  56.2 &  44.7 &  47.6 &  70.9 &  46.4 \\
5318.78 & CR1 &  37.5 &  50.3 &   --  &  51.0 &  53.5 &  37.4 &  37.0 &  77.7 &  39.9 \\
5329.12 & CR1 & 113.1 & 129.0 &  --   & 131.1 & 134.6 & 113.3 & 117.8 & 158.4 & 117.7 \\
5340.44 & CR1 &  38.1 &  49.5 &  43.9 &  49.4 &  53.6 &  47.3 &  32.4 &  65.8 &  39.5 \\
5348.32 & CR1 & 142.3 & 174.7 & 159.4 &  --   & 177.0 & 162.8 & 162.4 & 212.3 & 152.2 \\
5783.07 & CR1 &  66.7 &  75.8 &  72.1 &  80.2 &  83.8 &  69.0 &  70.9 &  93.5 &  64.6 \\
5783.87 & CR1 &   --  &   --  &   --  &   --  &   --  &   --  &   --  &   --  &   --  \\
5787.99 & CR1 &  77.5 &  90.6 &  85.0 &  93.1 &  95.6 &  78.4 &  87.6 & 100.8 &  81.9 \\
5788.39 & CR1 &  21.5 &  34.8 &  31.6 &  34.6 &  47.2 &  23.6 &  29.8 &  59.8 &  29.0 \\
5844.61 & CR1 &  17.6 &  31.5 &  25.0 &  32.8 &  34.5 &  23.0 &  20.7 &  49.6 &  21.5 \\
5863.96 & CR1 &   --  &  18.9 &   --  &  23.0 &  32.5 &   --  &   --  &  34.1 &   --  \\
6135.78 & CR1 &  31.8 &  42.5 &  41.7 &  41.8 &  44.2 &  37.0 &  35.8 &  55.7 &  32.7 \\
6501.21 & CR1 &  23.9 &   --  &   --  &   --  &   --  &  34.8 &  30.0 &  79.5 &  28.9 \\
6630.02 & CR1 &   --  &   --  &   --  &   --  &   --  &   --  &   --  & 109.7 &   --  \\
5305.87 & CR2 &  49.9 &  49.9 &  48.4 &  55.5 &  39.2 &  58.8 &  57.3 &  53.4 &  56.1 \\
5310.70 & CR2 &  29.5 &  27.3 &  32.4 &  30.0 &  23.5 &  34.5 &  41.5 &  32.3 &  32.2 \\
5313.59 & CR2 &  65.3 &  57.6 &  67.3 &  64.1 &  36.4 &  67.6 &  70.9 &  69.0 &  66.8 \\
5334.88 & CR2 &  58.1 &  61.5 &  58.1 &  59.0 &  55.3 &  67.0 &  69.8 &  60.3 &  65.1 \\
5133.69 & FE1 & 179.1 & 197.9 & 185.2 & 201.7 & 191.4 & 189.1 & 186.1 & 207.8 & 183.7 \\
5141.75 & FE1 & 122.5 & 159.4 & 128.3 & 159.2 & 163.3 & 150.4 & 150.3 & 183.2 & 144.6 \\
5143.73 & FE1 &  79.7 & 103.0 &  84.0 &  --   &  --   &  87.5 &  86.6 &  --   &  85.2 \\
5293.97 & FE1 &  62.1 &  73.6 &  68.2 &  75.2 &  64.0 &  68.1 &  70.6 &  81.9 &  65.3 \\
5294.55 & FE1 &  30.1 &  52.6 &  47.5 &  54.2 &  53.5 &  43.8 &  45.9 &  68.5 &  43.8 \\
5295.32 & FE1 &  57.1 &  61.5 &  59.8 &  65.1 &  61.7 &  59.1 &  61.6 &  73.1 &  62.7 \\
5307.36 & FE1 & 138.5 & 167.0 & 154.0 & 164.8 & 165.9 & 163.1 & 159.1 & 198.4 & 154.9 \\
5315.07 & FE1 &  67.6 &  78.3 &  74.4 &  71.1 &  74.6 &  76.6 &  --   &  --   &  66.5 \\
5320.05 & FE1 &  48.0 &  59.9 &  56.5 &  62.4 &   --  &  54.4 &   --  &  71.0 &  52.9 \\
5321.11 & FE1 &  --   &  --   &  --   &  --   &   --  &  --   &   --  &  90.1 &  74.6 \\
5322.05 & FE1 & 106.0 & 125.6 & 117.9 & 124.8 & 125.5 & 121.0 & 122.1 & 149.7 & 117.9 \\
5326.79 & FE1 &  40.7 &  --   &  42.5 &  --   &  --   &  37.0 &  37.0 & --    &  42.5 \\
5339.94 & FE1 & 160.8 & 185.5 & 163.2 & 175.1 & 184.9 & 181.2 & 181.6 & 202.2 & 171.6 \\
5358.10 & FE1 &  --   &  --   &  --   &  --   &  --   &   --  &  --   &  --   &  --   \\
5367.47 & FE1 &  --   & 154.4 & 155.2 & 160.7 & 155.2 & 157.5 & 163.0 & 157.8 & 152.6 \\
5369.97 & FE1 &  --   & 170.5 &  --   &  --   & 171.0 &  --   &  --   & 181.5 &   --  \\
5568.81 & FE1 &  33.3 &  44.1 &  42.5 &  42.0 &  51.5 &  38.4 &  39.0 &  54.1 &  37.4 \\
5759.27 & FE1 &  17.6 &  24.0 &  22.3 &  22.9 &  25.4 &  19.1 &  18.6 &  27.8 &  20.4 \\
5760.35 & FE1 &  48.3 &  62.1 &  59.6 &  60.4 &  66.5 &  54.8 &  63.9 &  71.8 &  53.4 \\
5775.09 & FE1 &  --   &   --  &  --   &  --   &  --   &   --  &  --   &  --   &  --   \\
5778.47 & FE1 &  65.1 &  79.8 &  72.1 &  79.8 &  82.7 &  71.3 &  70.5 &  99.3 &  67.4 \\
5784.69 & FE1 &  64.4 &  77.3 &  67.9 &  75.9 &  78.5 &  70.2 &  --   &  92.4 &  66.9 \\
5838.42 & FE1 &  42.9 &  55.0 &  56.8 &  57.7 &  58.1 &  51.7 &  53.4 &  68.9 &  48.7 \\
5849.70 & FE1 &  24.4 &  33.6 &  31.7 &  41.2 &  40.2 &  27.0 &  29.9 &  48.0 &  31.5 \\
5852.19 & FE1 &  71.3 &  --   &  78.8 &  --   &  --   &  79.4 &  80.8 &   --  &  73.9 \\
5853.18 & FE1 &  35.1 &  62.4 &  47.0 &  54.4 &  65.1 &  45.6 &  45.2 &  83.7 &  42.5 \\
5855.09 & FE1 &  43.6 &  52.8 &  48.9 &  53.2 &  52.5 &  46.1 &  --   &  58.5 &  44.3 \\
5856.08 & FE1 &  --   &  --   &  --   &  --   &  --   &  --   &  --   &  --   &   --  \\
5858.77 & FE1 &  33.5 &  43.2 &  31.1 &  45.3 &  45.0 &  36.6 &  33.4 &  52.6 &  39.0 \\
5859.61 & FE1 & --    &  --   &  --   &  --   &  --   &  --   &  --   &  --   &   --  \\
5862.36 & FE1 &  --   &  --   &  --   &  --   &  --   &   --  &   --  & --    &   --  \\
6003.03 & FE1 & 114.8 & 124.2 & 120.0 & 122.0 &  --   & 124.6 & 125.0 & 138.6 & 115.3 \\
6007.96 & FE1 &   --  &  --   &   --  & --    &  --   & --    &  --   &  --   &   --  \\
6008.58 & FE1 &   --  &  --   &   --  &  --   &  --   &  --   &   --  & --    &   --  \\
6015.25 & FE1 &  26.6 &  42.4 &  31.8 &  38.2 &  46.8 &  29.6 &  26.2 &  67.1 &  26.2 \\
6019.36 & FE1 &  23.2 &  30.9 &  31.8 &  34.7 &  37.7 &  24.0 &  20.8 &  47.0 &  20.8 \\
6024.07 & FE1 & 129.0 & 134.9 & 139.1 & 136.7 & 137.0 & 137.5 & 144.1 & 147.9 & 129.8 \\
6027.06 & FE1 & 101.2 & 106.5 & 104.2 & 105.3 & 106.9 & 111.7 & 114.0 & 117.1 & 102.6 \\
6034.04 & FE1 &  29.4 &  34.1 &  36.5 &  33.1 &  37.2 &  25.0 &  25.2 &  41.0 &  21.9 \\
6035.34 & FE1 &  20.8 &  22.8 &  24.4 &  25.1 &  30.1 &  21.4 &  18.6 &  36.8 &  19.4 \\
6054.10 & FE1 &  25.1 &  30.0 &  28.2 &  31.8 &  33.8 &  25.7 &  25.8 &  38.5 &  24.4 \\
6120.25 & FE1 &  32.7 &  65.1 &  43.3 &  50.6 &  68.8 &  46.6 &  42.3 &  91.0 &  37.5 \\
6151.62 & FE1 &  96.1 & 116.4 & 102.0 & 108.5 & 116.5 & 111.3 & 113.8 & 139.8 & 100.3 \\
6157.73 & FE1 & 104.4 & 120.2 & 112.0 & 117.4 &  --   &  --   &  --   &  --   & 110.0 \\
6165.37 & FE1 &  75.4 &  84.9 &  82.8 &  90.1 &  87.2 &  85.3 &  89.2 &  98.8 &  77.9 \\
6173.34 & FE1 & 121.0 & 144.9 & 128.2 & 141.2 & 144.4 & 140.0 & 144.2 & 168.9 & 128.3 \\
6475.63 & FE1 & 101.1 & 128.9 & 114.4 & 114.7 & 117.2 & 122.9 & 119.9 & 158.7 & 116.5 \\
6481.87 & FE1 & 111.7 & 139.4 & 118.5 & 129.5 & 135.9 & 133.2 & 133.7 & 164.2 & 117.2 \\
6483.94 & FE1 &    -- &  --   &  --   &   --  & --    &  --   &   --  &   --  &  --   \\
6495.74 & FE1 &  65.7 &  80.8 &  69.8 &  73.1 &  85.2 &  80.9 &  76.8 &  91.3 &  67.1 \\
6496.47 & FE1 &  86.0 & 101.1 &  97.9 &  98.9 & 102.0 &  97.5 &  97.5 & 116.7 & 100.1 \\
6498.95 & FE1 & 103.6 & 143.8 & 112.8 & 131.1 & 140.7 & 131.3 & 130.2 & 179.6 & 116.0 \\
6627.56 & FE1 &  51.3 &  65.1 &  59.6 &  64.7 &  65.1 &  --   &  60.8 &  72.8 &  58.2 \\
6633.42 & FE1 &  50.9 &   --  &  --   &  --   &  --   &  --   &  --   &  --   &  54.8 \\
6633.76 & FE1 &  95.0 &  98.2 &  95.8 &  97.0 &  99.3 & 102.5 &  --   & 104.7 & 100.0 \\
6646.98 & FE1 &  40.8 &  63.7 &  55.6 &  61.4 &  67.9 &  58.8 &  --   &  --   &  --   \\
6648.08 & FE1 & --    &  --   &   --  &  --   & --    &  --   & --    &  --   &  --   \\
5132.67 & FE2 &  52.8 &  53.8 &  53.9 &  56.6 &  55.9 &  61.1 &  65.7 &  51.6 &  55.9 \\
5256.94 & FE2 &  48.6 &  --   &  54.5 &  50.7 &  --   &  59.2 &  63.8 &  --   &  52.6 \\
5264.81 & FE2 &   --  &  71.1 &  75.1 &  70.8 &  67.9 &  87.6 &   --  &  68.0 &  78.0 \\
5325.56 & FE2 &  70.1 &  70.4 &  70.2 &  74.4 &   --  &  85.2 &  90.7 &  71.5 &  80.4 \\
5414.08 & FE2 &  52.1 &  48.3 &  49.6 &  45.8 &  43.5 &  61.3 &  64.5 &  44.5 &  53.0 \\
5425.26 & FE2 &   --  &  62.9 &  65.4 &  63.6 &  59.5 &  76.6 &   --  &  57.7 &  66.8 \\
6084.10 & FE2 &  45.3 &  46.4 &  46.1 &   --  &  --   &  59.4 &  61.7 &  46.7 &  50.5 \\
6113.33 & FE2 &  33.3 &  32.6 &   --  &  37.6 &  --   &  41.5 &    -- &  37.7 &  --   \\
6129.70 & FE2 &   --  &   --  &  --   &  --   &  --   &  --   &  --   &  --   &  --   \\
6149.24 & FE2 &   --  &  51.6 &  56.9 &  57.0 &  47.4 &  66.6 &  74.2 &  50.1 &  59.8 \\
6247.56 & FE2 &  78.2 &  69.9 &   --  &  68.0 &  62.5 &  91.2 &  95.3 &  60.7 &  76.8 \\
6369.46 & FE2 &  44.6 &  42.4 &  44.2 &  43.4 &  35.4 &  51.8 &  55.4 &  38.4 &  43.4 \\
6416.93 & FE2 &  61.7 &  59.6 &   --  &  61.9 &  56.6 &  72.2 &  74.8 &  62.7 &  62.8 \\
6456.39 & FE2 &  89.7 &  84.4 &  91.1 &  89.9 &  79.3 & 103.6 & 106.7 &  71.7 &  90.6 \\
5301.04 & CO1 &  67.9 &  95.0 &  82.7 &  88.6 &  96.8 &  80.7 &  80.9 & 121.5 &  75.7 \\
5325.28 & CO1 &  24.9 &  30.5 &  33.7 &  37.8 &  35.5 &  27.7 &  26.5 &  41.4 &  30.3 \\
5342.70 & CO1 &   --  &  63.6 &  57.9 &  63.7 &  63.8 &  59.3 &  58.8 &  73.9 &  57.7 \\
5352.05 & CO1 &  51.0 &  71.0 &  66.2 &  73.0 &  74.8 &  67.6 &  62.5 &  85.3 &  63.2 \\
5359.20 & CO1 &  18.1 &  24.5 &  --   &  24.1 &   --  &  21.2 &  21.1 &  30.9 &  20.2 \\
5369.59 & CO1 &   --  &   --  &   --  &   --  &  --   & 108.1 &   --  &   --  &  --   \\
6117.00 & CO1 &  28.7 &  54.9 &  39.1 &  48.4 &  60.9 &  42.0 &  41.3 &  82.2 &  33.2 \\
6490.34 & CO1 &  21.6 &  46.0 &  33.9 &  34.5 &  47.5 &  19.1 &  21.2 &  67.3 &  29.5 \\
6632.47 & CO1 &  36.3 &  57.4 &  42.2 &  54.2 &  59.9 &  49.9 &  44.2 &  80.7 &  43.5 \\
5137.08 & NI1 & 152.3 & 172.6 & 147.7 & 161.5 &  --   & 161.0 & 162.3 & 172.9 & 147.8 \\
5593.74 & NI1 &  67.3 &  75.1 &  75.6 &  76.5 &  76.4 &  76.9 &  78.5 &  80.1 &  70.3 \\
5760.83 & NI1 &  59.6 &  72.9 &  68.3 &  72.0 &  74.1 &  68.0 &  79.3 &  87.2 &  62.5 \\
5847.01 & NI1 &  65.8 &  91.7 &  81.1 &  89.3 &  92.0 &  82.9 &  81.7 & 114.7 &  73.5 \\
6007.31 & NI1 &  68.3 &   --  &  75.6 &  80.8 &  87.0 &  78.1 &   --  & 104.5 &  70.6 \\
6053.68 & NI1 &  39.1 &  45.5 &  48.6 &  50.3 &  53.2 &  48.2 &  49.8 &  58.9 &  42.1 \\
6111.06 & NI1 &  55.6 &  68.2 &  67.2 &  68.1 &  70.8 &  67.4 &  66.1 &  73.7 &  61.7 \\
6128.99 & NI1 &  67.4 &  93.1 &  78.0 &  89.4 &  95.0 &  83.5 &  88.8 & 119.0 &  75.7 \\
6130.13 & NI1 &  38.6 &  46.6 &  44.0 &  47.0 &  44.5 &  45.2 &  48.9 &  46.8 &  38.7 \\
6635.15 & NI1 &  49.2 &  56.3 &  57.0 &  58.5 &  57.9 &   --  &  58.2 &  60.0 &  55.3 \\
6643.64 & NI1 & 150.5 & 180.8 & 162.1 & 172.3 & 178.8 & 177.8 & 174.4 & 208.8 & 162.1 \\
5119.12 & Y2  &  52.9 &  60.4 &  51.8 &  58.1 &  57.7 &  63.6 &  67.4 &  71.2 &  56.5 \\
5289.82 & Y2  &  24.2 &  33.4 &  25.6 &  31.2 &  28.0 &  32.4 &  35.1 &  51.8 &  32.7 \\
5330.58 & CE2 &  27.8 &  37.5 &  25.8 &  29.9 &  30.7 &  31.3 &  33.1 &  61.8 &  33.6 \\
6043.39 & CE2 &  14.2 &  21.2 &  17.4 &  20.4 &  19.3 &  20.4 &  18.6 &  29.5 &  14.5 \\
6645.11 & EU2 &  24.0 &  37.8 &  29.8 &  33.9 &  35.2 &  47.6 &  46.0 &  55.4 &  38.1 \\
\hline
\end{longtable}

\begin{longtable}{ccccccccccc}
\caption{\label{tab:LE4} Equivalent widths of the additional \ion{Fe}{ii} lines 
used in the abundance analysis of the stars NGC 2360\_7, 50, 62, NGC 2447\_28, 34, 
and 41, previously analized by \citet{Ham00}. 
Lines with equivalent widths smaller than 10 m$\AA$ and larger than 150 m$\AA$ were 
not used.} \\ 
\hline\hline
$\lambda$ (\AA) & Elem. & $\chi$ (eV) & log gf & 2360\_7 & 2360\_50 & 2360\_62 & 2360\_86 
& 2447\_28 & 2447\_34 & 2447\_41 \\
\hline  
\endfirsthead
\caption{continued.} \\
\hline\hline
$\lambda$ (\AA) & Elem. & $\chi$ (eV) & log gf & 2360\_7 & 2360\_50 & 2360\_62 & 2360\_86 
& 2447\_28 & 2447\_34 & 2447\_41 \\
\hline
\endhead
\hline
\endfoot
5132.67 & FE2 & 2.79 & $-$4.110 & 46.1 & 51.9 & 49.5 & 44.7 & 51.2 & 46.6 & 48.5 \\
5256.94 & FE2 & 2.89 & $-$4.050 &  --  &  --  &  --  &  --  &  --  &  --  &  --  \\
5264.81 & FE2 & 3.23 & $-$3.200 &  --  &  --  & 59.3 & 63.2 &  --  & 76.9 &  --  \\
5325.56 & FE2 & 3.22 & $-$3.160 & 71.7 & 69.7 & 67.2 & 65.0 & 74.6 & 74.1 & 72.0 \\
5414.08 & FE2 & 3.22 & $-$3.650 & 48.8 & 49.0 & 40.6 & 44.3 & 58.8 &  --  & 45.5 \\
5425.26 & FE2 & 3.20 & $-$3.220 & 70.4 & 58.5 & 55.0 &  --  & 70.0 & 62.6 & 61.3 \\
6084.10 & FE2 & 3.20 & $-$3.760 & 49.2 & --   & 34.2 &  --  & 52.3 & 45.5 & 44.1 \\
6113.33 & FE2 & 3.22 & $-$4.110 & 30.6 & 28.5 & 24.9 & 26.9 & 35.6 & 33.6 & 33.5 \\
6129.70 & FE2 & 3.20 & $-$4.601 &  --  &  --  & 14.8 &  --  &  --  &  --  &  -- \\
6149.24 & FE2 & 3.89 & $-$2.700 & 58.0 & 50.9 & 51.4 & 51.7 & 57.4 & 57.5 & 55.1 \\
6247.56 & FE2 & 3.89 & $-$2.310 & 73.0 & 74.6 & 62.6 & 69.8 & 82.1 & 77.9 & 75.9 \\
6369.46 & FE2 & 2.89 & $-$4.150 & 41.1 & 30.5 & 28.6 &  --  & 42.6 & 41.5 & 39.9 \\
6416.93 & FE2 & 3.89 & $-$2.720 & 57.8 & 54.7 & 55.7 &  --  & 62.2 & 57.5 & 60.6 \\
6456.39 & FE2 & 3.90 & $-$2.060 &  --  & 84.4 & 78.3 & 82.7 &  --  & 91.6 & 97.1 \\
\hline
\end{longtable}

\end{document}